\documentclass[a4paper,book,10pt,twoside]{combine}

\usepackage{a4wide, amssymb, epsfig, graphicx, fancyhdr}

\usepackage{amsmath} 
\usepackage{psfrag} 
\usepackage{color}
\usepackage[outercaption,wide]{sidecap} 

\newcommand{\Section}[1]{\section{#1}\vspace{-8pt}}
\newcommand{\Subsection}[1]{\vspace{-4pt}\subsection{#1}\vspace{-8pt}}

\begin{document}
\pagestyle{fancy}
\fancyhead[LO,RE]{B. Hoistadt, T. Pe\~{n}a and C. F. Redmer}
\fancyhead[LE,RO]{PrimeNet Workshop 2010,  Lisbon}
\fancyfoot{}
\fancyfoot[LE,RO]{\thepage}
\newcommand{\authortochead}[1]{%
\coltocauthor{#1}
\fancyhead[LO,RE]{#1}
}
\setlength{\toctitleindent}{1.5em}
\setlength{\tocauthorindent}{3em}

\pagestyle{empty}
\begin{center}
\vspace*{5cm}
\vspace {0.5cm} {\bf \Huge International PrimeNet Workshop}\\

\vspace{2.6cm} {\Large September, 16 -- 18, 2010, Lisbon, Portugal\\}

\vspace{5cm} {\huge Summary of Contributions}
\end{center}

\newpage
\mbox{  }\\
\newpage
\tableofcontents 
\clearpage

\pagestyle{fancy}
\setcounter{page}{1}
\setcounter{chapter}{1}
\section*{Introduction}\label{intro}
\addcontentsline{toc}{part}{Introduction}

This workshop is part of the activities in the project "Study of Strongly Interacting Matter"
(acronym HadronPhysics2), which is an integrating activity of the Seventh Framework Program of EU.
This HP2 project  contains several activities, one of them being the network PrimeNet having the
focus on Meson Physics in Low-Energy QCD.  This network is created to exchange information on
experimental and theoretical ongoing activities on mainly $\eta$ and $\eta^\prime$ physics at
different European accelerator facilities and institutes. 
\vspace{0.75cm}

\noindent Although by the end of the 1970's quantum chromodynamics (QCD) was established as the
theory of the strong interaction, we still have only little understanding of various possible forms
of confined quark states and their decays, and today there is vast area of research on hadrons, in
particular mesons and their interactions, still to be explored. The talks presented at the workshop
showed that experimental collaborations in this field have been crucial for recent advances. In
particular, the study of $\eta$ and $\eta^\prime$ decays is opening up a new era of precision to the
determination of the light quark mass difference, along with information on $\pi\pi$ and $\pi\eta$
interactions. From the theoretical side, the progress achieved through Chiral Perturbation Theory
(ChPT) and Large-N$_{\textrm{c}}$ ChPT may now be tested, while lattice calculations are already
providing the first ground-breaking results. On the other hand, strong and electromagnetic probing
of meson-baryon interactions progressed tremendously with new and powerful detecting techniques.
\vspace{0.75cm}

\noindent The present workshop included the three general topics:
\begin{enumerate}
 \item $\eta$ and $\eta^\prime$ decays from experimental and theoretical perspectives.
 \item Meson production in photo reactions and from light ion collisions.
 \item Interaction of $\eta$ and $\eta^\prime$ with nucleons and nuclei including $\eta$ bound
states.
\end{enumerate}

\noindent The different talks covered the very recent achievements in each field from the
experimental facilities KLOE at DAPHNE, Crystal Ball at MAMI, Crystal Ball and TAPS at Elsa and
\mbox{WASA-at-COSY} as well as from different theory institutes. Viewgraphs from each talk can be
found on
the PrimeNet homepage http://www.fz-juelich.de/ikp/primenet

\vspace{0.75cm}

\noindent The detailed program was arranged by a program committee having the members:
Reinhard~Beck, Johan~Bijnens, Simona~Giovannella, Dieter~Grzonka, Christoph~Hanhart, Bo~H\"{o}istad,
Bastian~Kubis, Andrzej~Kupsc, Hartmut~Machner, Pawel~Moskal, Eulogio~Oset, Micheal~Ostrick, 
Teresa~Pe\~{n}a, Susan~Schadmand. 

\vspace{0.75cm}

\noindent The workshop was held in Sept 16-18, 2010, at the campus of Instituto Superior
T\'{e}cnico, enjoying kind hospitality and support from IST, Lisboa, Portugal.

\vspace{0.75cm}

\noindent Financial support is gratefully acknowledged from the European Commission under the 7th
Framework Programme through the 'Research Infrastructures' action of the 'Capacities' Programme; 
Call: FP7-INFRASTRUCTURES-2008-1, Grant Agreement N. 227431.

\vspace{1.75cm}

{\Large Bo H\"{o}istad, Teresa Pe\~{n}a and Christoph Redmer}
\cleardoublepage

\pagestyle{plain}
\begin{center}
 \vspace*{0.3\textheight} {\Huge\textbf{\boldmath $\eta$ and $\eta^\prime$ Decays}}\\
 \vspace{2cm} {\Large\textbf{from Experimental and Theoretical Perspectives}}
\addcontentsline{toc}{part}{\boldmath $\eta$ and $\eta^\prime$ Decays from Experimental and
Theoretical Perspectives}
\end{center}
\cleardoublepage
\pagestyle{fancy}

\begin{papers}
\coltoctitle{\boldmath Results on $\eta/\eta'$ With KLOE}
\authortochead{S.~Giovannella}
\label{art5}

\setcounter{chapter}{1}
\setcounter{section}{0}
\setlength{\parindent}{0pt}
\setlength{\parskip}{10pt}
\newcommand{\ep}{\mbox{$e^+$}}
\newcommand{\el}{\mbox{$e^-$}}
\newcommand{\pip}{\mbox{$\pi^+$}}
\newcommand{\pim}{\mbox{$\pi^-$}}
\newcommand{\pio}{\mbox{$\pi^0$}}
\newcommand{\etppg}{\mbox{$\eta\to\pi^+\pi^-\gamma\ $}}
\newcommand{\etppp}{\mbox{$\eta\to\pi^+\pi^-\pi^0\ $}}
\newcommand{\etgg}{\mbox{$\eta\to\gamma\gamma\ $}}
\newcommand{\phippp}{\mbox{$\phi\to\pi^+\pi^-\pi^0\ $}}
\newcommand{\phietg}{\mbox{$\phi\to\et\gamma\ $}}
\newcommand{\et}{\mbox{$\eta$}}
\newcommand{\phietgppg}{\mbox{$\phi\to\gamma\et,\et\to\pi^+\pi^-\gamma\ $}}
\maketitle

\Section{Introduction}
From 2000 to 2006, the KLOE experiment collected 2.5 fb$^{-1}$ of \ep\el\ 
collisions at the $\phi$ peak and about 240 pb$^{-1}$ below the $\phi$ 
resonance ($\sqrt{s}=1$ GeV). The whole data set includes $1\times 10^8$
$\eta$'s and $5\times 10^5$ $\eta'$'s produced through the radiative decays
$\phi\to\eta\gamma/\eta'\gamma$ and tagged by means of the monochromatic 
recoil photon.

\Section{\boldmath $\eta\to\pi\pi\pi$}The dynamics of $\eta\to\pi\pi\pi$ is
sensitive to the $u$-$d$ quark 
mass difference. A precise study of this decay can lead to a very 
accurate measurement of $Q^2=(m_s^2-\hat{m}^2)/(m_d^2-m_u^2)$.
At KLOE, the background for both charged and neutral final states is 
at the level of few per mil \cite{KLOE_dalitz_mixed,KLOE_dalitz_neutral}.
The conventional Dalitz plot variables for the \pip\pim\pio\ final 
state are $X\propto T_+ - T_-$ and $Y\propto T_0$, where $T$ is the 
kinetic energy of the pions. The measured distribution is parametrized 
as: $|A(X,Y)|^2 = 1 + aY + bY^2 + cX + dX^2 + eXY + fY^3$.
As expected from $C$ parity conservation, the odd powers of $X$ are 
consistent with zero. 
The other parameters are measured with an accuracy between 0.5 to 10\%.
$C$ parity conservation has been tested also by measuring the
left-right, quadrants and sextants charge asymmetries: all of them 
are consistent with zero at $10^{-3}$ level, thus improving existent 
evaluations obtained combining different experiments.
In the $\eta\to\pio\pio\pio$ decay, the Dalitz plot density is
described by a single parameter $\alpha$: 
$|A(z)|^2\propto 1+2\,\alpha\, z$, where $z$ is related to the 
three pion energies in the $\eta$ rest frame. Photons are paired 
to form \pio's after kinematically constraining the total four-momentum 
to $M_\phi$ to improve the energy resolution.
Three samples, with different efficiency and purity on photon pairing 
have been analyzed. The resulting value for $\alpha$ is:
${\alpha = -0.0301 \pm 
  0.0035_{\rm stat}\,^{+0.0022}_{-0.0035}\,_{\rm syst} }\,$.
%

\Section{\boldmath $\eta\to\pi^+\pi^-\gamma$}
The $\eta \to \pi^+ \pi^- \gamma$ decay provide a good tool to 
investigate the box anomaly, a higher term of Wess-Zumino-Witten 
Lagrangian. The invariant mass of the pions is a good observable 
to disentangle this contribution from other possible resonant ones, 
e.g. from the $\rho$-meson. 
Recently the CLEO collaboration published their results on the ratio
$\Gamma(\eta\to\pi^+\pi^-\gamma)/\Gamma(\eta\to\pi^+\pi^-\pi^0)$, 
which differ more than 3\,$\sigma$'s from old results \cite{BrCLEO}. 
From the analysis of 1.2 fb$^{-1}$, the preliminary KLOE measurement 
of the same ratio is:
$\Gamma(\eta\to\pi^+\pi^-\gamma)/\Gamma(\eta\to\pi^+\pi^-\pi^0) = 
  0.2014\pm 0.0004_{\rm stat}\pm 0.0060_{\rm stat}$. 
The final systematic error, under evaluation, is expected to be less 
than 1\%. Our number is in agreement with the old results and differs 
significantly from recent CLEO measurement.

\Section{\boldmath $\eta\to\pi^+\pi^- e^+ e^- / e^+ e^- e^+ e^-$}
The $\eta \to \pi^+ \pi^- e^+ e^-$ decay allows to probe the structure 
of the $\eta$ meson, to compare the predictions of the branching ratio 
value based on Vector Meson Dominance model and Chiral Perturbation 
Theory and to study CP violation beyond the prediction of the Standard 
Model by measuring the angular asymmetry between pions and electrons 
decay planes \cite{Gao:2002gq}.
The analysis has been performed on 1.73 fb$^{-1}$ \cite{PLBetappee}.
Background contribution is evaluated performing a fit on the sidebands
of the $\pi^+ \pi^- e^+ e^-$ invariant mass with background shapes. 
For the signal estimate the $\eta$ mass region ($[535,555]$ MeV) is
considered, performing the event counting after background subtraction.
The resulting number of signal events, $N_{\pi\pi ee}= 1555 \pm 52$, is 
used to extract the branching ratio:
${\rm BR}(\eta \to \pi^+ \pi^- e^+ e^- \gamma) = 
  (26.8 \pm 0.9_{\rm stat} \pm 0.7_{\rm syst}) \times 10^{-5}$.
The decay plane asymmetry $A_\phi$ is defined as the sign asymmetry of 
the quantity $\sin\phi\cdot\cos\phi$, $\phi$ being the angle between the 
pion and the electron planes in the $\eta$ rest frame. 
It has been evaluated for the events in the signal region after background 
subtraction. The value obtained is:
$A_{\phi} = ( -0.6 \pm 2.5_{\rm stat} 
                     \pm 1.8_{\rm syst}) \times 10^{-2}$.
This is the first measurement of this asymmetry.
The distribution of the $\sin \phi \cos \phi$ variable is shown in the
left panel of Fig.\ \ref{Fig:eta4ch}.

With the same data sample, the $\eta\to e^+ e^- e^+ e^-$ decay has been 
studied. Events with four electrons are selected using time of flight 
information provided by the calorimeter. 
The number of signal events is obtained by fitting the data distribution 
of the four electron invariant mass, M$_{eeee}$, with signal and background
shapes (Fig.\ \ref{Fig:eta4ch}-right).
From the fit we extract $N_{\eta\to e^+ e^- e^+ e^-} = 413 \pm 31$.
This constitutes the first observation of this decay.

\begin{figure}[!t]
 \epsfig{file=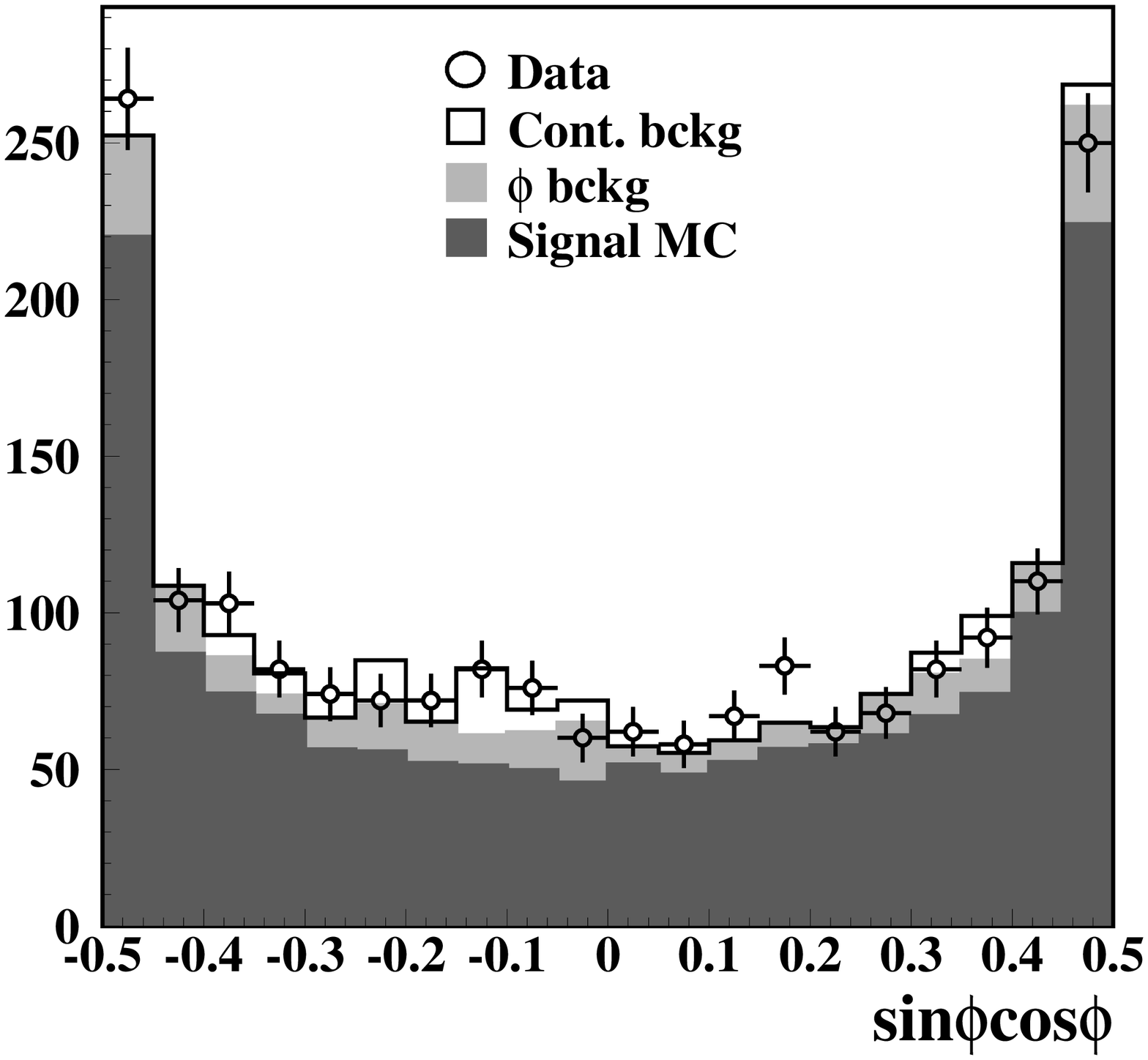,width=0.5\textwidth}
 \epsfig{file=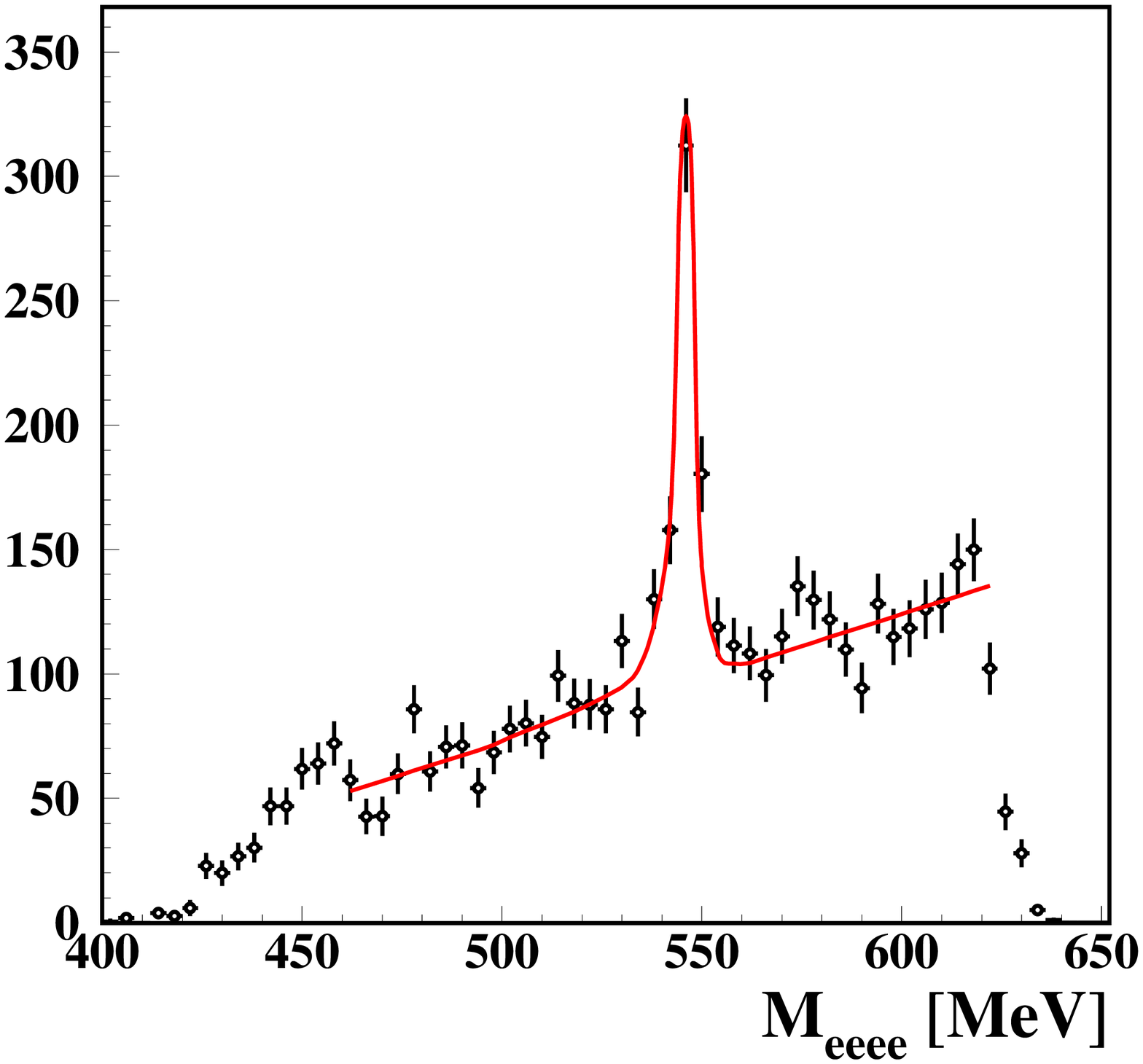,width=0.5\textwidth}
  \caption{Left: angular asymmetry between pions and electrons decay 
    planes for $\eta\to\pi^+\pi^- e^+ e^-$ events. Data (dots) are
    compared with the expected distribution in case of null $A_\phi$ 
    value for the signal.
    Right: fit to the four electron invariant mass for events 
    $\eta\to e^+ e^- e^+ e^-$.}
  \label{Fig:eta4ch}
\end{figure}

\Section{\boldmath Search for gluonium in $\eta'$}
The $\eta'$ meson, being almost a pure SU(3) singlet, is a good candidate 
for a sizeable gluonium content. In this hypothesis, the $\eta$ and $\eta'$
wave functions can be written in terms of the $u$,$d$ quark wave function
the strange component and the gluonium \cite{Rosner}.
From the study of $\phi\to\eta'\gamma\to\pip\pim 7\gamma$'s and 
$\phi\to\eta\gamma\to 7\gamma$'s decays, the ratio 
$R_\phi = BR(\phi\to\eta'\gamma)/BR(\phi\to\eta\gamma)$ has been 
measured \cite{PLBmixing}: $R_{\phi} = ( 4.77 \pm 0.09_{\rm stat} \pm 
0.19_{\rm syst} ) \times 10^{-3}$.
Combining our result with other experimental constraints and using the 
corresponding SU(3) relations between decay modes, we found a $3\,\sigma$ 
evidence for gluonium content in $\eta'$ \cite{JHEPmixing}.

\Section{\boldmath $\phi\to\eta e^+ e^-$}
Pseudoscalar production at the $\phi$ factory associated to internal 
conversion of the photon into a lepton pair allows the measurement of 
the form factor ${\cal F}_P(q_1^2=M_\phi^2, q_2^2 > 0)$ of pseudoscalar 
mesons $P$ in the kinematical region of interest for the VMD model.
The only existing data on $\phi \to \eta e^+ e^- $ come from the SND 
experiment at Novosibirsk which has measured the M$_{ee}$ invariant mass 
distribution on the basis of 213 events \cite{SND_phietaee}.
At KLOE, a preliminary study of this decay has been performed on 
213 pb$^{-1}$ (about 1/10 of the total luminosity) using the 
$\eta\to\pip\pim\pio$ final state. 
Simple analysis cuts provide $\sim 2000$ signal events with very small 
residual background contamination (Fig.~\ref{Fig:phietaee}). 
The extraction of the branching fraction and the study of the form
factor is under way. The analysis will be extended to the whole data set.

\begin{figure}[hbt]
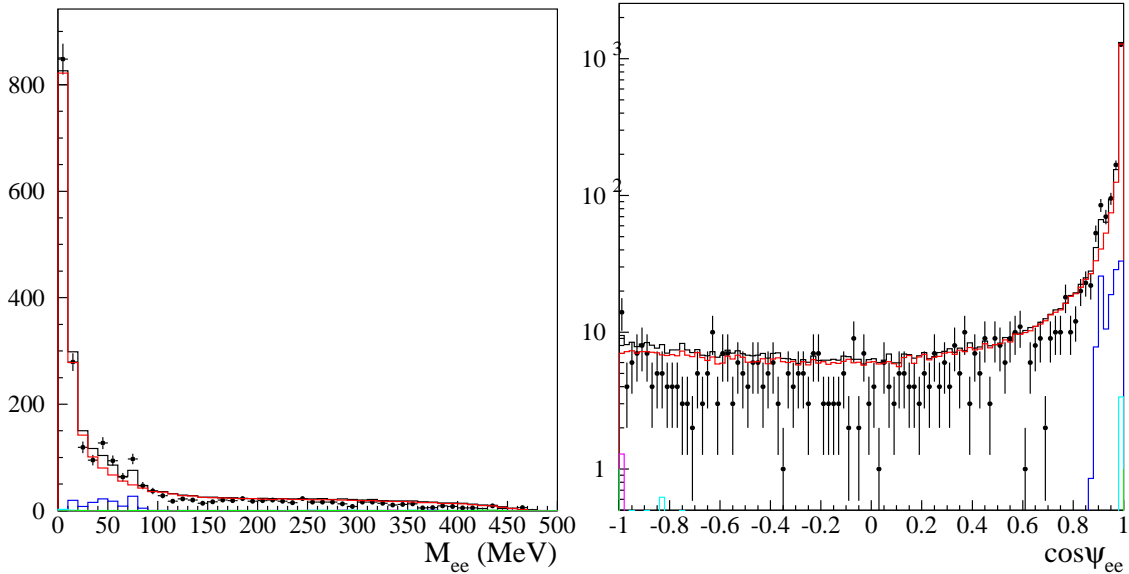

 \epsfig{file=./GiovannellaSimona/Giovannella3.epsi,width=0.5\textwidth}
 \epsfig{file=./GiovannellaSimona/Giovannella4.epsi,width=0.48\textwidth}
  \caption{Invariant mass (left) and 3-dimensional opening angle (right) 
    of \ep\el\ pairs for $\phi\to\eta\ep\el$ events. Dots: data, black:
    total MC shape, red: MC signal, other colours: residual background.}
  \label{Fig:phietaee}
\end{figure}


\coltoctitle{\boldmath Possibilities for $\eta'$ Physics at JLab}
\authortochead{A.~Starostin}
\label{art8}

\maketitle

Decay $\eta' \to \eta \pi \pi$ is a unique source of information on $\eta-\pi$ interactions; $\eta'
\to e^+ e^- \gamma$ gives access to the electromagnetic transition form factor of $\eta'$, which is
an essential contribution to the value of the muon anomalous magnetic moment, $g-2$, predicted by
the Standard Model. There is a long list of $\eta'$ decays allowing to test discrete symmetries $C$,
$P$ and $CP$ in search for physics beyond the Standard Model. Our present knowledge of $\eta'$
decays is based on experimentas with very limited statistics, therefore new high statistics
measurements will be important.

The $\eta'$ photoproduction cross section has a strong maximum between 1.6~GeV and 2.3~GeV photon
beam energy. A high intensity tagged photon beam in this energy range is available at the JLab
Hall-B, where the CLAS detector is installed~\cite{clas}. Currently, CLAS is a magnetic spectrometer
with toroidal field equipped with high resolution time-of-flight system, several layers of drift
chambers, forward electromagnetic calorimeter and a gas Cerenkov detector in front of the
calorimeter. It provides good acceptance and excellent momentum resolution for charged particles and
some acceptance for photons. Electrons and positrons can be separated from $\pi^-/\pi^+$ only for
forward angles  covered by the Cerenkov detector. There are two CLAS data sets which can be used to
extract information on some decay modes of $\eta'$. The $g11$ and $g12$ runs were taken by the CLAS
collaboration in 2004 and 2009 using the high intensity beam of tagged photons. In these two runs
the beam with maximum energy about 5.7~GeV was incident on 40~cm long liquid hydrogen target
installed slightly upstream in respect to the center of the CLAS detector. The data allow reliable
identification of $\eta' \to \eta \pi^+ \pi^-$, $\eta' \to \pi^0 \pi^+ \pi^-$, $\eta' \to \pi^+
\pi^- \pi^+ \pi^-$ and some other decay modes. In the cases when one neutral particle is involved
this particle can be identified by its missing mass. A preliminary analysis of the $g11$ data set
indicates that about $10^5$ $\eta'$ were produced with about $2 \times 10^5$ $\eta' \to \eta \pi^+
\pi^-$ and 5000 $\eta' \to \pi^0 \pi^+ \pi^-$ events reconstructed. This statistics potentially
allows to investigate the Dalitz plot of the decays although the detector acceptance is highly
asymmetric in respect to the positively and the negatively charged particles.
A substantial drawback of the $g11$ data set is that the gas Cerenkov detector was switched off for
the run. This problem was fixed in the $g12$ run allowing identification of the $\eta'$ Dalitz
decays, specifically $\eta' \to e^+ e^- \gamma$. The statistical accuracy of the existing data for
this decay mode is extremely poor so the new CLAS data can improve the existing data sample
substantially allowing a detailed investigation of the transition from factor of $\eta'$. The
analysis of the data is underway.

A new experimental hall, Hall-D, will be available at JLab after completion of the 12~GeV upgrade
in 2015. The hall will be equipped with GlueX detector and a beam of tagged photons with maximum
energy up to 12~GeV. The detector will provide high acceptance and resolution for both, charged and
neutral particles, and will be used as a universal setup for meson spectroscopy at JLab. This setup
can also be used for $\eta/\eta'$ physics and we anticipate that significant amount of new data will
became available once the GlueX will start its operation.

\coltoctitle{\boldmath A Preliminary Chiral Analysis of $\eta'\rightarrow\eta \pi \pi$}
\authortochead{P.~Masjuan}
\label{art9}

\setcounter{chapter}{1}
\setcounter{section}{0}
\maketitle

\Section{Introduction}

The decays $\eta^\prime\to\eta\pi\pi$ are interesting because, due to the quantum numbers of the
pseudoscalar mesons involved, if the decay proceeds through resonances these can be mostly of scalar
nature, $G$-parity prevents vectors to contribute, intended for the analysis of the properties of
the $f_0(600)$ (or $\sigma$) resonance (even though the $a_0(980)$ is also present and indeed
dominant) \cite{paper}. Also, the presence of $\eta$ and $\eta^\prime$ in this reaction is ideal for
studying the mixing properties of both mesons. Finally, this decay allows to test Chiral
Perturbation Theory (ChPT) and its possible extensions such as Large-$N_c$ ChPT and Resonance Chiral
Theory (RChT). For all that, precision measurements on $\eta$ and $\eta^\prime$ would be very
helpful and provide useful information in our understanding of low energy QCD.

Chiral Perturbation Theory (ChPT) \cite{GL85} is the low energy effective theory of Quantum
Chromodynamics (QCD). It is described in terms of an octet of pseudoscalar bosons appearing in the
theory as a result of the spontaneous breaking of the chiral symmetry of QCD which are identified
with the lightest hadronic states $(\pi, K, \eta)$. To include the pseudoscalar singlet $\eta_1$ one
should apply the large $N_c$ limit. Then, a simultaneous expansion in $p^2$, $m_q$ and $1/N_c$ is
possible and the interactions among the $(\pi, K, \eta, \eta^\prime)$ mesons are in principle well
described \cite{U3-ChPT}. In addition, large $N_c$ ChPT does not include resonances as external
states. The effects of these resonances are then virtual and encoded in the low energy constants
(LECs) of the chiral Lagrangian. However, when the energy of the process is of the order of the
resonance mass, the perturbative expansion of ChPT fails and the resonant effects must be taken
explicitly. This is accounted for in Resonance Chiral Theory (RChT) \cite{RChTa}, where the
interactions of the pseudoscalar mesons are supplemented with new interactions among these and
nonets of vectors, axials, scalars, representing the $\rho$, $a_1$, $\sigma$, etc., in a minimal
way.

From the experimental side, large effort has been done to obtain an accurate value for the Branching
ratio of this decay \cite{PDG} and also the Dalitz plot parameters are of interest, see Table
\ref{tablepar}.

\Section{The Role of Final State Interactions}
At Leading Order (LO) in large--$N_c$ ChPT, the prediction of the Branching ratio is completely off
the experimental value: the LO predicts $0.9\%$ to be compared with the PDG value $43.2\%$,
\cite{PDG}. This result can be improved by studying the Next to Leading Order (NLO) that predicts
$64.4\%$, still off but in the right direction. It turns out that at NLO, the amplitude is basically
dominated by a certain LECs combination, $3L_2+L_3$. This combination can be predicted by using the
experimental Branching ratio and then, reexpanding, obtain the Dalitz plot parameters. One could
estimate what is missing when computing the amplitude only at NLO by a unitarization process,
\cite{paper}. This process evaluates the role of final state interactions indicating the presence of
the $\sigma$ meson in the $s-$channel. The $N/D$ unitarization method \cite{Oset-ND} for this decay
produces very accurate results for the Dalitz plot parameters, shown in
Table~\ref{tab.Bexpdalitzplot}.

\begin{table}
\centering
{\scriptsize
\begin{tabular}{ccc}
\hline
Parameter &  Exp. $[\eta'\to \eta \pi^0\pi^0]$ &  Exp. $[\eta'\to \eta \pi^+\pi^-]$ \\
& GAMS-4$\pi$ \cite{GAMSexp} & VES \cite{VESexp} \\
\hline
$a$ & $-0.066\pm 0.016\pm 0.003$ & $-0.127\pm 0.016\pm 0.008$  \\
$b$ & $-0.063\pm 0.028\pm 0.004$ & $-0.106\pm 0.028\pm 0.014$  \\
$c$ & $-0.107\pm 0.096\pm 0.003$ & $+0.015\pm 0.011\pm 0.014$  \\
$d$ & $+0.018\pm 0.078\pm 0.006$ & $-0.082\pm 0.017\pm 0.008$  \\
\hline \\
\end{tabular}
\caption{Experimental Dalitz slope parameters for $\eta^\prime\to\eta\pi^0\pi^0$ (second column) and
$\eta^\prime\to\eta\pi^+\pi^-$ (fourth column), respectively.}
\label{tablepar}
}
\end{table}

\begin{table}
\centering
\begin{tabular}{cccc}
  \hline
  & large--$N_c$ChPT & large--$N_c$RChT & large--$N_c$RChT \\
    & $N/D$ method & $N/D$ method  & $J=2$ reson. $+ N/D$ method \\
  \hline
  $\mathfrak{B}_{\eta'\to\eta\pi^+\pi^-}$  &  $43.2\%^\dagger$  & $43.2\%^\dagger$ & $43.2\%^\dagger$\\
  a            $[Y]$           & $-0.098(48)^{\dagger}$  &  - 0.098(48)$^\dagger$     &   - 0.098(48)$^\dagger$  \\
  b          $[Y^2]$           & -0.047(8)  & -0.0297(8) &  - 0.0332(5)  \\
  d          $[X^2]$           & -0.093(5)  & -0.0573(4) &  -0.0718(3) \\
  $\kappa_{21}$    $[X^2Y]$    &  0.002(3)  & -0.010(1)  &  -0.009(2)  \\
  $\kappa_{40}$    $[X^4]$     & 0.0023(2)  & 0.0009(1)  &  0.0013(1) \\
  \hline
  $(3L_2+L_3)\times 10^{3}$ & $1.03(5)$ & --- & ---\\
  $c_d$ (MeV)  & --- & 26.6(7)     & $24.8(8)$ \\
  $C$-constant & $-1.24 \leq C \leq -0.12$ & $2.06 \leq C \leq 3.37$ &$ 1.38 \leq C \leq 2.74$\\
  \hline
\end{tabular}
\caption{{\small Dalitz parameters of the decay $\eta^\prime\to\eta\pi^+\pi^-$ for the
$N/D$--unitarization of the large--$N_c$ $\chi$PT (second column) and large--$N_C$ R$\chi$T
amplitude (third and fourth columns). $3L_2+L_3$, $c_d$ and $C$ are fixed using the experimental
Branching Ratio and the Dalitz plot parameter $a$, marked with $\dagger$. The last column shows the
result if one also takes into account the the impact of $J=2$ resonances in the R$\chi$T framework.
To compare with experimental data, see Table \protect\ref{tablepar}.
}}
\label{tab.Bexpdalitzplot}
\end{table}

\Subsection{Acknowledgments}
I would like to thank the organizers for the nice atmosphere during the conference. This work has
been supported by the EU contract MRTN-CT-2006-035482 (FLAVIAnet), by MICINN, Spain (FPA2006-05294)
and the Spanish Consolider-Ingenio 2010 Programme CPAN (CSD2007-00042) and by Junta de
Andaluc\'{\i}a (Grants P07-FQM 03048 and P08-FQM 101).

\coltoctitle{\boldmath $\pi^0,\,\eta\to\gamma\gamma$ and $\eta\to3\pi$ at Two Loops}
\authortochead{K.~Kampf}
\label{art14}

\maketitle

All three presented processes $\pi^0\to\gamma\gamma$, $\eta\to\gamma\gamma$ and
$\eta\to3\pi$ (in fact four processes if we distinguish $\pi^0\pi^0\pi^0$ and
$\pi^0\pi^+\pi^-$ in the latter) represent important tools for studying basic
phenomena of the underlying theory: quantum chromodynamics (QCD). The first two
played an important role in understanding a symmetry pattern of the theory as
they are directly connected with the so-called $U(1)_A$ anomaly. QCD enlarged
by photons possesses, however, two such anomalies. The first one, internal,
connected with QCD only, proportional to gluonic term $G_{\mu\nu}\tilde
G^{\mu\nu}$, dubbed $U(1)$-problem and the resulting strong $CP$ problem is still
an open issue. As a remnant of the anomaly, the $\eta'$
plays a more important role than naively expected and has to be included in a
theoretical consideration. The second anomaly, external, in our case connected with
electromagnetic interaction (or $F_{\mu\nu}\tilde F^{\mu\nu}$) explains why
$\pi^0\to\gamma\gamma$ can decay so quickly even it should be suppressed due
to Sutherland's theorem. On the other hand, the two last processes represent 95\% of
all $\eta$-decay modes and are thus perfectly suited to study directly
properties of $\eta$ which is one of the main tasks of PrimeNet. 
Simultaneous treatment of two-photon $\pi^0$ and $\eta$ decays, apart from
testing or fixing our understanding of $\eta'$, can provide valuable information
on the decay constants $F_\pi$ and $F_\eta$ or quark mass ratios (e.g.
$R=\frac{m_d-m_u}{m_s-\hat{m}}$). It is well known that there are
no chiral logarithms for these decays generated by one-loop
diagrams~\cite{donbij} which motivate the necessity to go to two-loop order
\cite{kmb}. This order is also inevitable for $\eta\to3\pi$ as was performed in
the framework of chiral perturbation theory (ChPT) in \cite{bg}. The discrepancy
with present measurements \cite{exhere} is a motivation for a new dispersive
study (see \cite{disp} and references therein). There are two remarks to make.
The first, connected with the fact that $\eta\to3\pi$ as forbidden in isospin
limit is a good candidate for extracting of $m_d-m_u$ parameter (e.g. $R$).
The dispersive method, however powerful based only on very general
assumptions, has to rely at some point on ChPT to extract such a parameter.
First analyses show that how we perform this matching is very crucial and has
big impact on $R$ (or other physical parameters). Another remark concerns
the absolute value of the partial decay width
for $\eta\to3\pi$. Experimentally it is obtained via normalization to
$\eta\to\gamma\gamma$. Change in one decay width has thus influence in other
(a change by 1\% in $\Gamma(\eta\to\gamma\gamma)$ input shifts $R$ by $\approx0.2$).

To summarize all three decay modes of $\pi^0$ and $\eta$ are important for
basic properties of QCD and as they are closely interconnected it is
desirable to study them simultaneously at appropriate order. First results of
these calculations were presented.

\coltoctitle{\boldmath Towards a Better Understanding of the Slope Parameter in $\eta\to3\pi^0$}
\authortochead{S.~Schneider}
\label{art15}

\maketitle

The isospin-breaking decay $\eta\to3 \pi$ is an ideal tool to extract information on light quark
mass ratios from experiment. For precise determinations, however, a detailed theoretical and
experimental understanding of the Dalitz plot distribution is essential. While experimentally agreed
upon, the slope parameter $\alpha$ of the neutral decay channel raises some question marks from a
theoretical point of view. Chiral perturbation theory (ChPT) at next-to-leading~\cite{GLeta3pi} (one
loop, $\mathcal{O}(p^4)$) and next-to-next-to-leading~\cite{BGeta3pi} (two loops,
$\mathcal{O}(p^6)$) predicts a positive sign, in strong disagreement with the world
average~\cite{pdg} of $\alpha=-0.0317\pm0.0016$ (see also Fig.~\ref{fig:slopes}). A dispersive
approach~\cite{KWWeta3pi} predicts the correct sign, but still misses the experimental value
substantially. We use the modified non-relativistic effective field theory (NREFT)
approach~\cite{K3pi} as a diagnostic tool in order to understand these
discrepancies~\cite{SKDeta3pi}.

In Dalitz plot analyses experimental data is fitted to the squared amplitude, which for the charged
and neutral $\eta\to3\pi$ decay channels is expanded in polynomial terms according to
\begin{eqnarray}\label{eq:ampsq}
 |\mathcal{M}_c(x,y)|^2\!\!\!\!&=&\!\!\!\!|\mathcal{N}_c|^2\big\{ 1+ay+by^2+dx^2 + \ldots  \big\}~,\qquad |\mathcal{M}_n(z)|^2=|\mathcal{N}_n|^2 \big\{1+2\alpha z + \ldots \big\}~,\nonumber\\
 x\!\!\!\!&=&\!\!\!\!\frac{\sqrt{3}(u-t)}{2M_\eta Q_\eta}~, \qquad y=\frac{3(s_n-s)}{2M_\eta Q_\eta}~, \qquad z=x^2+y^2~, \qquad Q_\eta=M_\eta-3M_\pi~,
\end{eqnarray}
where the amplitudes are related by isospin symmetry (in Condon-Shortley phase convention),
\begin{eqnarray}\label{eq:isospinrel}
 \mathcal{M}_n(s,t,u)=-\mathcal{M}_c(s,t,u)-\mathcal{M}_c(t,u,s)-\mathcal{M}_c(u,s,t)~.
\end{eqnarray}
In order to determine the coefficients of the polynomial in Eq.~(\ref{eq:ampsq}), we use NREFT,
which is ideally suited to describe final-state interactions among the pions and has been applied
to the precise extraction of $\pi\pi$ scattering lengths from experiment~\cite{K3pi}. One of the
great advantages of NREFT is the fact that the decay amplitude can be parameterized directly in
terms of $\pi\pi$ threshold parameters, so that we can implement physical scattering lengths,
effective ranges, etc., instead of expanding them in terms of quark masses (as is the case in ChPT).
Furthermore, we obtain fully covariant results with the correct thresholds in the physical region.
In that respect the term ``non-relativistic'' only summarizes the fact that we neglect
inelasticities outside the physical region.

\begin{figure}
 \centerline{\includegraphics[width=0.53\linewidth]{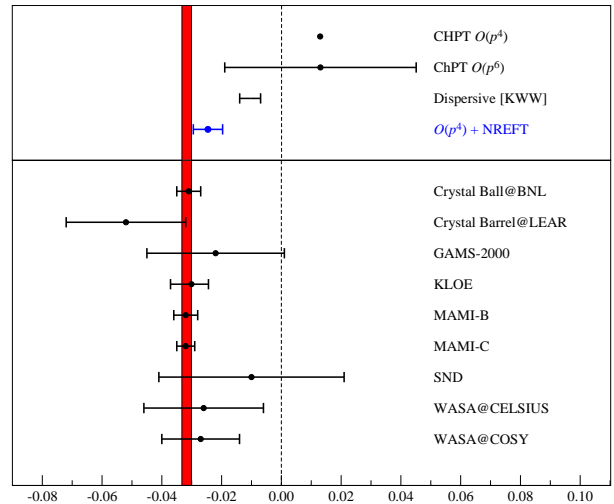}}
 \caption{Comparison of values for the slope parameter $\alpha$. Top:
theoretical predictions. Bottom: experimental determinations. The red shaded
area is the particle data group average~\cite{pdg}.
 \label{fig:slopes}}
\end{figure}

The low-energy coupling constants of the non-relativistic theory are not determined {\it a priori} and have to be determined otherwise. For the tree-level decay constants this is done by matching to ChPT at $\mathcal{O}(p^4)$. 
To fix the $\pi\pi$ final-state interactions we match to the low-energy representation of the $\pi\pi$ scattering amplitude. Numerically we thus obtain for the slope parameter~\cite{SKDeta3pi},
\begin{equation}\label{eq:alpha}
 \alpha=-0.025\pm0.005~,
\end{equation}
where the error stems from varying between two sets of threshold parameters from phenomenological analyses~\cite{scatt} and from estimating higher-order contributions by partial unitarization. Our result is considerably closer to the experimental determination
than previous analyses and it is interesting to note that while tree-level and one-loop diagrams give a positive contribution to $\alpha$, a large negative shift is induced by the inclusion of the two-loop bubble diagrams.
Our result is compared to several other theoretical and experimental determinations in Fig.~\ref{fig:slopes}.

Within our approach it is also possible to understand the ChPT determination of $\alpha$. For that we ``simulate'' the ChPT result by inserting $\pi\pi$ rescattering vertices only up to $\mathcal{O}(p^2)$ in the two-loop
graphs and up to $\mathcal{O}(p^4)$ in the one-loop graphs. We find $\alpha_{\rm ChPT}=-0.0011$, which is very close to zero and shifted strongly from our initial result in Eq.~(\ref{eq:alpha}). 
This discrepancy can be attributed to the significantly weaker $\pi\pi$ rescattering.

By means of the isospin relation~(\ref{eq:isospinrel}), one can relate charged and neutral Dalitz plot parameters,
\begin{equation}
 \alpha=\frac{Q_n^2}{4Q_c^2}\Bigl(b+d-\frac{a^2}{4}\Bigr)-\zeta_1(1+\zeta_2a)^2~,\qquad Q_{c(n)}=M_\eta-M_{\pi^0}-2M_{\pi(\pi^0)}~,
\end{equation}
where the coefficients $\zeta_{1/2}$ are exclusively determined by $\pi\pi$ phases and therefore not subject to uncertainties from matching to ChPT at $\mathcal{O}(p^4)$. Using the precise KLOE determinations for the charged Dalitz plot 
parameters as input~\cite{KLOEcharged}, we find
\begin{equation}\label{eq:KLOE}
 \alpha_{\rm KLOE}^{\rm theo}=-0.062\pm0.003({\rm stat})_{-0.006}^{+0.004}({\rm stat})\pm0.003({\pi\pi})~,
\end{equation}
which strongly disagrees with the world average and the collaboration's own experimental finding. The $\Delta I=1$ determination of $\alpha$ from a fit of the charged data in~\cite{KLOEcharged}, which is in agreement with 
the world average, can be understood using chiral one-loop phases. An ongoing dispersive analysis~\cite{CLPeta3pi} indicates that the discrepancy in Eq.~(\ref{eq:KLOE}) might be slightly overpredicted.

\coltoctitle{\boldmath Determination of Quark Masses: The Contribution of $\eta \to 3 \pi$}
\authortochead{G.~Colangelo}
\label{art16}

\maketitle

In this talk I have discussed the current status of the determination of
the light quark masses, concentrating in particular on those based on
lattice calculations. I have presented a summary of the available lattice
calculations, as it has now appeared in the FLAG (FLAVIAnet Lattice
Averaging Group) review \cite{Colangelo:2010et}. Current lattice
determinations have reached a remarkable level of precision: about 10\% for
$m_s$ and the average of the up and down quark masses $m_{ud}$, and about
4\% for the ratio $m_s/m_{ud}$. Including isospin breaking effects on the
lattice, and in particular photons, is however still problematic. A first
calculation with two dynamical flavours (but with photons still treated in
the quenched approximation) has recently appeared \cite{Blum:2010ym}, but
the precision is not yet comparable to what has been achieved in the
isospin limit. In particular the control of systematic effects is not yet
fully satisfactory.

The analysis of the process $\eta \to 3 \pi$, on the other hand, allows one
to extract clean information about the isospin-violating mass ratio $Q$:
\[
\frac{1}{Q^2}=\frac{m_d^2-m_u^2}{m_s^2-\hat m^2} \; \; ,
\]
as it has already been shown by dispersive analyses of this process
\cite{Anisovich:1996tx,Kambor:1995yc}. The current experimental activity
aiming at a better measurement of the decay rate and the Dalitz plot of
this process, which has been extensively reported at this Workshop, and
recent as well as forthcoming theoretical analyses
\cite{Lanz,Schneider,Schneider:2010hs} will indeed allow a precision
determination of this quark mass ratio. This information is complementary
to the one provided by the lattice in the isospin limit and will lead to a
precise determination of all three light quark masses.

\coltoctitle{\boldmath A new Dispersive Analysis of $\eta \to 3 \pi$}
\authortochead{S.~Lanz}
\label{art17}

\maketitle

The decay $\eta \to 3 \pi$ is of particular theoretical interest because it can only proceed through
isospin breaking. If the strongly suppressed electromagnetic contributions are neglected,
the decay amplitude is proportional to $(m_u - m_d)$ or, alternatively, to the quark mass double ratio
\begin{equation}
	\frac{1}{Q^2} = \frac{m_d^2 - m_u^2}{m_s^2 - \hat{m}^2} \;, \quad \mbox{with} \quad \hat{m} = \frac{m_u+m_d}{2} \;.
\end{equation}
Furthermore, this process involves two theoretical puzzles. The first one is the fact that the
results for the decay width from current algebra, but also from one-loop chiral perturbation theory (ChPT)
fail to reproduce the experimental value. It is understood that this is due to large final state
rescattering effects that can be ideally treated using dispersion relations.
The second puzzle is related to the slope parameter $\alpha$ in the neutral
channel, where ChPT, but also an older dispersive analysis \cite{Kambor+1996} fail to reproduce the experimental finding.

In 1996 two dispersive analyses of $\eta \to 3 \pi$ have been published \cite{Kambor+1996,Anisovich+1996}.
Our analysis follows closely the method proposed in the second reference, where a detailed description can be found.
Many developments in recent years, as well as current activity by many groups, have made a new analysis
worthwhile. Several groups provide accurate descriptions of the $\pi \pi$ scattering phase shifts \cite{Ananthanarayan+2001,Descotes-Genon+2002,Kaminski+2009}.
Non-relativistic effective field theories allow for yet another way of treating this process and, in particular, explicitly including
isospin breaking in the pion masses \cite{NREFT}. There is a lot of experimental activity that will lead to several
precise measurements of the Dalitz plot in the near future. In ChPT, the decay amplitude has been calculated
to two-loop order \cite{Bijnens+2007} and effects of $O(e^2 m)$ have been completely analysed \cite{Ditsche+2009}. In addition to
our work, there is a second on-going dispersive analysis following a somewhat different approach \cite{Kampf+2009a}.

We rely on a decomposition of the amplitude into isospin amplitudes $M_I(s)$ that are functions of one variable only \cite{Stern+1993},
\begin{equation}
	M(s,t,u) = M_0(s) + (s-u) M_1(t) + (s-t) M_1(u) + M_2(t) + M_2(u) - \frac{2}{3} M_2(s) \;.
	\label{eq:MIsospin}
\end{equation}
From unitarity and analyticity follow dispersion relations for the function $M_I(s)$,
\begin{equation}
	M_0(s) = \Omega_0(s)\; \left\{ \alpha_0 + \beta_0 s + \gamma_0 s^2 + \frac{s^2}{\pi} \int \limits_{4 m_\pi^2} ^{\infty}
							\frac{ds^\prime}{s^{\prime 2}} \frac{\sin \delta_0(s^\prime) \hat{M}_0(s^\prime)}{|\Omega_0(s^\prime)|
							(s^\prime - s -i \epsilon)} \right\} \;,
\label{eq:dispRel}
\end{equation}
and similarly for $M_1(s)$ and $M_2(s)$. The functions $\hat{M}_I(s)$ are angular averages over all of the $M_I$ such that
the dispersion relations are coupled. The Omn\`es function is given by
\begin{equation}
	\Omega_I(s) = \exp \left\{ \frac{s}{\pi} \int \limits_{4 m_\pi^2}^{\infty} \frac{\delta_I(s^\prime)}{s^\prime (s^\prime - s)}\; ds^\prime \right\} \;.
\end{equation}

The dispersion relations contain a total number of four subtractions constants: $\alpha_0$, $\beta_0$ and $\gamma_0$ in the equation for
$M_0$ and one more, $\beta_1$, in the equation for $M_1$. These are left free by the dispersion relations and have to be determined otherwise.
The dispersion relations are solved numerically by an iterative procedure, starting at some initial configuration for the $M_I(s)$.
If the subtraction constants are determined by a matching to the one-loop result from ChPT (as in Ref.~\cite{Anisovich+1996}), it turns out
that there is some deviation from the experimental result in Ref.~\cite{Ambrosino+2008} for large $s$ (see Fig.~\ref{fig:dalitz}).
Alternatively, we can determine the subtraction constants by a combined fit to the measured Dalitz plot and to one-loop ChPT around the Adler zero,
where the series is expected to converge best. Because the amplitude is proportional to $Q^{-2}$, the normalisation from experiment
can not be used. Instead, it has to be taken from ChPT. The results from a preliminary fit are also shown in Fig.~\ref{fig:dalitz}.

Comparing the decay width that is calculated from our dispersive amplitude with the present PDG value of $\Gamma_{\pi^0 \pi^+ \pi^-} = 295 \pm 20~\mbox{eV}$,
we can extract a value for $Q$. Using the subtraction constants from the matching to one-loop we get $Q = 22.6 \pm 0.4$, thus updating the analysis from
Ref.~\cite{Anisovich+1996}. The preliminary result using the subtraction constants from the fit to the KLOE data is $Q = 22.0 \pm 0.4$. In both cases the given
error comes from the uncertainty on the decay width only and does not yet include an estimate for the theoretical uncertainty.
We plan to extend our analysis to include additional contributions as well as a complete error analysis, resulting in an accurate estimate
for $Q$ and other physical quantities.

\begin{figure}[ht]
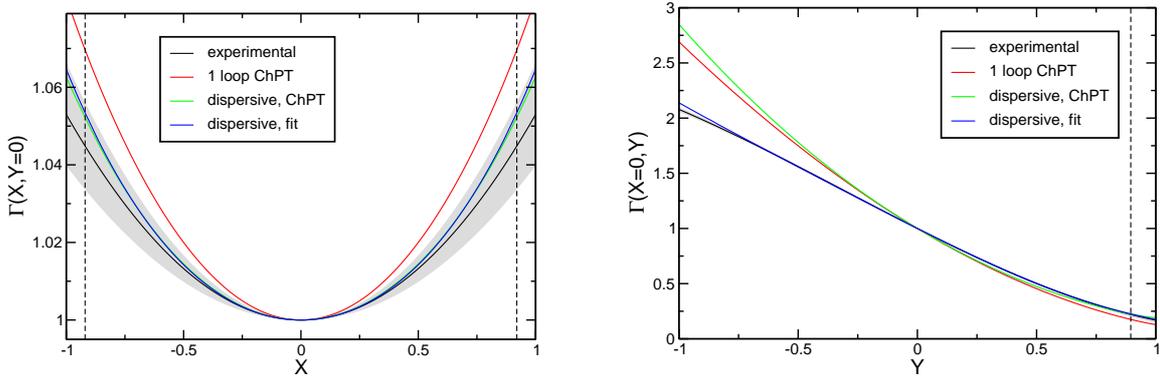

	\includegraphics[clip,width=7cm]{./LanzStefan/Lanz1.eps}
	\hspace{1cm}
	\includegraphics[clip,width=7cm]{./LanzStefan/Lanz2.eps}
\caption{The squared amplitude for the decay $\eta \to \pi^0 \pi^+ \pi^-$ along the lines $Y = 0$ (left) and $X = 0$ (right). $X$ and $Y$ are the usual Dalitz plot
variables, the dashed lines represent the limits of the physical region.
One can clearly see the disagreement between the dispersive result and experiment for $Y < 0$ (which corresponds to large $s$).}
\label{fig:dalitz}
\end{figure}

\coltoctitle{\boldmath The $\eta\rightarrow\pi^+\pi^-\pi^0$ Decay with WASA-at-COSY}
\authortochead{P.~Adlarson}
\label{art24}

\setcounter{chapter}{1}
\setcounter{section}{0}
\maketitle

\Section{Introduction}
The isospin violating strong decay $\eta\rightarrow\pi^+\pi^-\pi^0$ is an important decay since it
allows access to light quark mass ratios. At lowest order of chiral perturbation theory (ChPT) the
amplitude is given by
\begin{equation}
A \propto \frac{m_d-m_u}{F_{\pi}^2}\left( 1 + \frac{3(s-s_0)}{m_{\eta}^2-m_{\pi}^2} \right),
\end{equation} 
where $F_{\pi}$ is the pion decay constant, $s=(p_{\pi^+}+p_{\pi^-})^2=(p_{\eta}-p_{\pi^0})^2$,
$s_0=\frac{1}{3}(m_{\eta}^2+3 m_{\pi}^2)$ and $m_{\pi^+}=m_{\pi^0}=m_{\pi}$. Electromagnetic
corrections are small \cite{DS66}. The tree level decay width, $\Gamma_{tree}\approx 70$ eV
\cite{WB67}, deviate more than a factor of four from the PDG value $\Gamma=296\pm16$ \cite{PDG}.
Higher order diagrams are needed which, among other things, take into account $\pi\pi$ final state
pion interactions \cite{CR81,JG85,JB07}. Alternatively, $\pi\pi$ final state re-scattering can be
implemented in dispersive approaches \cite{AA96,JK96,JB02,SL10}. The decay width scales as
\begin{equation}
\Gamma=\left(\frac{Q_D}{Q}\right)^4 \bar{\Gamma},
\end{equation}
where $Q^2=(m_s^2-\hat{m}^2)/(m_d^2-m_u^2)$, $\hat{m}=\frac{1}{2}(m_u+m_d)$,
and the decay width $\bar{\Gamma}$ and $Q_D=24.2$ are calculated in the Dashen limit \cite{RD69}.
$Q$ gives access to light quark mass ratios and serves also as an important input for lattice QCD
\cite{GC10}. To derive $Q$, $\bar{\Gamma}$ has to be known reliably from theory which can be tested
by comparing with the experimental Dalitz plot distributions. As Dalitz plot variables 
\begin{equation}
x=\sqrt{3}\frac{T_+-T_-}{Q_{\eta}}, \ y=\frac{3T_0}{Q_{\eta}}-1.
\end{equation}
\noindent are used. Here  $T_+$, $T_-$ and $T_0$ denote the kinetic energies of $\pi^+$, $\pi^-$
and $\pi^0$ in the $\eta$ rest frame, and $Q_{\eta}=m_{\eta}-2m_{\pi^+}-m_{\pi^0}$.
The standard way to parametrize the Dalitz plot density is a polynomial expansion around $x=y=0$
\begin{equation} 
\left|A(x,y)\right|^2 \propto 1+ay+by^2+dx^2+fy^3+gx^2y+...
\end{equation}
where $a, b, ...,g$ are called the Dalitz Plot parameters. 
The most precise experimental result is based on a Dalitz plot containing 1.34$\cdot$10$^6$ events
obtained by KLOE \cite{FA08}. Parameters $b$ and $f$ in \cite{FA08} deviate significantly from ChPT
predictions. Independent measurements are therefore important and  WASA-at-COSY aims at providing
two independent data sets with $\eta$ produced in $pp$ and in $pd$-reactions. 

\Section{Experiment}
\noindent  In 2008 and 2009  WASA-at-COSY \cite{AA04} collected $1\cdot10^7$ and $2\cdot10^7$
$\eta$ mesons respectively, in the reaction $pd\rightarrow \mbox{}^{3}\mbox{He} X$ at a kinetic beam
energy of 1~GeV. The missing mass with respect to  $\mbox{}^{3}\mbox{He}$ is used to tag the $\eta$
in the reaction $pd\rightarrow  \mbox{}^{3}\mbox{He} X$ (Fig.~\ref{MMHe} left). In addition, two
tracks of opposite charge in the central drift chamber and two photons with an invariant mass close
to $\pi^0$ in the electromagnetic calorimeter are required. Background from $pd\rightarrow
\mbox{}^{3}\mbox{He} \pi\pi$ reactions is reduced by imposing conditions on the missing mass
calculated for $\mbox{}^{3}\mbox{He} \pi^+\pi^-$ and for $\mbox{}^{3}\mbox{He} \pi^0$. The
preliminary analysis gives 149 000 $\eta\rightarrow\pi^+\pi^-\pi^0$ candidates from the 2008 data,
shown in Fig.~\ref{MMHe}~right.
\begin{figure}[h!]
\begin{center}
\parbox[c]{0.46\textwidth}{
\centering
\includegraphics[width=0.5\textwidth]{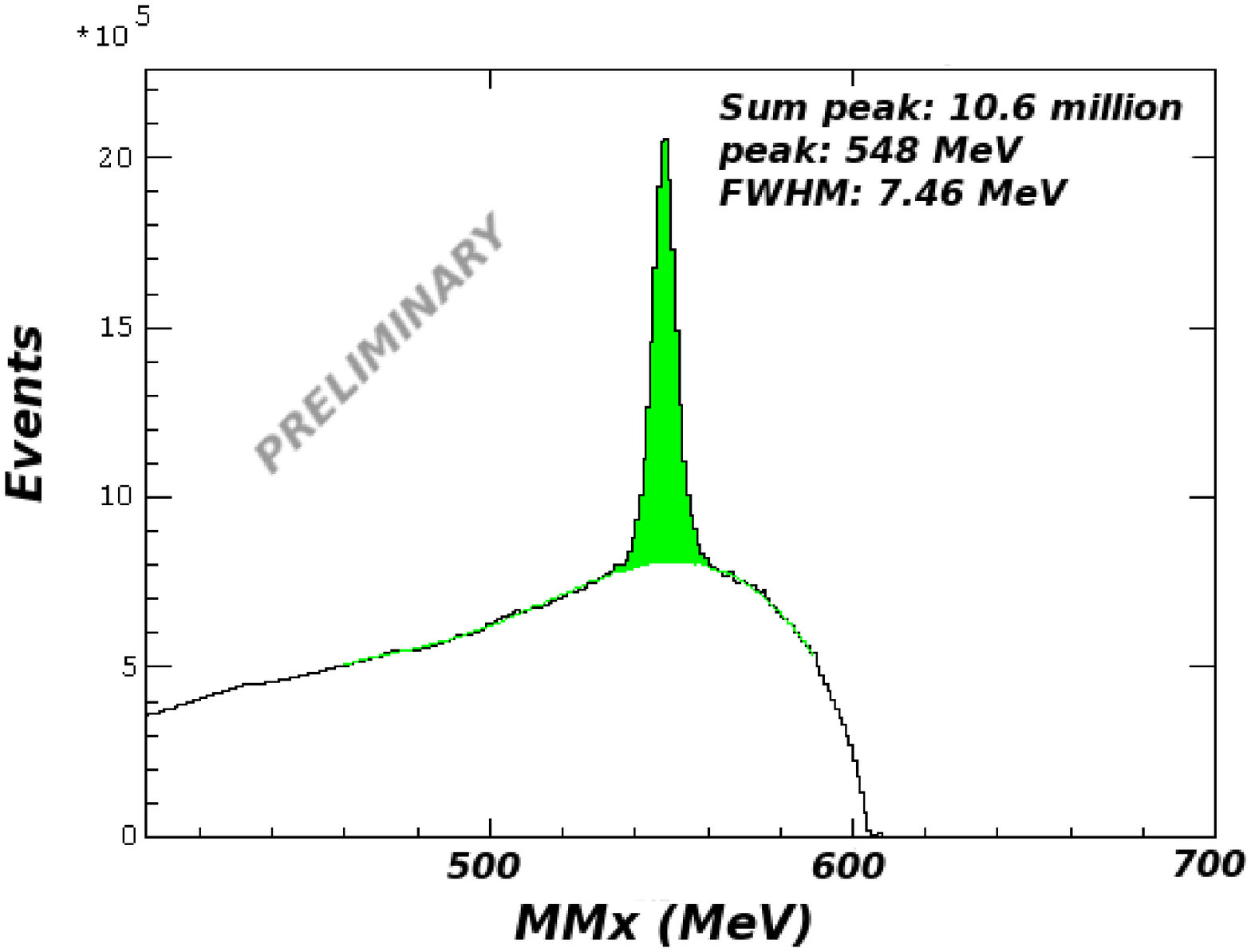}
}
\parbox[c]{0.5\textwidth}{
\centering
\includegraphics[width=0.5\textwidth]{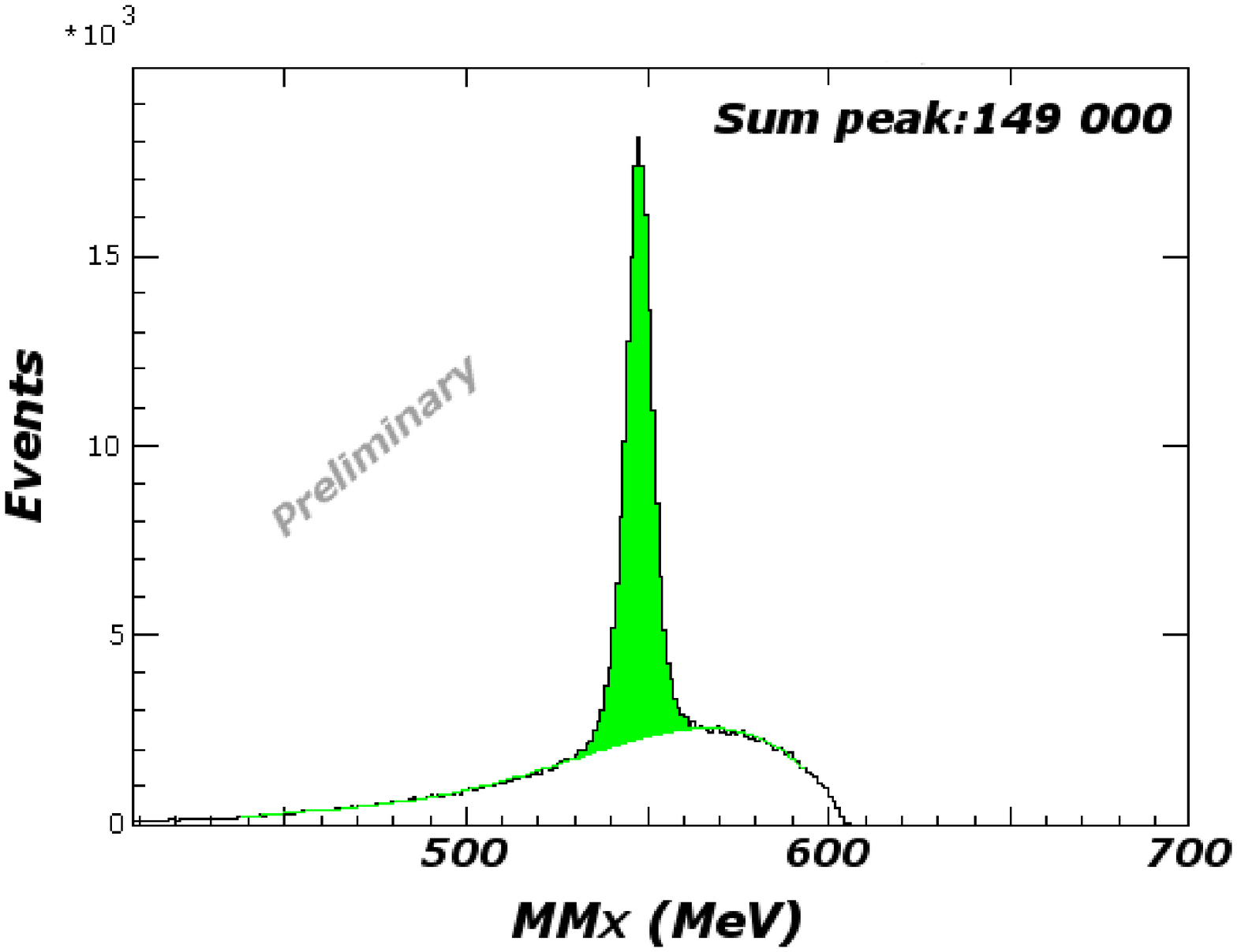}
}
\caption{\label{MMHe}Results from 2008 data. ({\bf left}) The missing mass calculated from the
identified $\mbox{}^{3}\mbox{He}$ $MM(\mbox{}^{3}\mbox{He})$ shows a distinct peak at the $\eta$
mass. ({\bf right})  Same after selecting $\pi^+\pi^-\pi^0$ signature in the central detector and
including a cut on the missing masses with respect to $\mbox{}^{3}\mbox{He} \pi^+\pi^-$ and
$\mbox{}^{3}\mbox{He} \pi^0$.}
\end{center}
\end{figure}

The experimental resolution is better for the $\eta$ four-momenta from $\mbox{}^{3}\mbox{He}$
compared to the information derived from the $\eta$ decay products. To improve the resolution for
the four-momenta of the $\eta$ decay products, a kinematical fit for the reaction $pd\rightarrow
\mbox{}^{3}\mbox{He}\pi^+\pi^-\gamma\gamma$ has been used with $\mbox{}^{3}\mbox{He}$ observables
fixed and a cut on the 1\% level of the probability density function. The $\eta$ content in each
Dalitz Plot bin is derived by performing a four-degree polynomial fit over the background region and
the fitted polynomial is subtracted from the signal region. The preliminary experimental results for
the $x$,$y$ projections of the Dalitz Plot are compared in Fig.~\ref{XYprojection} to Monte Carlo
simulations of the $\eta\rightarrow\pi^+\pi^-\pi^0$ weighted with the tree-level prediction
(equation 1).

\begin{figure}[h!]
\begin{center}
\parbox[c]{1.0\textwidth}{
\centering
\includegraphics[width=1.0\textwidth]{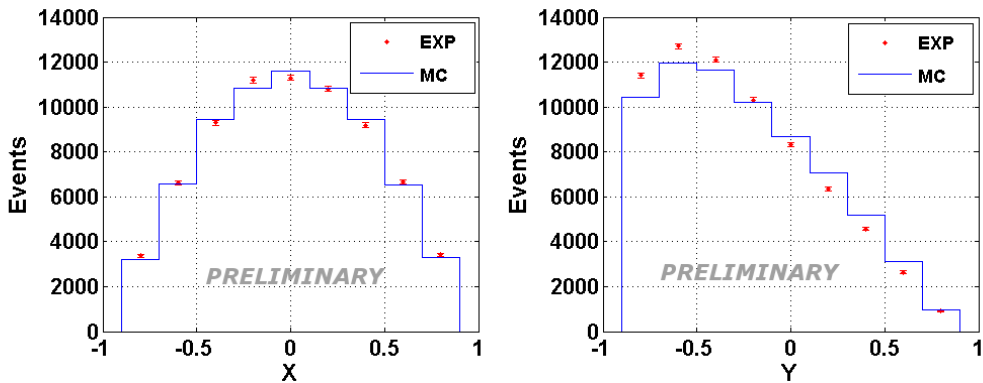}
}
\caption{Dalitz Plot projection from 2008 data, not corrected for acceptance. The error bars
represent only statistical errors of the experimental data on $x$ {\bf(left)} and $y$ {\bf(right)}
for MC (solid line) and experimental data.}
\label{XYprojection}
\end{center}
\end{figure}

\noindent In 2008 and 2010 measurements have been performed using the tagging reaction
$pp\rightarrow ppX$ at a beam kinetic energy of 1.4~GeV. From the 2010 data approximately $10^7$
$\eta\rightarrow\pi^+\pi^-\pi^0$ are recorded on disk and the analysis of this reaction is in
progress \cite{PA10}.

\section{Outlook}
The work for both \textit{pd} and \textit{pp} data will be continued in order to obtain two
independent determinations of the Dalitz plot parameters for $\eta\rightarrow\pi^+\pi^-\pi^0$.
Further improvements include the acceptance correction of the experimental data and an assessment of
systematical errors.

\coltoctitle{\boldmath Radiative Corrections in $K\to\pi\ell^+\ell^-$ and Related Decays}
\authortochead{B.~Kubis}
\label{art18}

\maketitle

We calculate radiative corrections to the flavor-changing neutral current processes
$K\to\pi\ell^+\ell^-$~\cite{KubisSchmidt}.  Such transitions are suppressed to the one-loop level
in the Standard Model and therefore rare; the CP-allowed $K^\pm$ and $K_S$ 
decays are expected to be dominated by one-photon exchange.
Furthermore these processes are also suppressed to one loop in the effective-theory description 
of chiral perturbation theory~\cite{Ecker}.

The process is characterized by the form factor $F(s)$, which can be extracted from the spectrum 
 with respect to the squared invariant mass of the dilepton pair $s$,
\begin{equation}
\frac{d\Gamma}{ds} = \frac{\alpha^2|F(s)|^2}{3(4\pi)^5M_K^7} \lambda^{3/2}(M_K^2,s,M_\pi^2)
\sqrt{1-\frac{4m_\ell^2}{s}} \left(1+\frac{2m_\ell^2}{s}\right)  (1+\Omega) ~.
\end{equation}
$\lambda(a,b,c) \doteq a^2+b^2+c^2-2(ab+ac+bc)$ is the K\"all\'en function.
Radiative corrections are included in the factor $\Omega$, 
which is of the order of the fine structure constant $\alpha$.
$\Omega$ is given as a simple sum of corrections to the hadronic and the leptonic current,
both of which are ultraviolet finite without requiring a counterterm.
The correction $\Omega_{K\pi}$ to the hadronic current, which only applies to the charged kaon decays ($K^\pm\to\pi^\pm$),
is a smooth function of $s$ that increases the spectrum by only 1\%--1.5\%.
We therefore only discuss the leptonic part $\Omega_{\ell^+\ell^-}$ in the following.
The infrared divergences that occur due to the effects of (massless) virtual photons
are canceled by the inclusion of real-photon radiation (bremsstrahlung), 
which is integrated up to a maximal photon energy cut $E_{\rm max}$.
The calculation of the bremsstrahlung contribution is often simplified by applying the 
``soft-photon approximation'', which accurately reproduces terms $\propto \log E_{\rm max}$
as well as constant parts, but differs from the exact result by terms of $\mathcal{O}(E_{\rm max})$.

For the muon final state, the correction factor $\Omega_{\mu^+\mu^-}$ is dominated by the Coulomb
singularity $\propto \alpha(s-4m_\mu^2)^{-1/2}$ close to threshold; apart from this enhancement
it is small, on the 1\%--2\% level.
The soft-photon approximation with a cut on the additional
photon energy in the bremsstrahlung contribution is very accurate compared to the exact result.
Employing the same expression for the electron final state leads to apparently huge electromagnetic
corrections of the order of 10\% or more, with furthermore 
a significant deviation of the soft approximation from the exact result.
The reason can be seen when expanding $\Omega_{e^+e^-}$ in the electron mass, 
\begin{eqnarray}
\Omega_{e^+e^-} \!\!\!&=&\!\!\! \frac{\alpha}{\pi} \bigg\{\frac{1}{4} - 2 \bigg[\log\epsilon 
+ \frac{(1-\epsilon)(3-\epsilon)}{4}\bigg]\log\delta  
+\frac{1-\epsilon}{2} \bigg[(3-\epsilon)\log\left(1-\epsilon\right) - 
\frac{11 - 3\epsilon}{2}\bigg] \nonumber\\
&&\quad- 2 \log\epsilon + \frac{\pi^2}{3}-2\,{\rm Li}(\epsilon) 
\bigg\} + \mathcal{O}(m_e)~, \qquad {\rm Li}(z)=-\int_0^z\frac{\log(1-t)}{t}dt ~,\label{eq:Sudakov} 
\end{eqnarray}
with $\epsilon = 2E_{\rm max}/\sqrt{s}$ and $\delta=m_e^2/s$.  
The appearance of mass singularities $\propto \log\delta$ in Eq.~(\ref{eq:Sudakov}), 
so-called Sudakov logarithms, is only seemingly in contradiction with the 
Kinoshita--Lee--Nauenberg theorem~\cite{Kinoshita,LeeNauenberg}, which guarantees
the absence of such terms only in total or \emph{inclusive} transition probabilities.
Indeed, in the inclusive limit $\epsilon\to 1$, the $\log\delta$ term in Eq.~(\ref{eq:Sudakov})
vanishes.  This cancellation, however, does obviously \emph{not} hold in the soft-photon approximation, 
in which terms of $\mathcal{O}(\epsilon)$ are neglected.

\begin{figure}
 \centering
\includegraphics[width=0.6\linewidth]{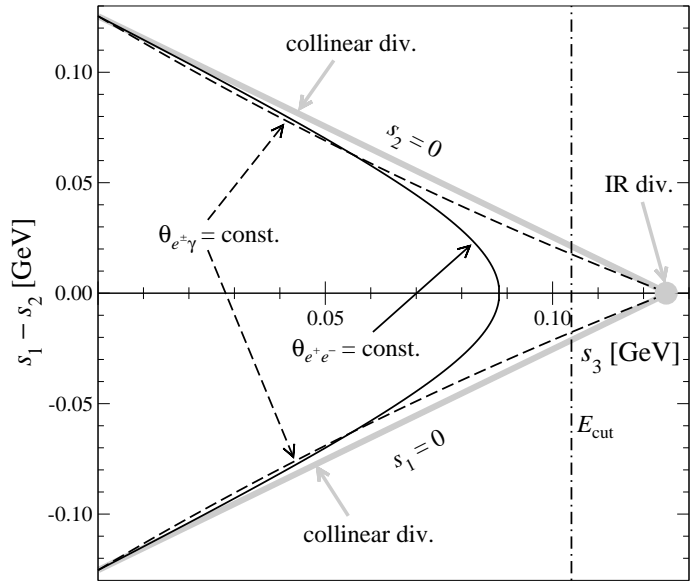}
 \caption{Dalitz plot for $\gamma^*\to e^+e^-\gamma$ at maximal $s=(M_K-M_\pi)^2$. 
  Shown are the positions of the collinear singularities for $s_1=0$ and $s_2=0$ (broad grey bands)
  as well as the infrared singularity at their intersection (thick grey dot).
  Drawn relative to those are various cut positions in $\theta_{e^\pm\gamma}$ (dashed line), $E_{\rm max}$ (vertical dash-dotted), 
  and $\theta_{e^+e^-}$ (full black). Figure taken from Ref.~\cite{KubisSchmidt}.}
 \label{fig:Dalitz}
\end{figure}
The origin of these mass singularities is to be found in the collinear radiation of photons
off very light particles (here: electrons).  
However, much as the inclusion of soft photons in a realistic measurement is an experimental
requirement due to finite detector energy resolution, so is the inclusion of collinear photons
due to finite angular resolution.
It is therefore the (unphysical) excision of hard collinear bremsstrahlung photons that leads to the survival of
logarithmic mass singularities in the electron channels.  
In realistic experiments, this requires the introduction
of additional angular cuts in the leptonic radiative corrections, 
either between bremsstrahlung photon and electron/position, or on the electron--positron angle.
The position of the various infrared and collinear singularities in the Dalitz plot of the
sub-decay process $\gamma^* \to e^+e^-\gamma$ is depicted in Fig.~\ref{fig:Dalitz},
together with the various energy and angular cuts.
Only a correction factor $\Omega_{e^+e^-}$ inclusive with respect to all singular regions
of the bremsstrahlung contribution is physically reasonable.
We provide closed analytic results for the cut dependence of these correction factors
in the limit of the electron mass going to zero, which is shown to be a very good approximation~\cite{KubisSchmidt}.
The mass singularities of Eq.~(\ref{eq:Sudakov}) are translated into phase space singularities therein.

Finally, we show that the universal correction factors for the dilepton spectra can be 
directly used also for other decays such as $\pi^0$ or $\eta$ Dalitz decays 
or vector meson conversion decays like $\omega\to\pi^0\ell^+\ell^-$~\cite{Terschluesen}, despite a totally different
form of the hadronic vector current involved.
The consequences of our results in particular for existing high-precision studies of
the Dalitz decay $\pi^0 \to \gamma e^+ e^-$~\cite{Kampf} ought to be investigated.

\coltoctitle{Electromagnetic Transitions from Vectors to Pseudoscalars}
\authortochead{S.~Leupold}
\label{art19}

\setcounter{chapter}{1}
\setcounter{section}{0}
\maketitle

\Section{Introduction}
Electromagnetic form factors are regarded as interesting quantities to learn about the structure of
hadrons. In the following we will apply a recently proposed effective field theory
\cite{Lutz:2008km,Leupold:2008bp} to determine the transition form factors of vector to pseudoscalar
mesons \cite{Terschluesen:2010ik}. In the present section we will motivate why an effective field
theory can give a more reliable answer than a hadronic model. We focus in particular on the
vector-meson-dominance model (see , e.g., \cite{Landsberg:1986fd}). A form factor is a decay or
reaction rate normalized to the corresponding process for point-like particles. For the decay of a
vector meson into a pseudoscalar meson and a dilepton the form factor, $F$, is a function of $q^2$
which is the four-momentum squared of the lepton-antilepton pair. It is normalized to the photon
point, i.e.\ $F(q^2 =0) = 1$. In the following we will discuss three cases: a) a model for $F$, b) a
very general field theory, c) an effective field theory.

In the vector-meson-dominance model (VMD) the form factor is given by
\begin{equation}
  \label{eq:vmd}
  F(q^2) = \frac{m^2}{m^2 -q^2}
\end{equation}
with the mass $m$ of the intermediate vector meson. (We assume for simplicity that there is only one
contributing vector meson.) The good thing about a model is that it has predictive power. In the
present case, the only quantity which enters is the mass of the intermediate vector meson which
ideally can be obtained from another experiment. On the other hand, if the model fails to describe
the data, it is unclear how to improve it in a systematic way. 

In contrast to (\ref{eq:vmd}) a general field theory yields for the form factor
\begin{equation}
  \label{eq:general}
  F(q^2) = \frac{{c_0} \, m^2}{m^2 -q^2} + (1-{c_0}) 
  + {c_1} \, \frac{q^2}{m^2} + {c_2} \, \frac{q^4}{m^4} 
  + \ldots
\end{equation}
It is assumed here that this general field theory contains vector mesons and arbitrary point
interactions between the initial hadron, the final hadron(s) and the virtual photon. The 
general-field-theory expression (\ref{eq:general}) has infinitely many parameters $c_i$. Some of
them might be related to other processes and therefore could be fixed. But, having infinitely many
parameters, such an approach does not have any predictive power. It is rather ineffective.

An effective field theory formally yields the same expression (\ref{eq:general}). However, it
assigns different degrees of importance to the various parameters $c_i$. For example, one might get
$c_0, \, c_2 \sim O(1)$, $c_1 \sim O(q^2/\Lambda^2)$ and all other $c_i \sim O(q^4/\Lambda^4)$ or
higher. Here we have introduced two scales: a) The ``small'' scale which characterizes the energies
and momenta at which the effective field theory is supposed to work. We have implicitly assumed here
that we work in the energy regime of the vector mesons. Thus the small scale is $\sim m \sim
\sqrt{q^2}$. b) The ``large'' breakdown scale $\Lambda$ which describes the energy regime where new
degrees of freedom become active which are {\em not} considered in the effective field theory. As
long as the energy regime of interest is sufficiently far away from the breakdown scale we can
determine the form factor in the following way: In our example, the leading-order approximation
involves only $c_0$ and $c_2$. This is a finite number of parameters and we assume that one can
determine them from other experiments. Then we gain predictive power for the form factor. If we want
to get a more accurate description, we can include the $c_1$ term. In the energy regime where the
effective field theory works, this latter term is less important than the $c_0$ and $c_2$ terms,
i.e.\ it provides a correction to the leading-order result. Still it is more important than all the
other $c_i$ terms with $i>2$. In general, an effective field theory provides predictive power,
because for a given accuracy only a finite number of terms are needed. In addition, it yields a
systematic way how to improve the results.

Of course, the previous example did not tell how to get such an effective field theory. In practice,
the challenge is a) to construct a theory which includes all relevant degrees of freedom, b) to
devise a power counting scheme which assigns the respective degrees of importance to the interaction
terms, and c) to figure out for which energy regime the scheme works. From the example above one can
see that the breakdown scale is roughly located at the energy where the next-to-leading-order terms
become as important as the leading-order terms. Thus, to determine the breakdown scale one has to
perform at least a next-to-leading-order calculation.

\Section{Transition form factors for vector mesons}

\begin{figure}[bt]
  \centering
   \includegraphics[keepaspectratio,width=0.44\textwidth]{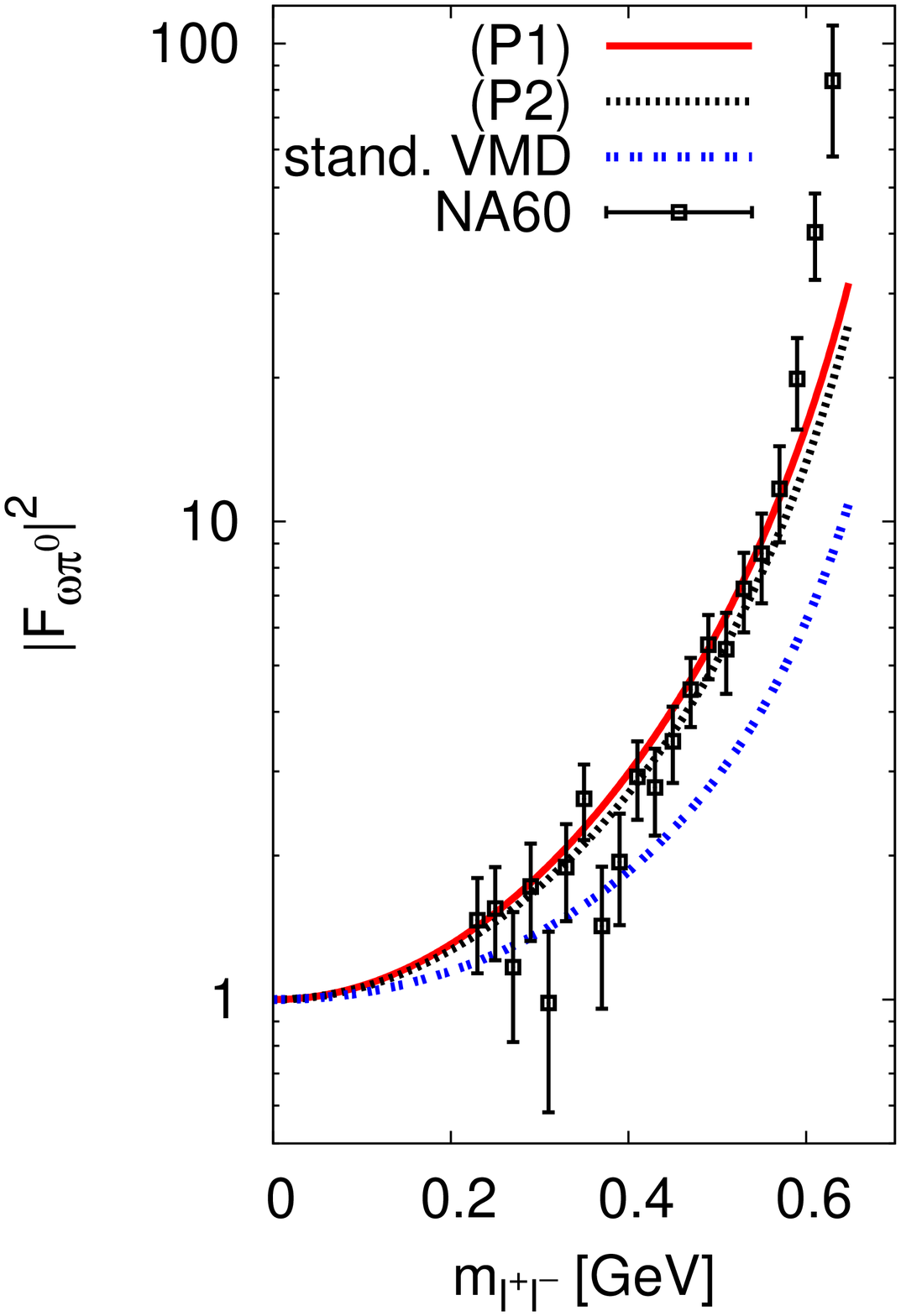}\hspace{1cm}%
   \includegraphics[keepaspectratio,width=0.44\textwidth]{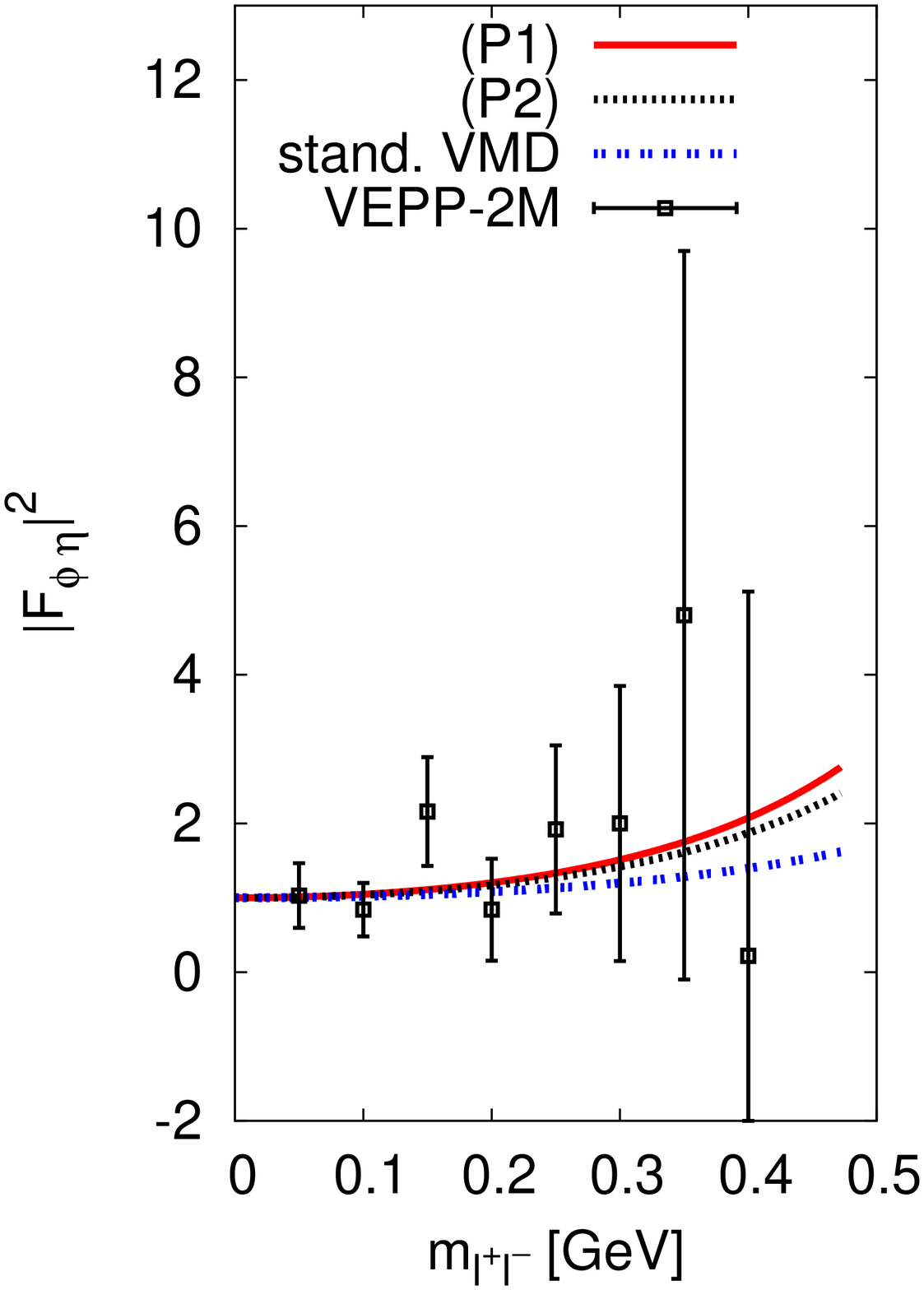}
  \caption{Transition form factors $\omega \to \pi$ (left) and $\phi \to \eta$ (right) calculated in
    VMD and in the effective field theory (full line) as compared to data from \cite{Arnaldi:2009wb}
    and \cite{Achasov:2000ne}, respectively. See \cite{Terschluesen:2010ik} for details.}
  \label{fig:omegapi}
\end{figure}

In \cite{Lutz:2008km,Leupold:2008bp} an effective field theory has been proposed which treats the
light pseudoscalar and vector mesons as active/relevant degrees of freedom. So far only
leading-order calculations have been performed. Therefore, it not clear yet for which energy regime
the scheme works. Nonetheless from a phenomenological point of view the results are very encouraging
\cite{Lutz:2008km,Leupold:2008bp,Terschluesen:2010ik}. 

For the transition form factor of the $\omega$ meson to the pion the proposed effective field theory
yields that only the parameter $c_0$ in (\ref{eq:general}) is of leading order. All other parameters
are subleading (and have not been determined yet). In addition, it turns out that $c_0$ can be
completely determined from the decays of vector mesons into pseudoscalar mesons and {\em real}
photons. Thus, the scheme has predictive power for the form factor. Numerically one finds that $c_0$
is very close to 2. Therefore, to a good approximation the form factor can be written as
\begin{equation}
  \label{eq:approx}
  F(q^2) \approx \frac{m^2 + q^2}{m^2 -q^2}
\end{equation}
which should be compared to the standard VMD formula (\ref{eq:vmd}). Obviously the results are
rather different. Both the results from the effective field theory \cite{Terschluesen:2010ik}
and from VMD are compared to data from NA60 \cite{Arnaldi:2009wb} in Fig.\ \ref{fig:omegapi}, left.
One observes that the effective field theory provides a much better description of the experimental
data. On the right hand side of Fig.\ \ref{fig:omegapi} the transition form factors for $\phi \to
\eta$ are compared. Here the experimental data are not accurate enough to discriminate between the
different theoretical results (and the results are much closer together as compared to the
$\omega$-meson case). The KLOE2 collaboration will study this $\phi$ decay with much better
statistics in the near future.

To summarize: The leading-order calculation within the proposed effective field theory provides a
parameter-free and much better description of the $\omega$ transition form factor than standard
vector-meson dominance. This provides a solution for a long-standing puzzle \cite{Landsberg:1986fd}.
In the future the impact of the vector mesons on the pseudoscalar decays into two real or virtual
photons will be examined. In addition, next-to-leading-order calculations must be carried out to get
an idea about the breakdown scale of the proposed effective field theory.

\coltoctitle{\boldmath New Results on the $\eta-\eta^\prime$ Transition FF's By BaBar}
\authortochead{A.~Denig}
\label{art10}

\setcounter{chapter}{1}
\setcounter{section}{0}
\maketitle

\Section{Introduction}
Meson transition form factors can be accessed at electron-positron colliders in $\gamma
\gamma$-processes, in which both incoming beam particles are emitting a t-channel photon. In this
paper we present recent BaBar results, in which a pseudoscalar mesons - either a $\pi^0$, $\eta$ or
$\eta^\prime$ - is produced in a $\gamma \gamma$-fusion reaction, where one of the photons is
required to be quasi-real, while for the second photon a four-momentum-transfer $Q^2 > 3$ GeV$^2$ is
required. As a consequence, the electron (or positron) emitting the quasi-real photon will have a
very small scattering angle and will not be detected by the BaBar detector, while the second
positron (or electron) will be tagged at large polar angles. 
\\
The $\gamma \gamma^*$ production cross section in such a so-called single-tag measurement is
proportional to the transition form factor $F(Q^2)$, which theoretically can be written as the
convolution of the hard scattering process into quarks ($\gamma\gamma^* \to q\bar{q}$) with the
meson distribution amplitude (DA) $\Phi(x,Q^2)$ after integration over $x$, where $x$ is the
fraction of the meson momentum carried by one of the two valence quarks. At present only
phenomenological models exist for the meson DAs, such that precision measurements can shed new light
on this important quantity in meson structure physics. 
\\
Measurements of the $Q^2$ dependence of the transition form factor can also be used to test models,
which are used in computations of the hadronic light-by-light scattering process for the standard
model predicition of the muon anomaly $(g-2)_\mu$.

\Section{\boldmath $\pi^0$ transition form factor}
BaBar results on the $\pi^0$ transition form factor~\cite{babar-pi}, multiplied by $Q^2$, are 
shown in Fig.~1(a). BaBar is in reasonable agreement with previous 
measurements by CELLO and CLEO, covering however a much wider $Q^2$ range up to
40 GeV$^2$. In total ca. 140.000 events of the kind $e^+e^-\to e^+e^-\pi^0$ 
could be selected in a total data sample of 500 fb$^{-1}$. The plot shows also
the theory prediction for $F(Q^2)$ of three different models for the pion DA: 
(i) ASY: asymptotic DA~\cite{asy}; 
(ii) CZ: Chernyak-Zhinitsky DA~\cite{cz}; 
(iii) BMS: Bakulev-Mikhailov-Stefanis DA~\cite{bms}. 
None of the models shows 
agreement with the BaBar experimental data. Especially, data exceed the 
asymptotic limit $Q^2 F(Q^2) = \sqrt{2} f_\pi$ above 10 GeV$^2$, 
which was previously thought to be the plateau value for this quantity.
It can be concluded, that the pion DA is poorly known at present.

\Section{\boldmath $\eta$ and $\eta^\prime$ transition form factors}
BaBar preliminary results on the $\gamma\gamma^* \to \eta$ and $\eta^\prime$ 
transition form 
factors, measured in the $e^+e^- \to e^+e^- \eta^{(\prime)}$ reactions, are 
shown in Figs.1~(b) and 1(c) in comparison with previous
CLEO measurements. Presented is again the meson transition form factor
multiplied with $Q^2$. 
The BaBar spectra contain in total 2.760 $\eta$ events and ca. 5.000 $\eta^\prime$ events
from a 500 fb$^{-1}$ data sample.
BaBar data significantly improve the precision and extend 
the $Q^2$ region for these form factor measurements. In the case of $\eta^\prime$, 
results from BaBar and CLEO are in good agreement; for the $\eta$ transition
form factor the agreement is worse.
Shown are also the asymptotic limits for the $\eta$ and $\eta^\prime$ 
transition form factors, as calculated in Ref.~\cite{fk}. We observe a completely different
$Q^2$ behaviour compared to the $\pi^0$ transition form factor. While for the
$\pi^0$ transition form factor a significant slope even at high $Q^2$ is
observed, for the $\eta$ and $\eta^\prime$ cases, BaBar data is flattening
at or below the asymptotic limit. An analysis, in which also high-$Q^2$ 
measurements timelike form factors have been taken into account, and in which in a
flavour decomposition, the non-strange and strange form factors have been
mapped out, do also not show conclusive results.

\begin{figure}[h]
\begin{center}
\epsfig{file=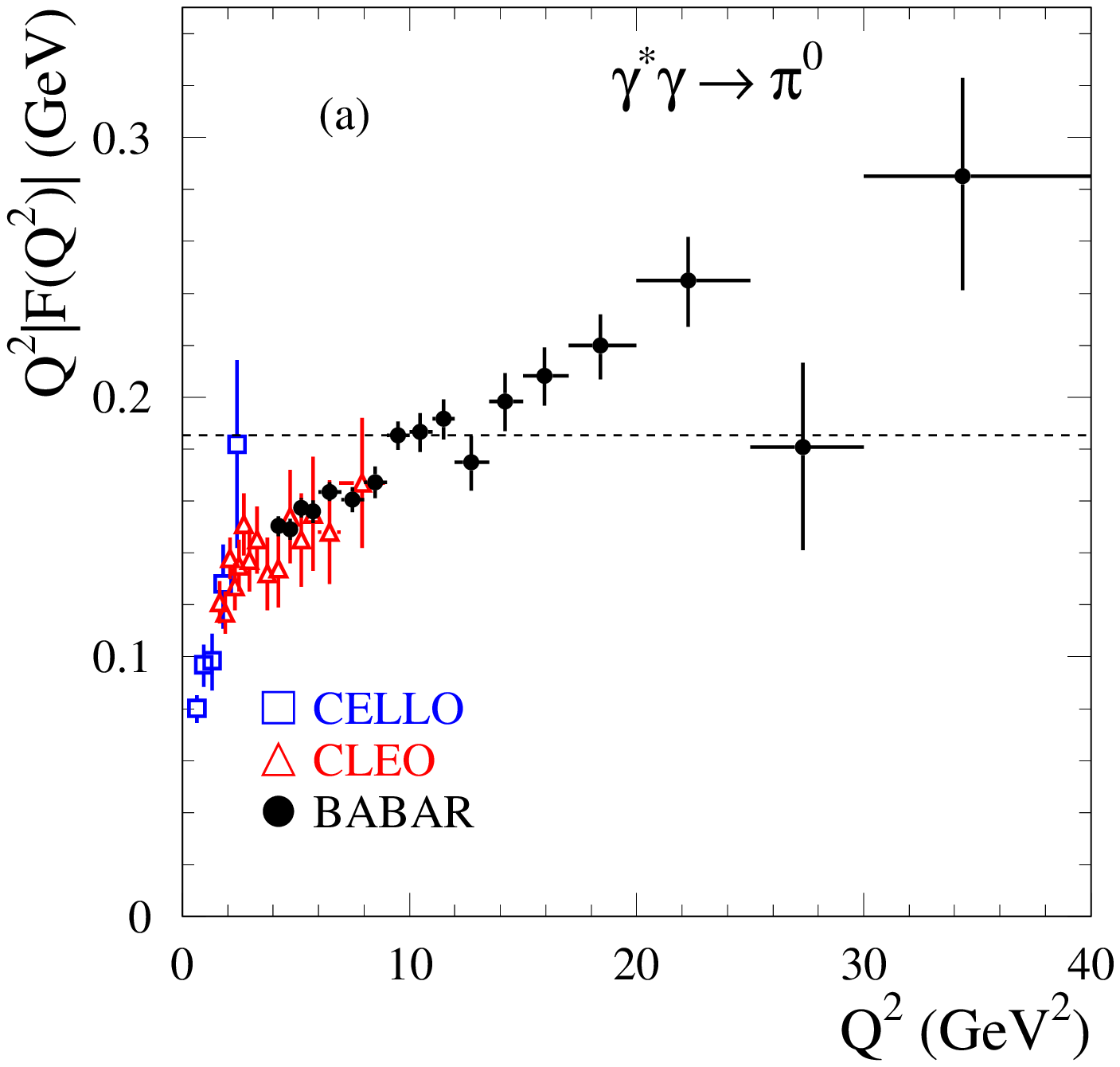, width=6.7cm}
\\
\epsfig{file=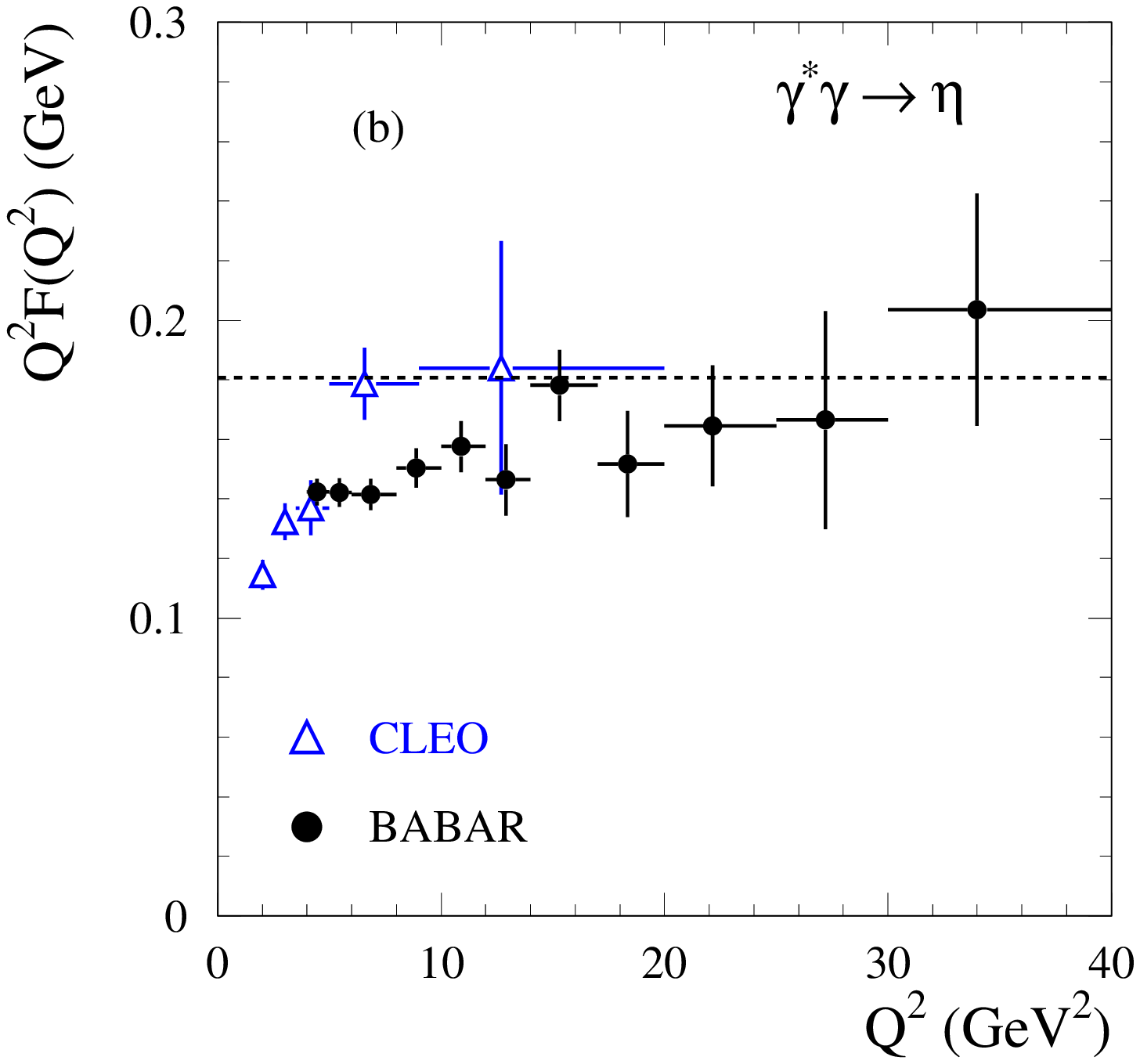, width=5.7cm}
\epsfig{file=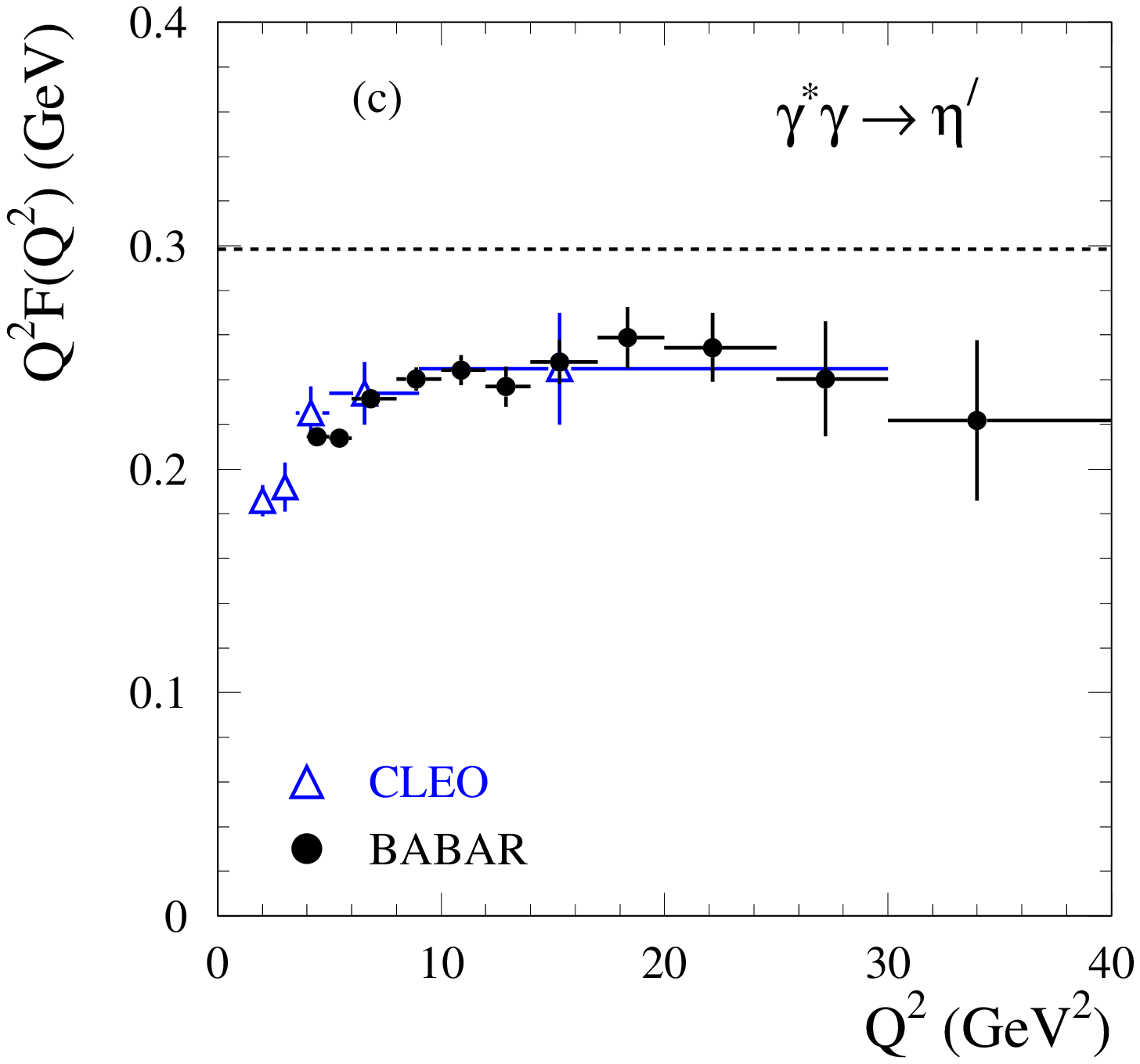, width=5.7cm}
\vspace{-0.75cm}
\caption{\small{BaBar results on the $Q^2$ dependence of meson transition form factors 
for 8a) $\pi^0$, (b) $\eta$ and (c) $\eta^\prime$ in comparison with previous measurements and
several models for the meson distribution amplitudes; results (b) and (c) are preliminary.}}
\label{diff}
\end{center}
\end{figure}

\coltoctitle{Meson Dalitz Decays with WASA-at-COSY}
\authortochead{J.~Klaja}
\label{art21}

\setcounter{chapter}{1}
\setcounter{section}{0}
\maketitle

\Section{Introduction}
The 4$\pi$ detector facility WASA~\cite{wasa_proposal} - an internal experiment at 
the COoler SYnchrotron COSY in J{\"u}lich, Germany - is designed for studies of 
light mesons decays, in particular (rare) $\eta$ decays.
Due to the unique capability of detecting neutral 
as well as charged particles a large number of the 
important decays modes can be measured in order to
study symmetries and symmetry breaking, hadron structure and hadron dynamics.\\
Here, we report the investigation of the electomagnetic transition form factor,
studied in meson Dalitz decays.

\Section{Meson Dalitz Decays}
Electromagnetic decay modes of light unflavored mesons
include Dalitz decays
$A \to B~\gamma^{\star} \to B l^{+}l^{-} $. Here, meson $A$
decays into an object $B$ (another meson or photon) and a lepton pair, formed
by internal conversion of an intermediate virtual photon with invariant mass $M$.
If the particles $A$ and $B$ were point-like, then  calculations for mass distributions
and decay widths would be straightforward using standard quantum electrodynamics 
(QED). However, the decay rate of this process is modified by the
electromagnetic structure arising at the vertex of the transition. Deviations of measured
quantities from the QED predictions are described by a transition form factor.
Vector mesons have  the same quantum numbers ($J^{P}= 1^{-}$) as the photon, 
and couple directly to real and virtual photons (the latter can be 
converted into dileptons). A virtual photon can interact with hadrons
not only directly but also after a transition to a virtual vector meson
- this type of interaction is called vector-meson dominance. 
The transition form factor is extracted by comparing the measured dilepton invariant
mass spectrum with the point-like QED predictions.\\ 
The form factor can be parameterized with a pole approximation ($\Lambda$): 
\begin{equation}
F={{1}\over{1-{q^2\over\Lambda^2}}} \approx 1 + { q^2\over\Lambda^2}.
\end{equation}
to get the parameter $\Lambda$ which can be then compared with theoretical 
model predictions.\\
Until now, significant results on transition form factors have been obtained
by the Lepton-G experiment~\cite{landsberg}. For the $\eta$ meson the $\Lambda^{-2}$
value has been found to be (1.9$\pm$0.4)~GeV$^{-2}$, which is compatible with standard VMD.
Experimental measurement of the transition form factor for $\omega \to \pi^{0}e^{+}e^{-}$ decay 
reveal a significant deviation with the fitted $\Lambda^{-2} = (2.36 \pm 0.21)$ GeV$^{-2}$ 
exceeding the VMD calculation by 3 standard deviations.\\
Recently, the heavy ion experiment NA60 has measured the electromagnetic transition
form factors of $\eta$ and $\omega$ mesons~\cite{na60}. The values for Lambda agree with older data,
comfirming the strong enhancement of the $\omega$ form factor beyond the expectation from VMD.
However, NA60 is a heavy-ion experiment and some additional assumptions enter in the analysis.   
On the theoretical side, a new counting scheme for the theory of pseudoscalar and vector mesons
has recently been introduced to go beyond ChPT to include the vector mesons in a systematic way~\cite{leupold1}. 
A leading-order calculations for the $\omega\pi$ transition form factor yield an improvement in the descrition
of the experimental data~\cite{leupold2}. However, the available experimental data is still too
scarce and additional data are needed.
In this context, forthcoming data from WASA-at-COSY will contribute to understanding the problem.

\Subsection{$\eta$ meson decay}
Before investigating the $\omega$ decays, it is relevant to study the $\eta$ Dalitz decay by
measuring electron-positron pairs. The $\eta \to \gamma e^{+} e^{-}$ decays
are studied with WASA-at-COSY  both in proton-proton~\cite{himani} and 
proton-deuteron~\cite{gosia} collisions. A thirty thousand $\eta \to \gamma e^{+} e^{-}$ events are expected 
from the combined {\it pp} and {\it pd} data sets. Dilepton pairs are selected by using the reconstructed 
momentum and charge state of the particle from the tracking detectors. The main difficulty in selecting 
$e^{+}e^{-}$ pairs is the discrimination against the dominant pion background. For the higher invariant
masses of $e^{+}e^{-}$ pairs, the main source of the background comes from multipion production, like
the $\eta \to \pi^{+}\pi^{-}\gamma$ and $\eta \to \pi^{+}\pi^{-}\pi^{0}$ decay channels. Also, the
$\eta \to \gamma\gamma$ decay channel has a significant contribution where one of the photons undergoes
external conversion. Applying particle identification and restrictions based on the $\eta \to \gamma e^{+} e^{-}$
kinematics, the amount of background is significantly reduced.\\
In the {\it pp} case, the analysis for the determination the transition form factor is in progress.\\
In the {\it pd} case, the background coming from channels with pions in the final state is still dominant. 
The double Dalitz decay $\eta \to e^{+}e^{-}e^{+}e^{-}$ is studied using the $pd \to~^{3}He \eta$ reaction at a beam energy 
of 1.0 GeV, with the goal to establish the branching ratio and to determine the $\eta$ transition form factor
$F(q_{1}^{2},q_{2}^{2})$. 30 events candidates are identified being consistent with the current upper 
limit of $< 6.9 \times 10^{-5}$~\cite{nakamura}.   

\Subsection{$\omega$ meson decay }
The WASA-at-COSY Collaboration has begun the program for $\omega$ decays. The first beam time, using the 
$pd \to~^{3}He \omega$ reaction with beam momentum  2.196 GeV/c, is scheduled for four weeks in spring 2011. The
$\omega \to \pi^{+}\pi^{-}\pi^{0}$ Dalitz plot will be investigated. Futhermore, the  
$\omega \to \pi^{0}e^{+}e^{-}$ Dalitz decay will be studied as well as $\omega \to \pi^{0}\gamma$ as reference channel.\\
In a separate one week of experiment, the $pp \to pp\omega$ reaction will be measured. We will test the possibility of studying
$\omega$ decays with smaller branching ratios, in particular the conversion decays $\omega \to \pi^{0}e^{+}e^{-}$ and
$\omega \to \eta e^{+}e^{-}$. By comparing the data quality of the {\it pp} and {\it pd} data sets we will choose the better
reaction for the determination of the $\omega\pi$ transition form factor.

\coltoctitle{\boldmath Study of Anomalous $\eta$ Decays}
\authortochead{T.~Petri}
\label{art20}

\setcounter{chapter}{1}
\setcounter{section}{0}
\maketitle

\Section{Introduction and Agenda}
We report about a study\,\cite{Petri} of  
all anomalous $\eta$  decays and the corresponding decays of the $\eta'$ and, if kinematically allowed, $\pi^0$ mesons 
in the framework of vector meson dominance (VMD) models of the `hidden gauge' class\,\cite{hidden,Benayoun}.
These decays are interrelated by  the Wess-Zumino-Witten action 
and mostly belong to the group of rare or even very rare
decays. 
While in the workshop 
we also discussed the double-Dalitz decays of the above-mentioned pseudoscalar mesons into two dilepton pairs, 
in this presentation we focus on the decay of the $\eta$  meson into $\pi^+\pi^- e^+ e^-$  pairs. This process 
is governed by the box anomaly in the chiral limit, whereas in the
framework of  the so-called {\em hidden gauge} VMD  the decay amplitude features  
a subtle interplay of a triangle-anomaly term involving two virtual vector mesons and contact interactions of box-anomaly type, see,
{\it e.g.},  Refs.\,\cite{Bijnens,Picciotto,Holstein}.
Moreover, these decays are remarkable since
 a  measurement of an asymmetry 
  of the dihedral angle $\phi$ between the pion- and the lepton-pair planes  would signal the presence of an unconventional 
 flavor-conserving CP violating mechanism  of electric-dipole type. Note that the latter is sensitive to the electromagnetic
 coupling to the $s$-quark--$\bar s$-antiquark content of the decaying meson 
 and is therefore neither tested by flavor changing transitions in the kaon or $B$ meson sectors of the Standard Model 
 nor by the empirical upper bound on the electric dipole moment of the neutron, see Refs.\,\cite{Geng,Gao}.
 
\Section{Results and Discussion}
In Table\,\ref{tab:etappee} we list our results  for the branching ratio ${\rm BR}(\eta \to \pi^+\pi^- e^+e^-)$
that we calculated in the framework of the  {\em hidden gauge} VMD model \,\cite{hidden} and the modern refinement of Ref.\,\cite{Benayoun} 
together with the prediction of unitarized chiral perturbation theory (UChPT) of Ref.\,\cite{Borasoy} and  the
most recent measurement by the KLOE collaboration\,\cite{KLOE}. Furthermore,  by comparing our predictions of the asymmetry $A_{\phi}$ of the hidden gauge 
model\,\cite{hidden} and
the modified VMD model\,\cite{Benayoun}
with the KLOE measurement\,\cite{KLOE} 
(listed in the 2nd row of Table\,\ref{tab:etappee}), we can constrain
the free  parameter $G$  of the unconventional CP-violating mechanism 
of Refs.\,\cite{Geng,Gao} as $-0.9 < G < 1.2$. Finally, given the constrained $G$ values, 
the row ${\rm BR}_{E_1}$ of Table\,\ref{tab:etappee} shows that the 
contribution of the electric dipole amplitudes of Refs.\,\cite{Geng,Gao} is by far too small to alter the value of the tabulated branching ratio(s) substantially.
\begin{table}[htb]
\begin{center}
\renewcommand{\arraystretch}{1.5}
\begin{tabular}{lc|c|c|c|c}
         &         & hidden gauge & modified VMD  & UChPT \cite{Borasoy}&
         KLOE@DA$\Phi$NE \cite{KLOE} \\\hline \hline
${\rm {\rm BR}}$ &{\tiny($10^{-4}$)}  &  $3.14 \pm
  0.17$ & $3.02\pm 0.12$ &  $2.99^{+0.08}_{-0.11}$&$2.68 \pm 0.09 \pm 0.07$\\\hline
$A_\phi$&{\tiny($10^{-2}$)}              &   $\left(-3.88 \pm
  0.11\right)G$ & $\left( -3.95 \pm 0.08\right) G$&&$-0.6 \pm 2.5 \pm 1.8$ \\\hline
  
${\rm {\rm BR}}_{E_1}$& {\tiny($10^{-6}$)}            &  \multicolumn{2}{c|}{1.6 $|G|^2$
  independent on VMD }&& 
\end{tabular}
\caption{Theoretical values and experimental data of the
    branching ratio and asymmetry  $A_\phi$ in the decay $\eta \to
    \pi^+ \pi^- e^+ e^-$ and the 
    contribution of the electric terms to the branching ratio.
    \label{tab:etappee} }
\end{center}
\end{table}

Note that the KLOE measurement of the branching ratio is two standard deviations below the  theoretical results listed in  Table\,\ref{tab:etappee}, which are all compatible.
By normalizing the branching ratio ${\rm {\rm BR}}(\eta \to \pi^+ \pi^- e^+ e^-)$ to the branching ratio ${\rm {\rm BR}}(\eta \to \pi^+ \pi^- \gamma)$ most of the systematical errors in 
the calculations  drop out. In fact, in the reported VMD calculations only the sensitivity to
the $\rho$ meson mass and to the relative strengths of vector meson dominance and contact terms  remains -- see the reduced errors for the relative branching ratios listed in the first three columns of Table\,\ref{tab:fracetappee}.  The comparison with the fourth row of Table\,\ref{tab:fracetappee} that  lists the relative ratio of  
the measured KLOE ${\rm BR}(\eta \to \pi^+ \pi^- e^+ e^-)$  to the ${\rm BR}(\eta \to \pi^+ \pi^- \gamma)_{\rm PDG}=(4.60\pm0.16)\times10^{-2}$ of the PDG\,\cite{PDG1,PDG2} is thus a problem for theory or for experiment or for both.
\begin{table}[!htp]
\begin{center}
\renewcommand{\arraystretch}{1.5}
\begin{tabular}{c|c|c|c|c|c}
 hidden gauge & modified VMD  & UChPT \cite{Borasoy}
  &${\displaystyle  \frac{{\rm KLOE}\,  \cite{KLOE}}{{\rm PDG}\, \mbox{\cite{PDG1,PDG2}}}}$
 &${\displaystyle \frac{{\rm KLOE}\, \cite{KLOE}}{{\rm CLEO}\,  \cite{CLEO}}}$
 &${\displaystyle \frac{{\rm PDG(2008)}\, \cite{PDG1}}{{\rm PDG}\, \mbox{\cite{PDG1,PDG2}}}}$\\[2mm]\hline \hline 
  $6.32 \pm 0.01$ & $6.33\pm 0.01$ &$6.39_{-0.11}^{+0.08}$
  &$5.83\pm0.31$
  &$6.77\pm0.44$
  &$9.13\pm2.63$ 
\end{tabular}
  \caption{The relative branching ratio $\Gamma(\eta\to \pi^+ \pi^- e^+ e^-)/ \Gamma(\eta\to \pi^+ \pi^- \gamma)$ times
$10^3$.
           \label{tab:fracetappee}}
\end{center}
\end{table}

Therefore, in Ref.\,\cite{KLOE} it was argued that the KLEO branching ratio, if
normalized to the smaller  
 ${\rm BR}(\eta \to \pi^+ \pi^- \gamma)_{\rm CLEO}=(3.96\pm0.14\pm0.14)\times10^{-2}$ of the CLEO collaboration\,\cite{CLEO} 
 than to the above-mentioned PDG value, 
would be compatible with
the  theoretical result of Ref.\,\cite{Borasoy},  see the fifth column of Table\,\ref{tab:fracetappee}. Only for completeness, 
we have also listed the rather inaccurate ratio of the 2008 PDG value\,\cite{PDG1} for ${\rm BR}(\eta \to \pi^+ \pi^- e^+ e^-)$ to the above mentioned 
$ {\rm BR}(\eta \to \pi^+ \pi^- \gamma)_{\rm PDG}$ \,\cite{PDG1,PDG2}. Note, however, that the preliminary
result of the KLOE collaboration for the relative  branching ratio $\Gamma(\eta \to \pi^+ \pi^- \gamma)/\Gamma(\eta\to\pi^+\pi^-\pi^0)$, which was reported in this  workshop\,\cite{KLOE_prem}, is compatible with  the 
PDG average\,\cite{PDG2},
such that the above mentioned problem remains.

Further details --  also on the other  $\eta,\eta'\to \pi^+\pi^- l^+ l^-$ decays and the double Dalitz decays --  can be found in the Diplom thesis\,\cite{Petri} and, partially, in the upcoming paper\,\cite{Petri_pub}.

\coltoctitle{\boldmath Analysis of $\eta\to\pi^+\pi^-\gamma$ Measured with the WASA Facility at
COSY}
\authortochead{C.~F.~Redmer}
\label{art25}

\setcounter{chapter}{1}
\setcounter{section}{0}
\maketitle

\section{Introduction}
In the chiral limit the process $\eta\rightarrow\pi^+\pi^-\gamma$ is described by the box-anomaly
term of the Wess-Zumino-Witten Lagrangian~\cite{Wess:1971,Witten:1983}. Meanwhile, the dynamic
range of the decay, $4m_\pi^2 \leq s_{\pi\pi} \leq m_\eta^2$, is well above the chiral limit,
giving rise to a significant difference between the predicted and the experimental value of the
decay rate. Higher order terms of the chiral Lagrangian become important in order to achieve a
correct description. Corrections at the one-loop level~\cite{Bijnens:1990} can reduce the
discrepancy between theory and experiment, however, they are not sufficient to reproduce
experimental results. Efforts have been made to incorporate final state interactions by unitarized
extensions to the box anomaly term~\cite{Picciotto:1992,Holstein:2002,Benayoun:2003,Borasoy:2004}.
The validity of the models has to be tested with differential distributions of the Dalitz plot
variables, but experimental data are scarce. Moreover, the two statistically most significant
measurements~\cite{Gormley:1970,Layter:1973} have been published without efficiency corrections.
Instead, the model descriptions presented along with the data have been folded with the detector
efficiency. Recent attempts to interpret the data~\cite{Benayoun:2003,Borasoy:2007} yield ambiguous
results from the two data sets. The missing efficiency corrections are held responsible for the
ambiguities. A new high statistics measurement is described in the following.

\section{Experiment}
The measurement was carried out with the WASA facility at COSY~\cite{WASA:2004}. $\eta$ mesons have
been produced in the fusion reaction pd~$\rightarrow ^3$He~X at a proton beam momentum of 1.7~GeV/c.
The meson production is tagged by the identification of the helium ions in the detection system. To
reconstruct the decay $\eta\rightarrow\pi^+\pi^-\gamma$, signals from one neutral and two charged
particles are required in coincidence with the $^3$He, which can be associated with pion and photon
candidates. Further conditions are applied to reduce contributions from two and three pion
production, including the decay $\eta\rightarrow\pi^+\pi^-\pi^0$, which are the major background
reactions. Finally, a kinematic fit has been performed using energy and momentum conservation as the
only constraints. Remaining background is subtracted bin-wise by determining the number of $\eta$
mesons from the corresponding invariant mass distribution of the $\pi^+\pi^-\gamma$ system. The
resulting $13740\pm140$ events of the decay $\eta\rightarrow\pi^+\pi^-\gamma$ form the largest
sample taken in an exclusive measurement so far.

Fig.~\ref{fig:final} shows the background subtracted and efficiency corrected angular distribution
of the $\pi^+$ relative to the photon in the two-pion rest frame and the photon energy distribution
in the $\eta$ rest frame. The systematic uncertainties, indicated in gray, are currently dominating
the total error.

\begin{figure}[t]
    \centerline{\includegraphics[width=0.43\linewidth]{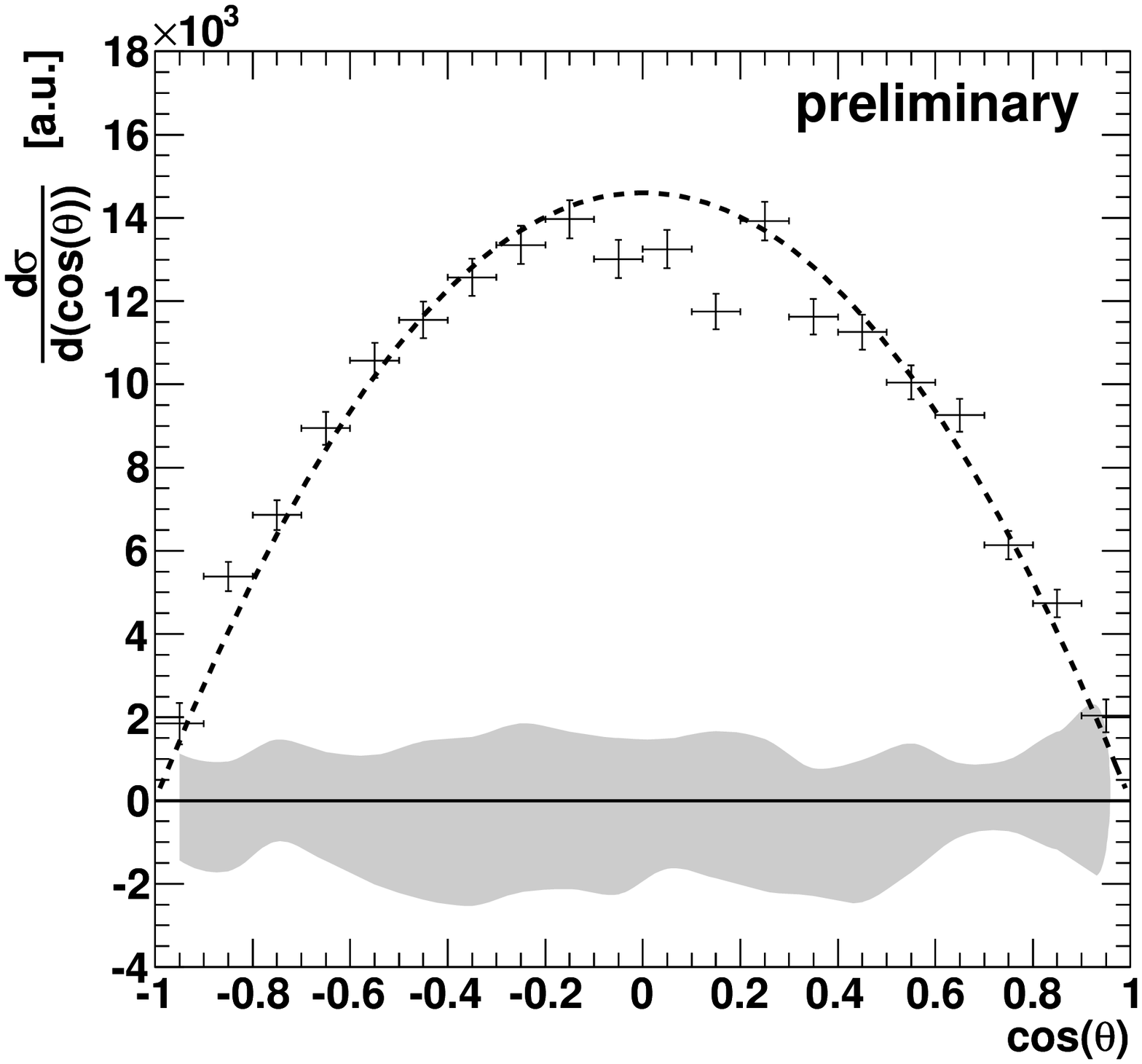}\hfill%
                \includegraphics[width=0.43\linewidth]{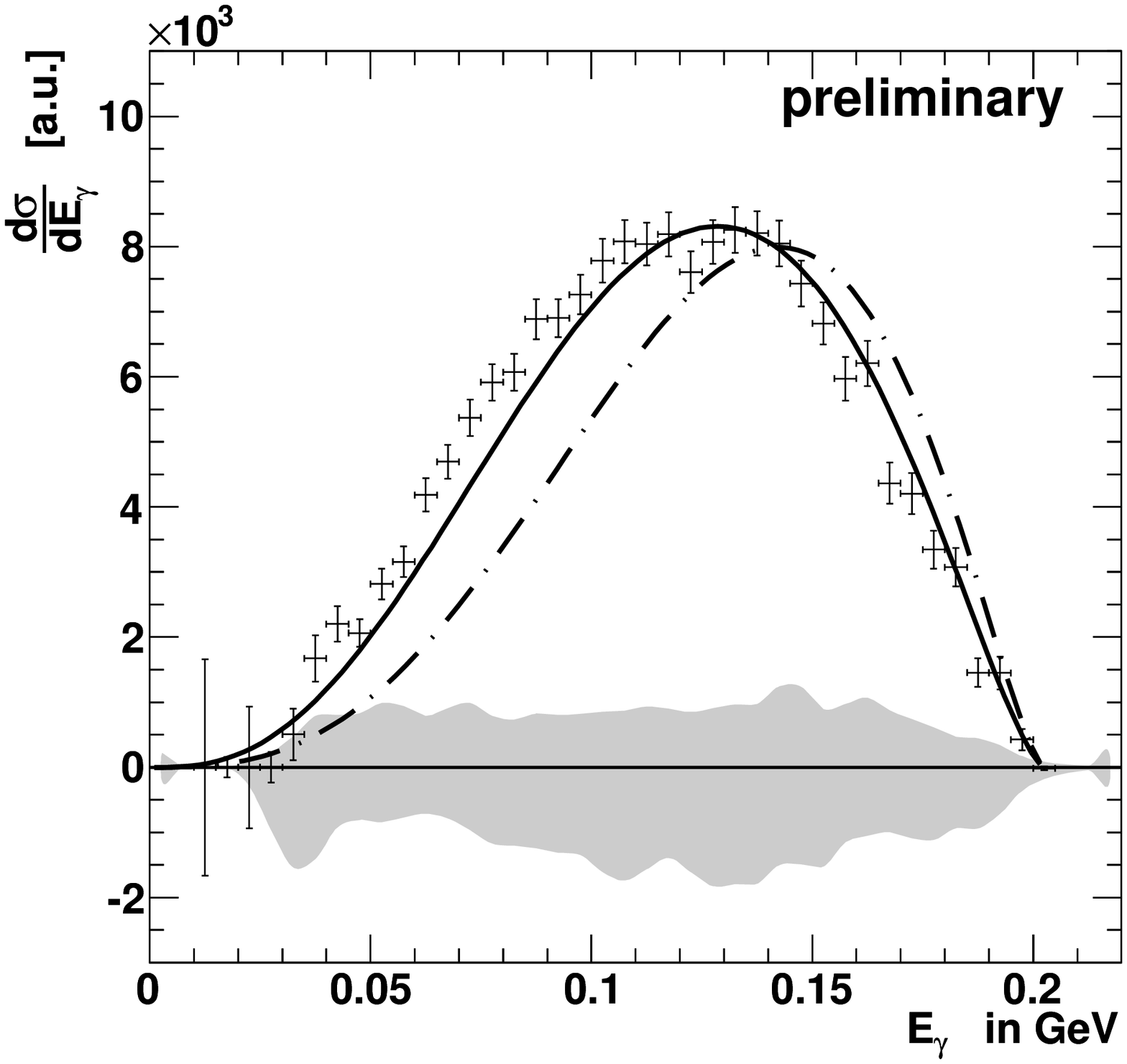}}
    \caption{\label{fig:final} The background subtracted and efficiency corrected angular
distribution of the pions (left) and the photon energy distribution (right). The shaded areas
indicate the systematic uncertainties. The angular distribution agrees with a relative {\it p} -
wave of the pions (dashed curve). The line shape of the photon energy distribution is compared to
predictions of the simplest gauge invariant matrix element (dot-dashed curve) and a VMD-based
calculation~\protect\cite{Picciotto:1992} (solid curve).}
\end{figure}

The angular distribution can be described by the function $\frac{d\sigma}{d \cos(\theta)}=
A\cdot\sin^2(\theta)$, which is shown with the dashed curve in Fig.~\ref{fig:final}. The dependence
corresponds to a relative {\it p} - wave of the pions. Evidence for angular momenta other than l=1
has not been found.
 
The line shape of the photon energy distribution cannot be described by the simplest gauge invariant
matrix element (dot-dashed curve), which confirms the observations of previous
experiments~\cite{Gormley:1970,Layter:1973}. A combination of the box-anomaly term with a VMD model
to describe final state interactions~\cite{Picciotto:1992} achieves a considerably better agreement.
The line shape prediction (solid curve) reproduces the measured distribution within the combined
statistical and systematical error.

\section{Outlook}
In order to perform more precise studies of the box-anomaly in the decay
$\eta\rightarrow\pi^+\pi^-\gamma$ the total error needs to be decreased. In recent production runs
with the WASA facility at COSY, further high statistics data sets on $\eta$ decays have been
acquired. The data will not only reduce the statistical error, but also help to reduce the
systematic uncertainties.

\end{papers}
\cleardoublepage

\pagestyle{plain}
\begin{center}
 \vspace*{0.3\textheight} {\Huge\textbf{\boldmath Meson Production}}\\
 \vspace{2cm} {\Large\textbf{in Photo Reactions and from Light Ion Collisions}}
 \addcontentsline{toc}{part}{Meson Production in Photo Reactions and from Light Ion Collisions}
\end{center}
\cleardoublepage
\pagestyle{fancy}

\begin{papers}
\coltoctitle{\boldmath $\eta$ Production in Nucleon-Nucleon Collisions}
\authortochead{C.~Wilkin}
\label{art2}

\maketitle

Preliminary results are reported on three new experiments that
measured the production of the $\eta$ meson in nucleon-nucleon
collisions near threshold. In addition, greater insight on the
existence of the $_{\eta}^{3}$He quasibound state is provided by
recent photoproduction data.

The $\eta d$ interaction was studied at CELSIUS in two different
kinematical regimes, \textit{viz} quasi-free $pd \to p_{\rm sp}
d\eta$~\cite{CAL1997} and the same reaction at much lower energies,
\textit{i.e.}\ well below the threshold in nucleon-nucleon
collisions~\cite{BIL2002}. A consistent \textit{FSI} description of
both data sets was achieved by dividing the $\eta d$ invariant mass
distribution by the corresponding phase space.

New data have been taken on $dp\to p_{\rm sp}dX$ at the maximum COSY
deuteron beam energy of 2.27~GeV and the meson $X$ identified from
the missing mass~\cite{DYM2010}. Here $p_{\rm sp}$ denotes a
spectator proton. Since the central neutron energy is below the
$d\eta$ threshold, only the higher Fermi momenta contribute to $\eta$
production. Below-threshold data provide a robust method for
background subtraction and the resulting $\eta$ signal is very clean
and the whole $\eta$ angular domain sampled. The analysis, which is
still in progress, suggests an $\eta d$ \emph{FSI} that is not as
strong as that found at CELSIUS. The prime reason for the difference
is the question whether the data should be compared to phase
space~\cite{BIL2002} or to a pure spectator model~\cite{DYM2010}.
Theoretical work~\cite{TEN2005} suggest that the truth lies between
these two extremes and this will have to be incorporated in the
ongoing analysis.

In order to study the effects of $s$-wave rescattering of the $\eta$
meson from a proton pair, and hence investigate the $\eta pp$
\textit{FSI}, it is important to know at what point higher partial
waves are needed for the description of the $pp\to pp\eta$ reaction.
The production of the $\eta$ in proton-proton collisions was
investigated by the CELSIUS-WASA collaboration at $Q = 40$ and 72~MeV
by detecting the $\eta$ through its $3\pi^0$ decay~\cite{PAU2006}.
However, the data from the two-photon decay branch have been
subjected to a much more refined analysis, which is now approaching
completion~\cite{PET2009}.

The $pp\to pp\eta$ Dalitz plots look similar when the meson is
detected through its $3\pi^0$~\cite{PAU2006} or $2\gamma$
decay~\cite{PET2009}. The distributions at both 40 and 72~MeV show a
deep valley along the diagonal where $m_{\eta p_1} \approx m_{\eta
p_2}$. This is probably due to the $\eta$ being able to form the
$N^*(1535)$ with only one nucleon at a time. The valley implies that
there must be higher partial waves in both the $pp$ and $\eta{pp}$
systems, at least $L_{pp}\ell_{\eta}=Pp$. In order to unfold the
acceptance, a partial wave fit was made to the combined 40 and 72~MeV
data with constant amplitudes. Although the $N^*(1535)$ is not
explicitly included, factors arising from the angular momentum
barriers and the $pp$ \textit{FSI} are.

The $pp$ and $\eta p$ invariant mass distributions are well
reproduced at 40~MeV but for the 72~MeV data, where the statistics
are higher, the description in the \textit{FSI} region is poorer. The
better description of the data at 40~MeV persists also for the $\eta$
angular distributions. In the simple approach used, deviations from
isotropy arise from $Ss$-$Sd$ interference and these must be too
small in the model at 72~MeV. This suggests that even higher partial
waves are required at this energy.

With parameters tuned to describe the 40 and 72~MeV data, the shapes
of the COSY-11 data at 15.5~MeV~\cite{MOS2004} are well described. In
the analysis, higher partial waves in the $pp$ system are vital for
the description of the data at the largest $m_{pp}$; it the
significant contribution from the $Ps$ wave that gives more events at
high $m_{pp}$ and hence low $m_{p\eta}$. Since there is no associated
angular dependence, this is NOT a proof and an unambiguous separation
of $Ps$ from $Ss$ would require a measurement of the initial
spin-spin correlation parameter.

Although the poorer description of the $\eta$ angular distribution at
72~MeV is probably a signal for the influence of even higher partial
waves, the constant amplitude \textit{ansatz} overestimates the
increase in the total cross section with energy. This was to be
expected and, despite this, the methodology is quite sufficient for
its primary purpose of performing acceptance corrections, which are
relatively small at WASA.

To study the $s$-wave $\eta pp$ \textit{FSI} experimentally, we need
to control the higher partial waves; the $\eta d$ case is simpler
because it is a two-body system. The ANKE spectrometer has only
limited acceptance, but it can measure well $pp \to \{pp\}_SX$, where
the diproton $\{pp\}_S$ has an excitation energy below 3~MeV, so that
it is dominantly in the $^{1\!}S_0$ state. Results have already been
published for $\eta$ production in quasi-two-body kinematics at
$Q\approx 55$~MeV~\cite{DYM2009} and similar data have been taken at
ANKE for the $\eta^{\prime}$, where about 500 events are expected.
Under such conditions, only the $Ss$ and $Sd$ partial waves are
important and so many uncertainties are removed.

Although outside the remit of the title, the new data on
$\gamma^3$He$\,\,\to\eta^3$He~\cite{PHE2010,KRU2010} are too
interesting for me not to mention them. The COSY data on
$dp\to\eta^3$He show clearly that there is a pole in the $\eta^3$He
sector for an excess energy $Q$ at
$|Q|<1$~MeV~\cite{MER2007,SMY2007}, though whether this is
quasi-bound or anti-bound remains ambiguous. The sign of the angular
asymmetry changes as a function of $Q$ and this probably reflects the
phase variation in the $s$-wave amplitude associated with the pole.

Analogous results on $\gamma^3$He$\,\,\to\eta^3$He confirm the pole
interpretation since the cross section jumps to one third of its
maximum within the first energy bin~\cite{PHE2010,KRU2010}. Even more
convincing evidence comes from a study of the angular dependence.
Away from threshold the data are described well by impulse
approximation, where the kinematics strongly favour the $\eta$
emerging in the beam direction. In the first two energy bins this
tendency is reversed, as it is for $dp\to\eta^3$He, and the reason is
the same. The strong $s$-wave $\eta^3$He \textit{FSI} changes the
$s$-$p$ interference and, using the same parameters as employed for
the COSY data~\cite{WIL2007}, the energy variation of the slope in
$\gamma^3$He$\,\,\to\eta^3$He can be understood. It is therefore
important to investigate the interaction of the $\eta$ with other
light nuclei to track more precisely how the pole moves with nucleon
number $A$. It is in this context that studies in the two-nucleon
sector assume an even greater importance.

\coltoctitle{Photon Induced Measurements}
\authortochead{B.~Krusche}
\label{art3}

\setlength{\parindent}{0pt}
\setlength{\parskip}{10pt}
\maketitle

Photoproduction of mesons has been studied in a series of experiments at the Bonn ELSA and the Mainz
MAMI accelerators. Both sets of experiments use very efficient 4$\pi$ electromagnetic calorimeters
(Crystal Barrel and TAPS in Bonn, Crystal Ball and TAPS in Mainz) for the detection of meson decay
photons as well as neutral (neutrons) and charged (protons, charged pions) particles. These setups
allow for example the measurement of the production of neutral mesons (e.g. $\eta$, $\eta '$) or
even pairs of neutral mesons such as $\pi^o\pi^o$, $\pi^o\eta$ in coincidence with recoil neutrons,
which is unique and nicely complements the experimental program at other facilities like the CLAS
experiment at Jlab, which is more focused on (at least partly) charged final states. 

So far, only very few precise experimental data are available for the photoproduction of mesons off 
the neutron. However, it is clear that this is the only direct access to the iso-spin dependence of
the electromagnetic excitation of N${^\star}$  resonances and due to SU(3) selection rules
\cite{Moorehouse_66} at least for some states the electromagnetic coupling to the proton is strongly
suppressed. The non-availability of free neutrons as a target requires of course the study of
quasi-free photoproduction reactions off neutrons bound in light nuclei, in particular the deuteron.
Apart from the technical complications stemming from the coincident detection of recoil neutrons,
also nuclear effects like e.g. Fermi motion and final state interaction (FSI) processes have to be
considered. However, fortunately both types of problems can be systematically investigated. This has
been demonstrated in detail for our results for quasi-free photoproduction of $\eta$ and $\eta '$
mesons off the neutron. The technical problem of controlling the notoriously difficult detection
efficiency of recoil neutrons can be solved by a comparison of cross section data measured in
coincidence with recoil neutrons ($\sigma_n$) to the difference of the inclusive cross section
($\sigma_{np}$) (without any condition for recoil nucleons) and the cross section with coincident
recoil protons ($\sigma_p$). Since for both reactions coherent photoproduction is negligible, the
simple relation $\sigma_{np}=\sigma_p+\sigma_n$ must hold as long as proton {\em and} neutron
detection efficiencies are correct. Agreement was indeed found within statistical uncertainties,
demonstrating that no significant systematic problems are involved. Nuclear effects can be tested
via a comparison of quasi-free proton cross sections with free proton cross sections. Here, it was
found for $\eta$ and $\eta '$ production that no significant effects beyond Fermi motion must be
considered. This is not too surprising since the momentum miss-match of participant and spectator
nucleons is large and the overlap of the final state wave function of the two-nucleon system with
the deuteron wave function is small (both reactions have dominant contributions from S$_{11}$
resonances excited via the $E_{0^+}$ spin-flip multipole). In case of the $\eta '$ channel even
Fermi motion does not cause large effects since the excitation functions have no fast varying
structures. The results for this channel, which have been already accepted for publication
\cite{Jaegle_etap} show some difference between the proton and neutron data in particular at
intermediate incident photon energies around $E_{\gamma}\approx$ 1.8 GeV, which might point to
different resonance contribution. However, so far model analyses of both the proton as well as the
neutron data are inconclusive since they are not sufficiently constrained by the angular
distributions alone. The measurement of polarization observables is clearly necessary. 

So far, the most interesting results are related to the $n\eta$ final state, which has been studied
in much detail. Already the results reported in \cite{Jaegle_08} have confirmed, that the neutron
excitation function shows a narrow structure of unknown nature around incident photon energies of 1
GeV. These data have now been further analyzed using a different concept \cite{Jaegle_eta}.
The excitation functions for protons and neutrons have been constructed in dependence on the final
state invariant mass rather than the incident photon energy. The advantage of this analysis is, that
it does not suffer from Fermi smearing. This analysis was possible since although the kinetic
energy of the recoil neutrons was not directly measured, the neutron angular information (together
with incident photon energy and measured four-momentum of the $\eta$) is sufficient for a full
reconstruction of the final state kinematics. The proton data reconstructed this way agree
excellently with free proton data. The neutron data show now an even much more pronounced structure
with a width around 25 MeV (still including some contribution from experimental resolution). New
data from MAMI show this structure with even much better statistical quality \cite{Werthmueller} 
and even in an experiment using $^3$He target nuclei, it is clearly seen in the excitation function
with coincident neutrons \cite{Witthauer}. Altogether, the existence of this structure is by now
established beyond any doubt but it's nature is still controversially discussed. Currently,
experiments with polarized deuteron targets (frozen spin deuterated buthanol) and polarized photon
beams at MAMI and ELSA try to identify the responsible partial waves via the measurement of double
polarization observables. 

Finally it should be mentioned, that the analysis of many other channels for quasi-free
photoproduction off the deuteron is almost finished and results will be published soon. These are
for example data for single $\pi^o$-, $2\pi^o$-, $\pi^o\pi^{\pm}$-, $\pi^o\eta$-, $\pi^{\pm}\eta$-,
and $K^o\Lambda$-production. In case of the double meson channels, not only results for differential
cross sections but also for the helicity asymmetry $I^{\circ}$ (circularly polarized photon beam,
unpolarized target) have been already obtained. Also here it could be demonstrated by a comparison
to free proton results that the effects from nuclear Fermi smearing can be well controlled in the
analysis. This observable promises very interesting additional information of the reaction dynamics.
First measurements of it for double pion production off the free proton \cite{Zehr_09} had
demonstrated it's large sensitivity but also an almost complete failure of available reactions
models.   

More recently, we started also with an investigation of coherent photoproduction of light nuclei,
which can be used to extract additional information for the iso-spin dependence of reaction cross
sections. This channel is strongly suppressed for $\eta$ and $\eta '$, however, larger cross
sections have been found not only for single $\pi^o$ but also for multiple meson final states like
$\eta\pi^o$, $2\pi^o$ and $3\pi^o$. These reactions can be used as iso-spin filters. For example, in
single $\pi^o$ production off the deuteron only $\Delta$-resonances contribute, while coherent
double $\pi^o$ production projects out the $N^{\star}$ states. Of particular interest is the
$\eta\pi^o$ channel, which might also give a new way of access to the search for $\eta$-mesic
states.
 
The other important topic in meson photoproduction off nuclei is the meson - nucleus interaction,
and for light nuclei in particular the search for meson - nucleus bound states like $\eta$-mesic
nuclei. Here the question is, whether the properties of the strong interaction allow the formation
of meson - nucleus bound states, which depend very sensitively on the strength of the interaction.
Pionic atoms are well established and deeply bound pionic states \cite{Geissel_02} yielded important
results \cite{Kienle_04}, directly related to the predicted decrease of the chiral condensate in
nuclear matter. In this case the superposition of the repulsive s-wave $\pi^-$ - nucleus interaction
with the attractive Coulomb interaction gives rise to bound states. Neutral mesons on the other hand
can form bound states only in case of a sufficiently attractive strong meson - nucleus interaction.
Due to the strong s-wave $\eta$-nucleus interaction at threshold (caused by the coupling to the
S$_{11}$(1535) nucleon resonance) $\eta$-mesic states are among the best candidates for such
systems.    

The investigation of the threshold behavior of $\eta$-photoproduction off light nuclei is one
promising approach for the search for $\eta$-mesic states. Previous experiments have investigated
quasi-free and coherent photoproduction off $^2$H, $^{3,4}$He in the threshold region. In the case
of $^3$He Pfeiffer et al. \cite{Pfeiffer_04} found a very strong threshold enhancement, although
the experiment was somewhat limited in counting statistics. Once a (quasi-)bound state has been
formed the $\eta$-meson may also be re-captured by a nucleon, exciting it to the S$_{11}$-state
which decays with a $\approx$50~\% branching ratio to a nucleon - pion pair. This decay channel
should result in a peak-like structure in the excitation function of nucleon - pion pairs, emitted
back-to-back in the photon - $^3$He center-of-momentum (cm) system. Some indication for this
effects had also been reported by Pfeiffer et al. \cite{Pfeiffer_04}.

A new experiment at MAMI using the Crystal Ball/TAPS detector setup covering almost 4$\pi$ of the
solid angle \cite{Pheron_10} has confirmed the extremely strong rise of the coherent
$\eta$-production reaction at threshold with much better statistical quality. However, it has also
shown, that the weak signal observed in \cite{Pfeiffer_04} for the peak-like structure in the $\pi^o
- p$ excitation function, although existent, can be attributed to contributions from $N^{\star}$
resonances to single $\pi^o$ production and is not related to an $\eta$-mesic state.

\coltoctitle{Meson Decays at MAMI--C - Results and Perspectives}
\authortochead{M.~Unverzagt}
\label{art4}

\setcounter{chapter}{1}
\setcounter{section}{0}
\maketitle

\Section{Experimental Setup}
The electron accelerator MAMI--C~\cite{Her83,Kai08} provides a very stable electron beam (energy
drift $\delta E < 100$~keV) with a maximum energy of $E_0 = 1604$~MeV. Within the A2 collaboration,
experiments are performed with a real photon beam. The photon beam for the production of the light
mesons is derived from the production of Bremsstrahlung photons during the passage of the MAMI
electron beam through a thin radiator. The Glasgow photon tagging spectro\-meter~\cite{Ant91,Hal96}
at Mainz provides energy tagging of the photons, through $E_\gamma = E_0 - E_{e^-}$, by detecting
the post--radiating electron energy.

The central $4\pi$--detector system, surrounding the liquid hydrogen target, consists of the
Crystal Ball calorimeter (CB)~\cite{Ore82,Sta01} combined with a barrel of scintillation counters
for particle identification and two layers of Multi--Wire Proportional--Chambers (MWPC). At forward
polar angles the TAPS detector~\cite{Nov91,Gab94} made of BaF$_2$--crystals provides acceptance for
particle detection.

\Section{Recent Results}
The experimental setup at MAMI--C is ideally suited for studying the neutral decays of the lightest
mesons. The key--investigation in this field was the determination of the Dalitz Plot Parameter
$\alpha$ from the isospin--breaking decay $\eta \to 3\pi^0$. Two MAMI results for $\alpha$ were
obtained from different data sets with different electron beam energies, $E_{e^-} =
883$\,MeV~\cite{Unv09} and $E_{e^-} = 1508$\,MeV~\cite{Pra09}, respectively. Thus, they can be
considered as independent measurements. These results, with $1.8 \cdot 10^6$ and $3 \cdot 10^6$
events, represent the two world highest statistics for the $\eta \to 3\pi^0$ decay. The two totally
independent values, $\alpha = -0.032 \pm 0.002_{\mathrm{stat}} \pm 0.002_{\mathrm{syst}}$ and
$\alpha = -0.032 \pm 0.003_{\mathrm{tot}}$, from MAMI agree perfectly with each other and are
consistent with other experiments. Recently, an analysis of MAMI--C data from 2009 confirmed these
results.

The rare, double--radiative decay $\eta \to \pi^0 \gamma \gamma$ has been investigated as there are
contradictions in its experimental decay width and its theoretical predictions. Although the
statistics gained at MAMI so far is lower than in the recently published result of the Crystal Ball
at AGS~\cite{Pra08}, a much better signal to background ratio could be achieved. Nevertheless, a
preliminary branching ratio $BR(\eta \to \pi^0 \gamma \gamma) = (2.25 \pm 0.46_{stat} \pm
0.17_{syst}) \cdot 10^{-4}$ was determined. The Crystal Ball at MAMI result agrees with the value
obtained at BNL within one standard deviation. Both are presently the most precise data for this
decay.

Another work was dedicated to the investigation of the Dalitz decays $\pi^0 \to e^+ e^- \gamma$ and
$\eta \to e^+ e^- \gamma$~\cite{Ber10}. The aim was to determine the electromagnetic transition form
factors, which provides significant information on the electromagnetic properties of the mesons and
is of crucial importance for testing hadronic models used in light--by--light contributions to
$(g-2)_\mu$. In the exclusive analysis of $\eta \to e^+ e^- \gamma$ 827 events were reconstructed,
which is an improvement by a factor 8 with respect to an old measurement by the SND collaboration,
the slope parameter of the associated transition form factor was determined to be consistent with
other experiments. The analysis of the $\pi^0$--Dalitz decay showed no deviation from the QED
prediction, as expected. This work also showed that it is possible with the A2--setup at MAMI to
investigate charged channels including electrons and positrons.

New upper limits for the C--violating decays $\omega \to \eta \pi^0$ and $\omega \to 3\pi^0$ were
calculated~\cite{Sta09}. Also for the first time a value for the upper limit of the branching ratio
of the $\omega \to 2\pi^0$ was determined. 

\Section{Detector Developments}
Since the $\eta'$ photoproduction threshold is at $E_{thr} \approx 1447$\,MeV, almost the full
photon energy range accessible at MAMI for $\eta'$ production is not covered by the Glasgow photon
tagging spectrometer. Therefore, a new tagging device (End--point tagger) is currently constructed.
At the maximum electron beam energy, $E_0 = 1604$\,MeV, photon energies between $E_\gamma \approx
1590$\,MeV and $E_\gamma \approx 1450$\,MeV can be tagged with the End--point tagger. To study the
decays of the $\eta'$ meson, the End--point tagger is essential for a clean trigger. The assembly of
the End--point tagger is underway and mid of 2011 first tests will be performed.

The MWPCs integrated in the Crystal Ball--setup have problems to work at high-rates meson--decay
experiments. Therefore, a Time Projection Chamber (TPC) with GEM read--out is currently investigated
as central tracking device. Not only the rate capability will be improved, but also the resolution
and the track reconstruction. In addition, the TPC could contribute to the particle identification.
First tests with an old TPC from Karlsruhe, Germany, were successful and 2011 a first prototype will
be constructed.

\Section{Perspectives at MAMI--C}
The A2--collaboration will continue to study light meson decays, especially the $\eta'$, after the
assembly of the End--point tagger. Within 600 hours of beam time, 9 million $\eta'$ will be
produced. With the roughly 400,000 $\eta' \to \eta \pi^0 \pi^0$ and 4,000 $\eta' \to 3\pi^0$ events
expected after analysis of these data, we will investigate the Dalitz plot parameters, the cusp
effect and the $\pi \pi$--scattering lengths. Within 800 hours of beam time, we will produce 250
million $\eta$--mesons. This will allow us to improve our results on the $\eta \to 3\pi^0$ and $\eta
\to \pi^0 \gamma \gamma$ decays by at least one order of magnitude. Furthermore, as the Crystal Ball
at MAMI setup could also be considered as an $\omega$--factory, we will investigate the transition
form factor of the $\omega \to e^+ e^- \pi^0$ with improved statistics.

\Section{Conclusions}
The Crystal Ball at MAMI--C experiment is ideally suited for meson--decay physics with high rates.
The worlds most precise results on $\eta \to 3\pi^0$ and $\eta \to \pi^0 \gamma \gamma$ already come
from this experiment. Other results as the $\eta \to e^+ e- \gamma$ transition form factor will be
published soon. The construction of the new detector devices (End--point tagger, TPC) will provide
high--rate production of $\eta'$--mesons and an efficient detection of charged decay channels. In
near future many results on the light mesons will be improved by the Crystal Ball at MAMI--C
experiment.

\coltoctitle{\boldmath HADES Results from NN Reactions and Perspectives for $\pi$N}
\authortochead{B.~Ramstein}
\label{art7}

\maketitle

Experiments with the High Acceptance Dielectron Spectrometer (HADES) \cite{Agakichiev09_techn} 
explore hadronic matter at high densities ($\rho \sim $ 2-3 $\rho_0$) and moderate temperatures (T
$\le$ 80 MeV) using the dilepton probe in heavy-ion reactions. These studies aim at testing 
predictions of in-medium modifications of vector meson spectral functions \cite{Leupold10}. This
requires as a prerequisite a detailed understanding of  dilepton production in the 1-2 AGeV range,
which motivates the HADES experimental program with elementary reactions. Baryonic resonances play
an important role in dilepton emission. This is first due to their mesonic decays and the subsequent
direct  dilepton   (e.g. $\rho / \omega \rightarrow $ \epem ) or Dalitz (also called conversion)
decay ($\pi^0 / \eta  \rightarrow \gamma$\epem\ or $\omega \rightarrow \pi^0 $\epem ) modes of these
mesons.  Baryonic resonances are also expected to contribute  directly to dilepton emission via
their  Dalitz decay modes (N$^{\star}$/\del$\to$ N\epem ).  All these baryonic transitions  involve
Time-Like electromagnetic form-factors which are of fundamental interest in the context of
electromagnetic structure of hadrons in the non-perturbative regime. Following the vector dominance
model (VDM), the Dalitz decay of baryonic resonances is intimately connected to off-shell ($\rho$,
$\omega$ and $\phi$) meson production, which is an important issue  for the description of dilepton
production in heavy-ion collisions.  
 \par
In the last years, the HADES collaboration has measured dilepton production in C+C collisions at 1
and 2 AGeV \cite{Agakichiev_CC1GeV_CC2GeV},  Ar+KCl at 1.75 AGeV and p+Nb at 3.5 GeV
\cite{Galatyuk10}.
In parallel, pp (at 1.25 GeV, 2.2 GeV and 3.5 GeV) and dp reactions at 1.25 AGeV have been studied.
The analysis tool for these studies is the PLUTO event generator\cite{Froehlich10}, which allows 
the implementation of different hadronic physics models, and can take into account the experimental
conditions (resolution, acceptance) in a very flexible way. Similarly to transport model
calculations, dilepton cocktails can be built up  from different processes added incoherently.
Differential cross-sections provided by theoretical calculations can also be used to generate the
events. \par 
  The inclusive pp$\to$\epem X dilepton production in pp reactions at 1.25 GeV (i.e. just below the
$\eta$ threshold) can be described quantitatively using only two dilepton sources: the \piz\ and the
\del\ Dalitz decays, which have known cross-sections.  An improvement of the description of the
\epem\ invariant mass spectra  is obtained when a VDM-like N-\del\ transition form factor is used,
which shows the sensitivity of our data to the electromagnetic structure of hadrons. In addition,
the exclusive pp$\to$ pp\epem\ measurement reveals the excitation of the \del\ resonance, which
confirms that the contribution of pp bremsstrahlung is small.  
 These results can therefore be considered as the first direct measurement of the \del\ Dalitz
decay; the branching ratio is found to be in agreement with QED calculations (4.2$\times$10$^{-5}$
within 20 $\%$). At higher energies, other resonances are contributing  and the issue is to get a
consistent description of dilepton emission including Dalitz decays of baryonic resonances with
corresponding electromagnetic form-factors and vector meson decays.  The \piz\ and $\eta$ Dalitz
decays could be isolated in exclusive pp$\to$pp\epem channels at  2.2 GeV, but the sensitivity to
form-factors is small. More significant results should be obtained at 3.5 GeV.\par  Using
"quasi-free" pn reactions at 1.25 GeV, a very strong isospin dependence of the inclusive dilepton
emission could be observed \cite{Agakichiev10_elem}. Calculations within the One Boson Exchange
model \cite{OBE}, which take into account graphs involving nucleons and baryonic resonances in a
coherent way, were used to describe the data obtained in pp and pn reactions. In pn interactions,
graphs involving only nucleons  have a more important contributions than in pp interactions, but the
dilepton production at large invariant mass is still underestimated by the calculations. As shown
recently in \cite{Shyam10}, an improvement of the description can be achieved by using time-like
pion electromagnetic form-factors for the in-flight emission graph, which hints again to the
sensitivity of dilepton emission to hadronic electromagnetic structure.\par
 Exclusive meson (\piz , $\pi^+$ and $\eta$) production in the pp reactions at 1.25 GeV and 2.2 GeV
was also analysed. In addition to detector and analysis efficiency tests, these channels  are very
useful to check that the production of meson and baryonic resonances is consistent with the inputs
of the PLUTO event generator used for the analysis of dilepton channels. In these reactions, the
Delta resonance is dominating although N(1440), N(1535) and N(1520) are also contributing. \par
HADES has just been upgraded to increase the granularity at forward angles and the data acquisition
rate.  Heavy systems (Au+Au, Ni+Ni) are planned to be studied. We are also preparing pion beam
experiments (p$_{\pi^{-}}\sim$ 0.6-1.5 GeV/c) with nuclear and hydrogen targets. Simultaneous 
studies of hadronic and dilepton channels in the $\pi^-$+p reaction with HADES should allow
dedicated investigations of resonances heavier than  \del(1232)  and give a unique opportunity to
probe their time-like electromagnetic structure.

\coltoctitle{\boldmath Comparative Studies of $\eta$ and $\eta^{\prime}$ Mesons at
COSY-11 Detector Setup}
\authortochead{P.~Klaja}
\label{art11}

\maketitle

The proton-proton and proton-$\eta^{\prime}$ invariant mass distributions have
been determined for the $pp \to pp\eta^{\prime}$ reaction at an excess energy of
Q = 16.4 MeV. The shapes of the determined invariant mass distributions
presented in Figure \ref{fig:norm_pp_peta} are similar to those of the $pp \to
pp\eta$ reaction and reveal an enhancement for large relative proton-proton
momenta. This result, together with the fact that the proton-$\eta$ interaction
is much stronger that the proton-$\eta^{\prime}$ interaction, excludes the
hypothesis that the observed enhancement is caused by the interaction between
the proton and the meson. Calculations assuming a significant contribution of
P-wave in the final state, 
and models including energy dependence of the production amplitude, 
reproduce the data within error bars equally well. Therefore, on the basis of
the presented in Figure \ref{fig:norm_pp_peta} invariant mass distributions, it
is not possible to disentangle univocally which of the discussed models is more
appropriate.
Future measurements of the spin correlation coefficients should help disentangle
these two model results in a model independent way.

For the very first time, the correlation femtoscopy method is applied to a
kinematically complete measurement of meson production in the collisions of
hadrons. A two-proton correlation function was derived from the data for the
$pp\to ppX$ reaction, measured near the threshold of $\eta$ meson production. A
technique developed for the purpose of this analysis permitted to establish the
correlation function separately for the production of the $pp+\eta$ and of the
$pp+pions$ systems. The shape of the two-proton correlation function shown in
left panel of Figure \ref{corr:double_quasi} for the $pp\eta$ differs from that
for the $pp(pions)$ and both do not show a peak structure opposite to results
determined for inclusive measurements of heavy ion collisions. \\
At present it is not possible to draw a solid quantitative conclusion about the
size of the system since it would require to solve a three-body problem where
$pp$ and $p\eta$ 
interactions are not negligible and both contribute significantly to the
proton-proton correlation. However, based on semi-quantitative predictions 
one can estimate that the system must be unexpectedly large with a radius in the
order of 4~fm. This makes the result interesting in context of the predicted
quasi-bound $\eta NN$ state 
and in view of the hypothesis 
that at threshold for the $pp\to pp\eta$ reaction the proton-proton pair may be
emitted from a large Borromean like object whose radius is about 4~fm.

The upper limit of the total cross section for the $pn \to pn\eta^{\prime}$
reaction has been determined near the kinematical threshold in the excess
energy range from 0 to 24~MeV. The measurement was performed using a deuteron
cluster target, and the proton beam with a momentum of 3.35~GeV/c. The energy
dependence of the upper limit of the cross section was extracted exploiting the
Fermi momenta of nucleons inside the deuteron. The comparison of the determined
upper limit of the ratio $R_{\eta^{\prime}}~=~{{\sigma(pn \to pn\eta^{\prime})}
/ {\sigma(pp \to pp\eta^{\prime})}}$ with the corresponding ratio for $\eta$
meson production is presented in right panel of Figure \ref{corr:double_quasi}
and disvafours the dominance of the $N^*(1535)$ resonance in the production
process of the $\eta^{\prime}$ meson and suggests non-identical production
mechanisms for the $\eta$ and $\eta^{\prime}$ mesons. To pin down the detailed
reaction mechanism further theoretical as well as experimental investigations
with better statistics are required.

\begin{figure}[t]
  \includegraphics[height=.5\textwidth]{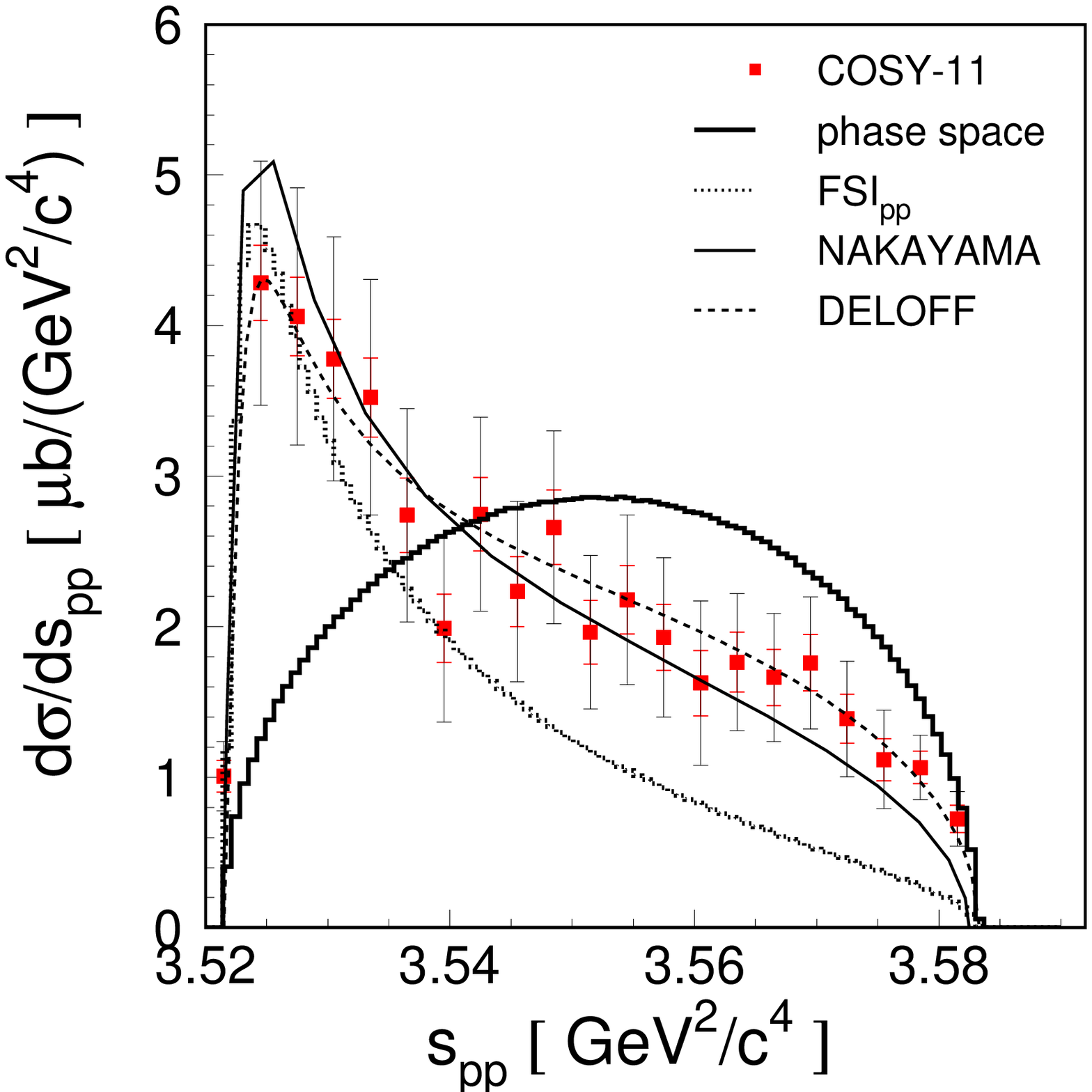}
  \includegraphics[height=.5\textwidth]{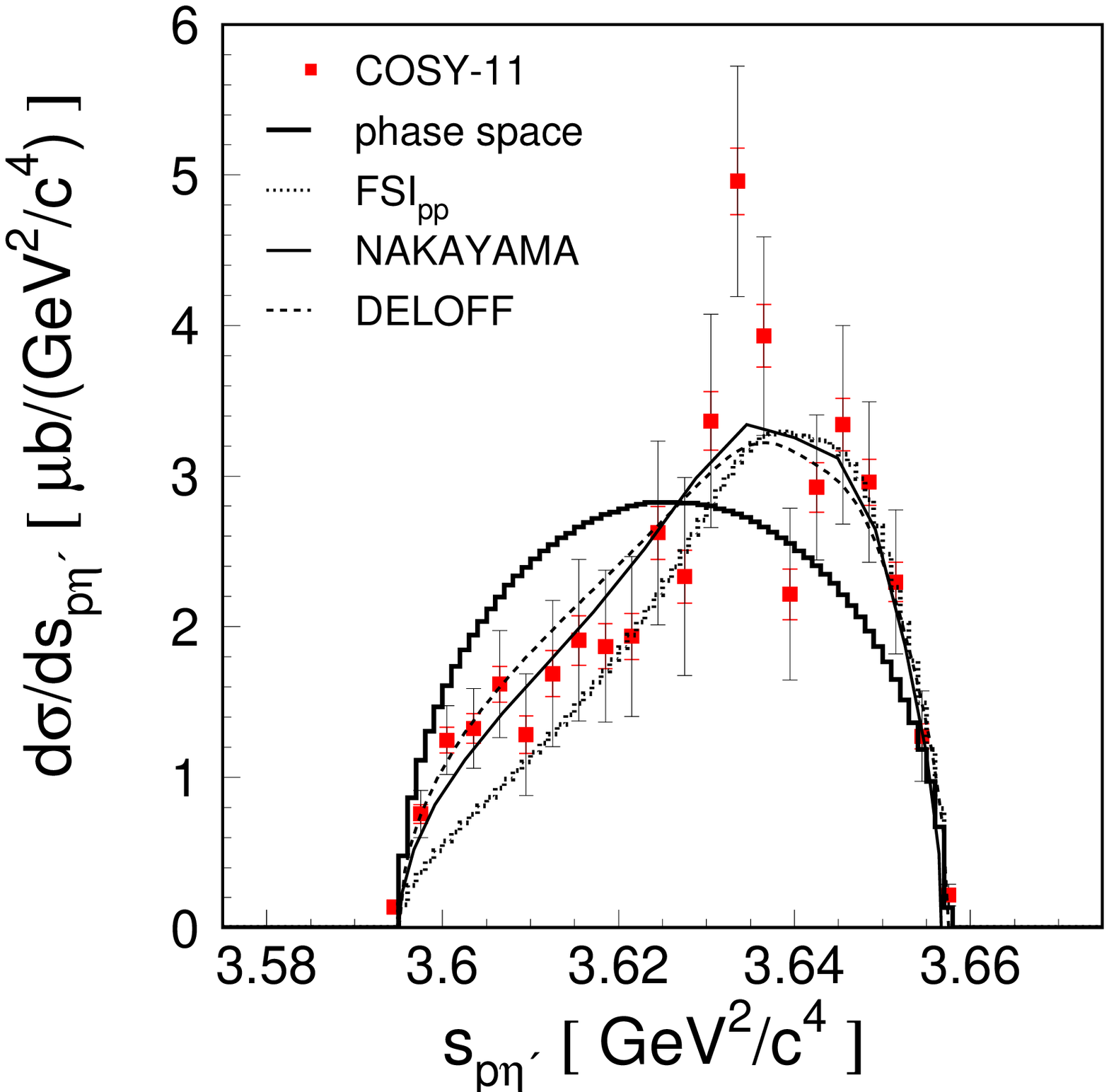}
        \caption{Distributions of the squared proton-proton ($s_{pp}$) and
        proton-$\eta^{\prime}$ ($s_{p\eta^{\prime}}$) invariant masses, for the
        $pp \to pp\eta^{\prime}$ reaction at the excess energy of Q = 16.4 MeV.
        The experimental data (closed squares) are compared to the expectation
        under the assumption of a homogeneously populated phase space (thick
        solid lines) and the integrals of the phase space weighted by the
        proton-proton scattering amplitude - FSI$_{pp}$ (dotted histograms). The
        solid and dashed lines correspond to calculations when taking into
        account contributions from higher partial waves 
	and allowing for a linear energy dependence of the $^{3}P_{0} \to
        ^{1}\!\!S_{0}s$ partial wave amplitude, 
	respectively.
         }
        \label{fig:norm_pp_peta}
\end{figure}

\begin{figure}[b]
  \begin{center}
  \includegraphics[width=.49\textwidth]{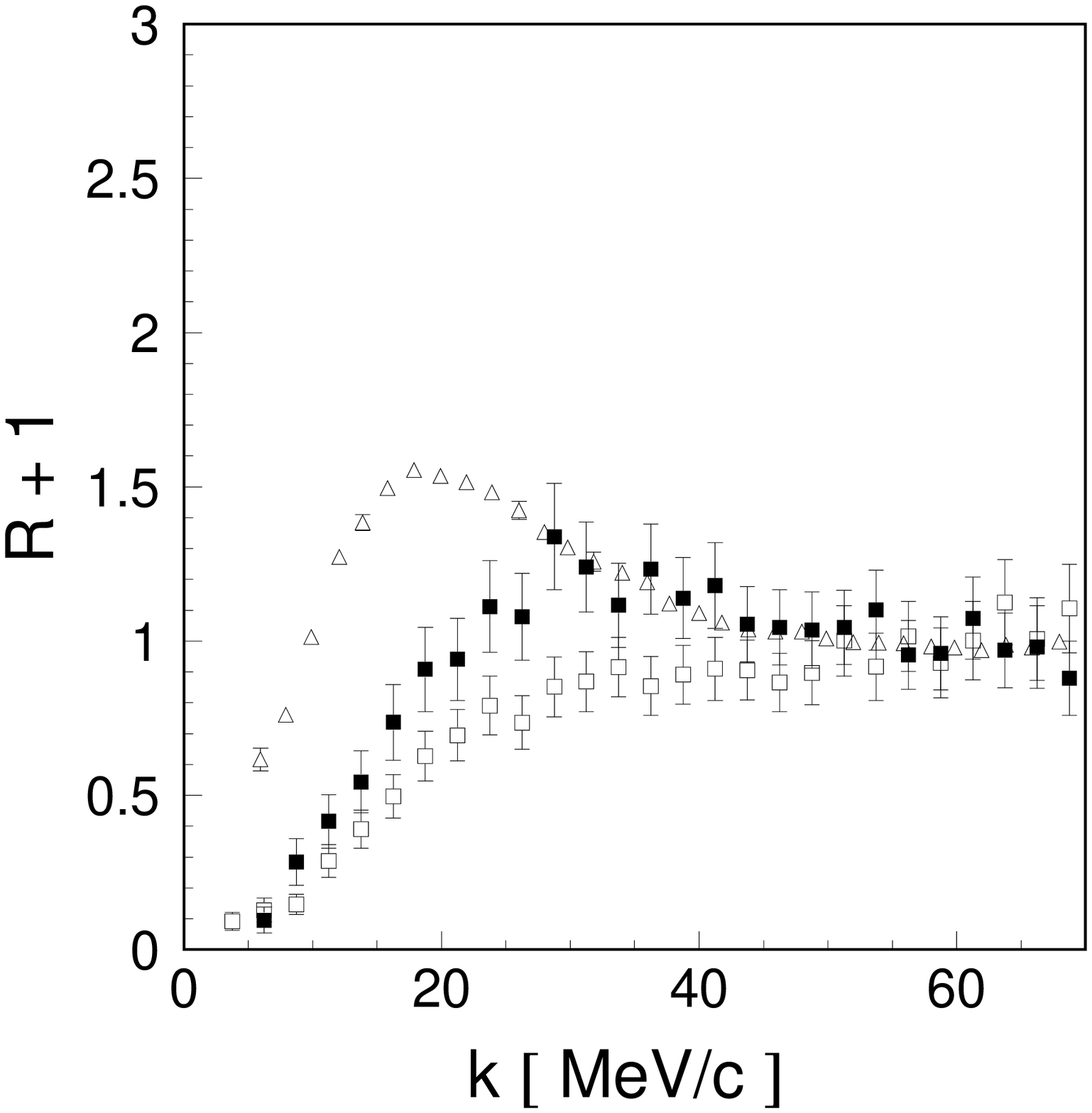}
  \includegraphics[trim=12mm 6mm 12mm 5mm, clip,width=.49\textwidth]
                  {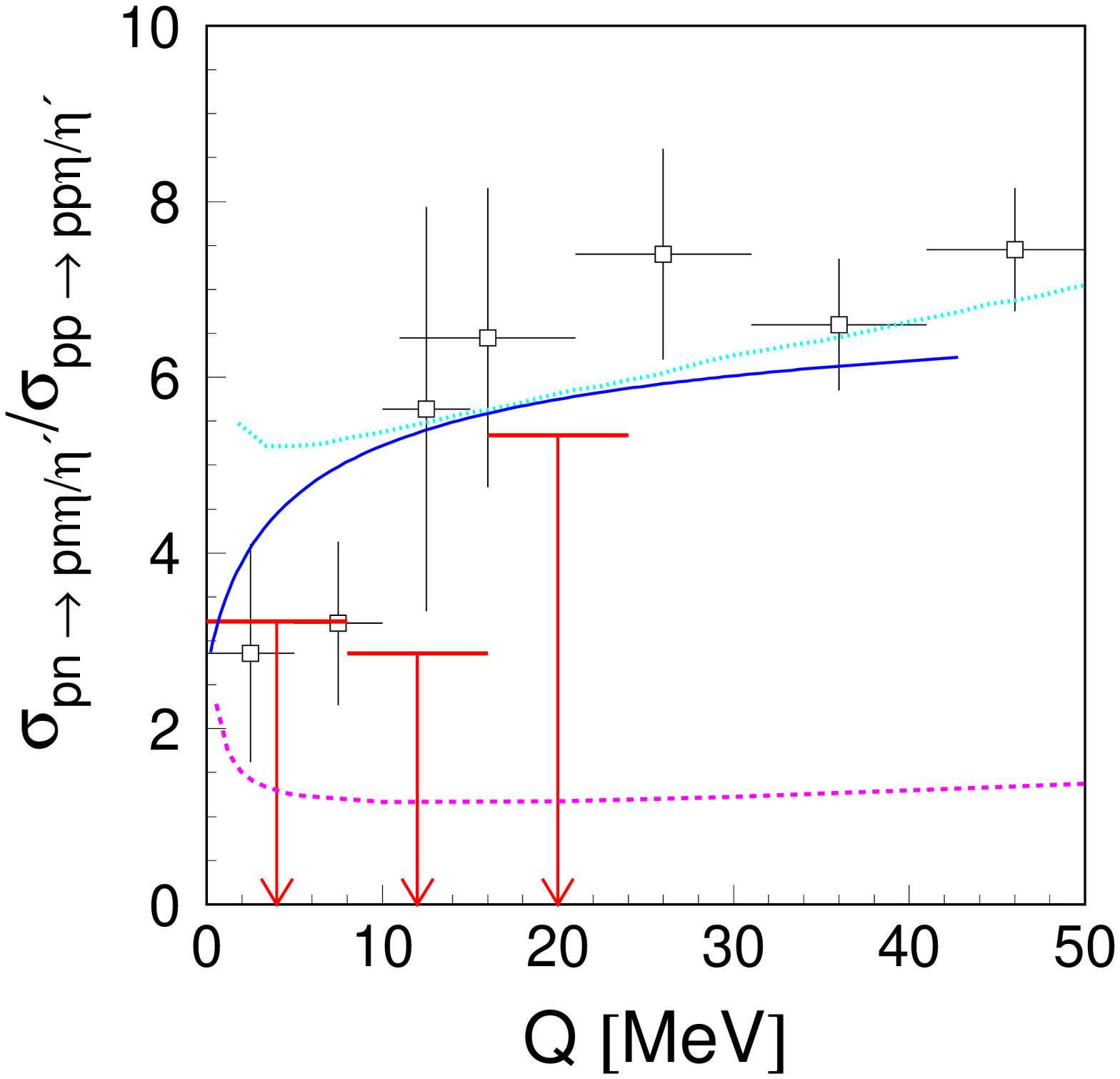}
  \end{center}
  \caption{{\bf (Left panel)} The two-proton acceptance corrected
  correlation functions normalized to
  the corresponding simulated correlation function for a point-like source.
  Results  
  for the $pp \to pp\eta$ (full squares) and $pp\to pp+pions$ (open squares)
  are compared to the two-proton correlation function determined 
  from heavy ion collisions (triangles).
   {\bf (Right panel)} Upper limit of the ratio $(R_{\eta^{\prime}})$ of 
         the total cross sections for the
         $pn \to pn\eta^{\prime}$ and $pp \to pp\eta^{\prime}$ reactions (arrows)
         in comparison with the ratio $(R_{\eta})$ determined for the $\eta$ 
         meson 
	 (open squares).
         The superimposed solid line indicates a result of a fit to the $R_{\eta}$ data
         taking into account the final state interaction of 
         nucleons.
         The dotted line presents the result of calculations performed 
         under assumption of a dominance of the 
         S$_{11}$(1535) resonance in the production process 
         and the dashed line denotes the result obtained within  a covariant effective meson-nucleon
         theory including meson and nucleon currents with
         the nucleon resonances S$_{11}$(1650), P$_{11}$(1710) and P$_{13}$(1720).
  \label{corr:double_quasi}}
\end{figure}

\coltoctitle{\boldmath Direct Determination of the $\eta'$ Meson Width}
\authortochead{E.~Czerwi\'nski}
\label{art12}

\maketitle

In the review by the Particle Data Group (PDG)~\cite{pdg}, two values
for the total width of the \ep\ meson are given.  One of these values, (0.30$\pm$0.09)~MeV/c$^2$, results from the average of
two measurements~\cite{Wurzinger,Binnie}, though
only in one of these experiments was $\Gamma_{\eta'}$ extracted directly
based on the mass distribution~\cite{Binnie}.
The second value (0.205$\pm$0.015)~MeV/c$^2$, recommended by the PDG,
is determined by fit to altogether 51 measurements of partial widths, branching ratios, and
combinations of particle widths obtained from integrated cross sections~\cite{pdg}.
The result of the fit is strongly correlated with the value of
the partial width $\Gamma(\eta'\to\gamma\gamma)$,
which causes serious difficulties when the total and the partial width have to be used at the same time,
like e.g. in studies of the gluonium content of the \ep\
meson~\cite{Biagio,Ambrosino2}. This encourage us to perform precise direct measurement of the total width of the \ep\ meson.

The reaction $pp~\to~pp\eta^{\prime}$ measured at five different beam momenta using \cc\ detector setup~\cite{Brauksiepe} was used
for the determination of the total width of the $\eta'$ meson, which
was directly derived from the mass distributions
independently of other properties of this meson, like e.g.
partial widths or production cross sections.
The $\eta'$ meson was produced in proton-proton collisions via the $pp\to pp\eta'$ reaction and its
mass was reconstructed based on the momentum vectors of protons taking part in the  reaction.
The schematic view of the \cc\ detector setup is presented in Fig.~\ref{c11}.
\begin{figure}[h]
\begin{center}
    \includegraphics[width=0.45\textwidth]{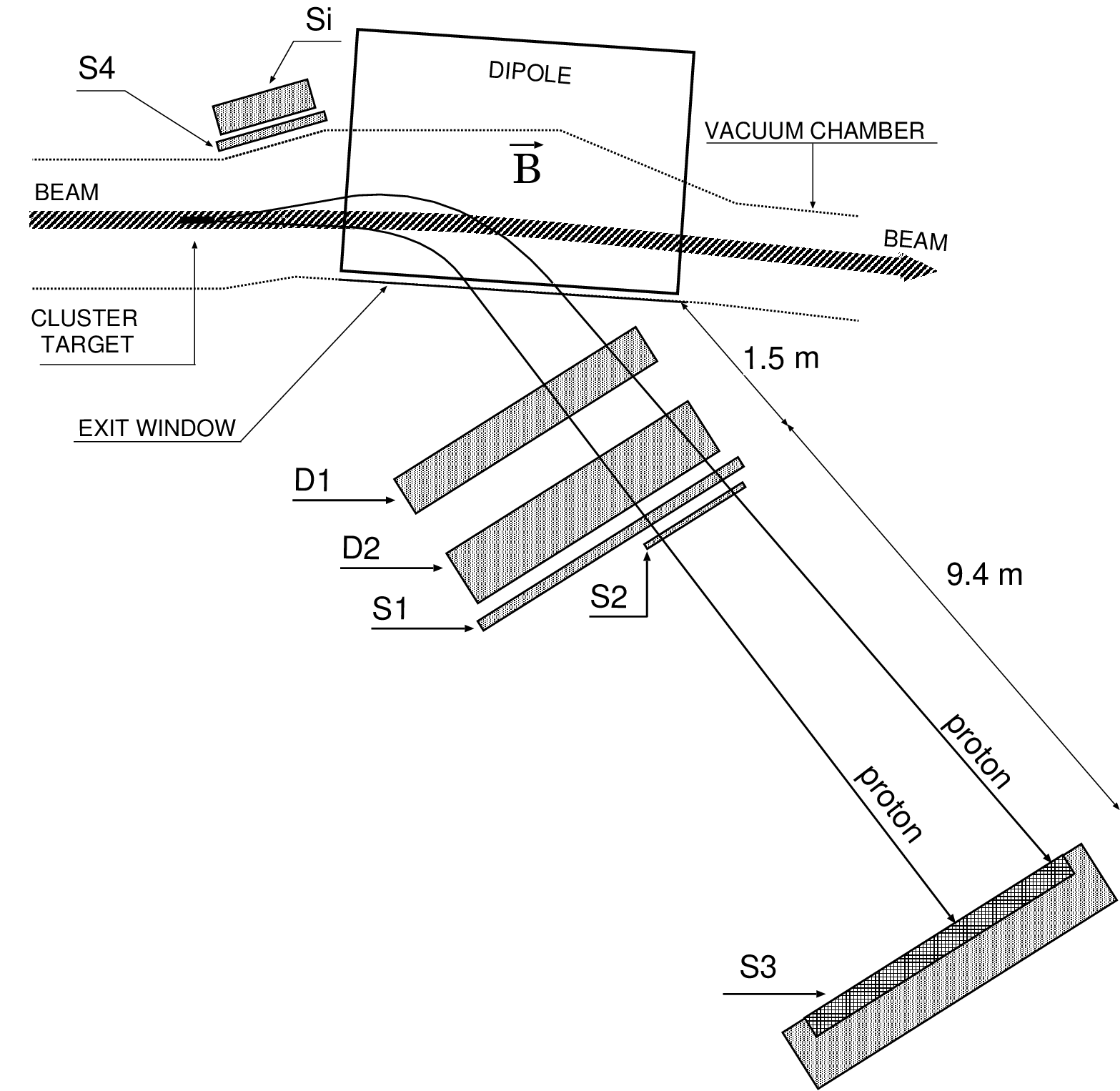}
\end{center}
 \caption{
         Schematic view of the \cc\ detector setup (top view).
         S1, S2, S3 and S4 denote scintillator detectors, D1 and D2 indicate drift chambers and  Si stands
         for the silicon-pad detector.
         }
 \label{c11}
\end{figure}
The reader interested in the description of the detectors and analysis procedures can find detailed informations
in Ref.~\cite{ErykPhD,eryk_prl}. The momentum of the COSY beam~\cite{Maier} and the dedicated zero degree \cc\ facility
enabled the measurement at an excess energy  of only a fraction of an MeV above the kinematic threshold
for the $\eta^{\prime}$ meson production.
This was the most decisive factor in minimizing uncertainties of the
missing mass determination, since at threshold the partial derivative of the missing mass
with respect to the outgoing proton momentum tends to zero.
In addition, close to threshold
the signal-to-background ratio increases due to the more rapid reduction of the phase space  for
multimeson production than for~the~$\eta^{\prime}$.

Based on the  $\eta'$ meson mass spectra reconstructed for 2300 events with the experimental resolution
of 0.3~MeV/c$^2$ the total width of the $\eta'$ meson was determined.
The systematic error was estimated by studying the sensitivity of the result to the
variation of parameters describing the experimental conditions
in the analysis and in the simulation~\cite{ErykPhD,eryk_prl}.
The final missing mass spectra are presented in Fig.~\ref{mm}.
\begin{figure}[h]
  \begin{center}
    \includegraphics[width=0.28\textwidth]{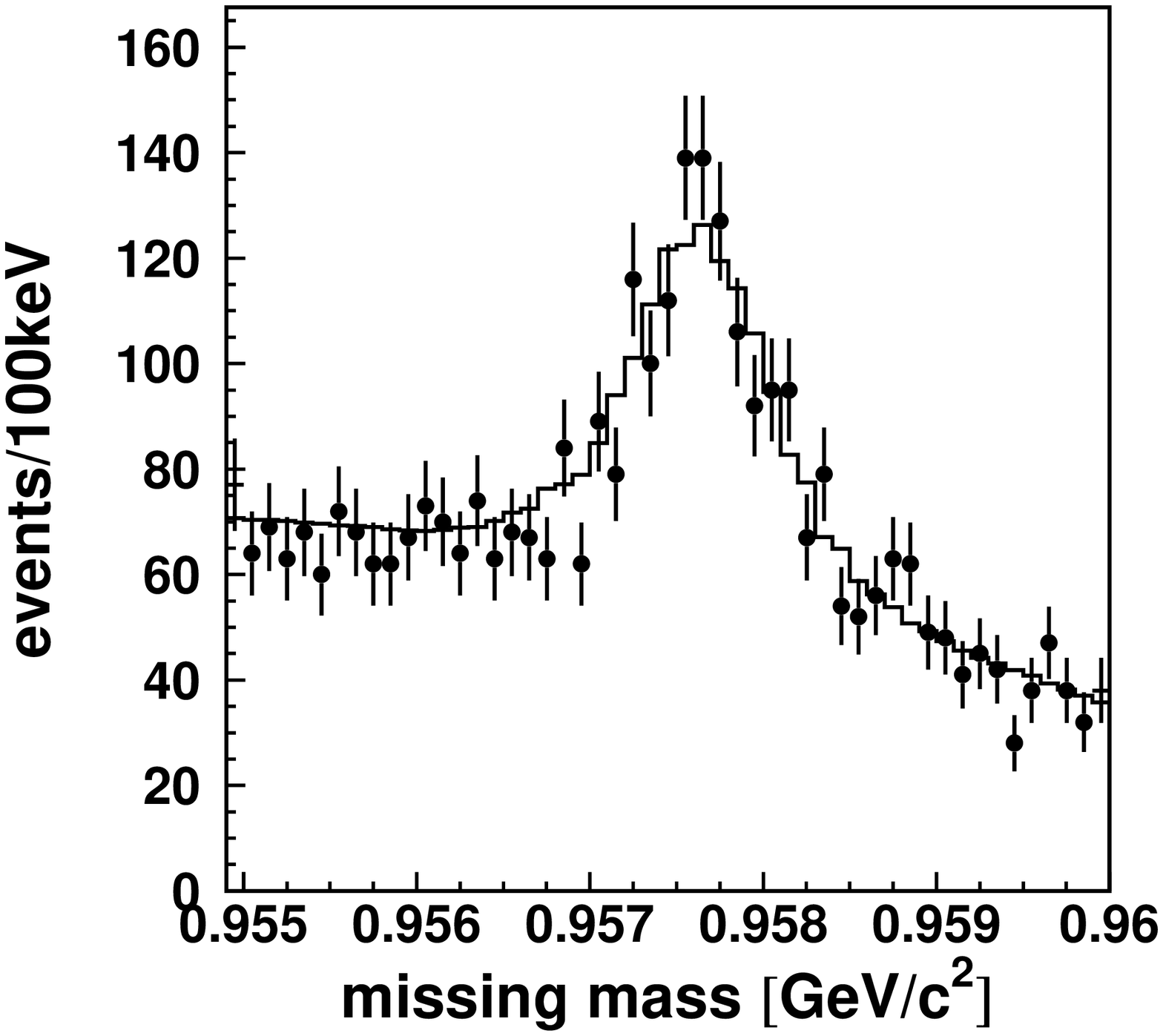}
    \includegraphics[width=0.28\textwidth]{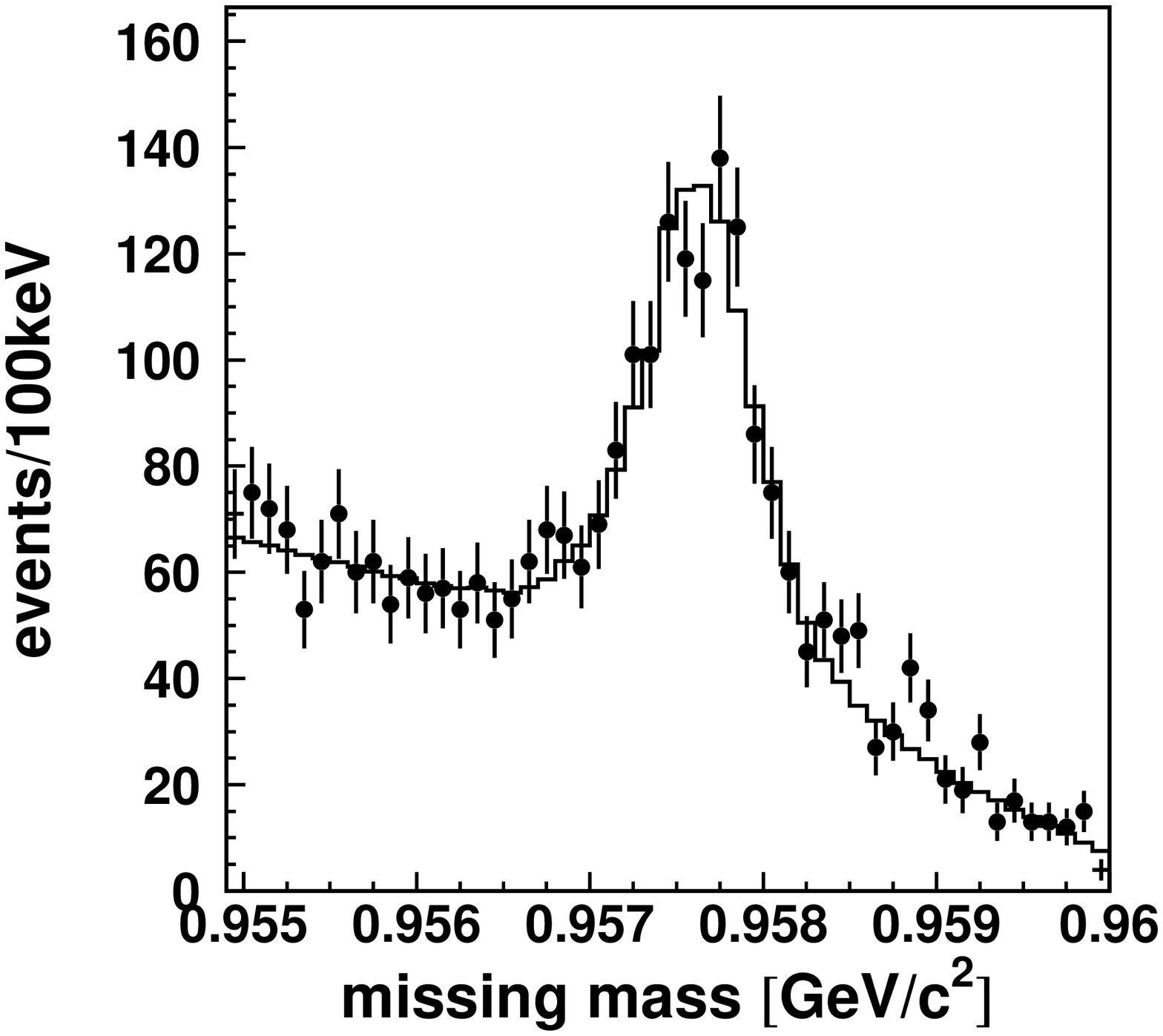}
    \includegraphics[width=0.28\textwidth]{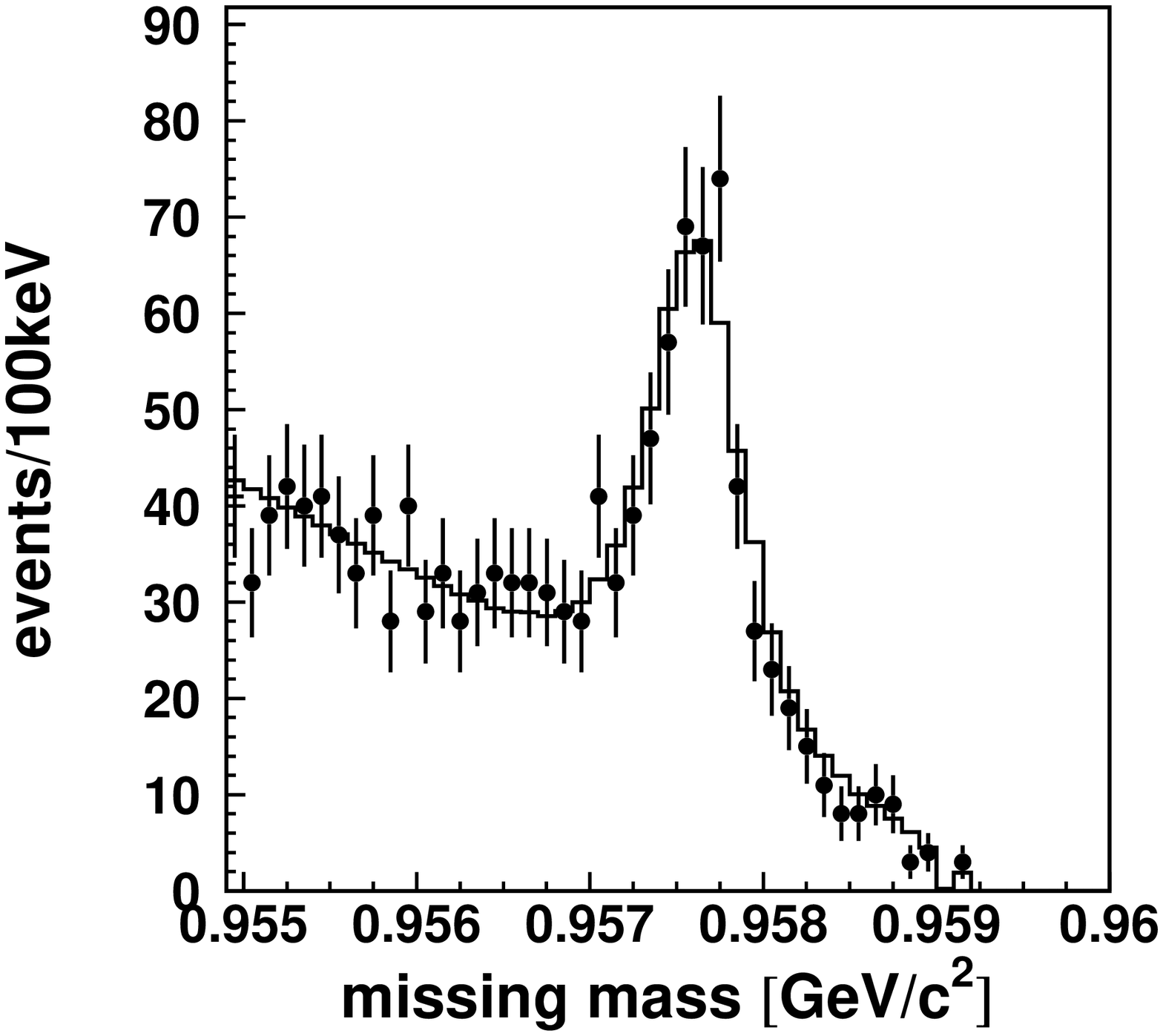}
    \includegraphics[width=0.28\textwidth]{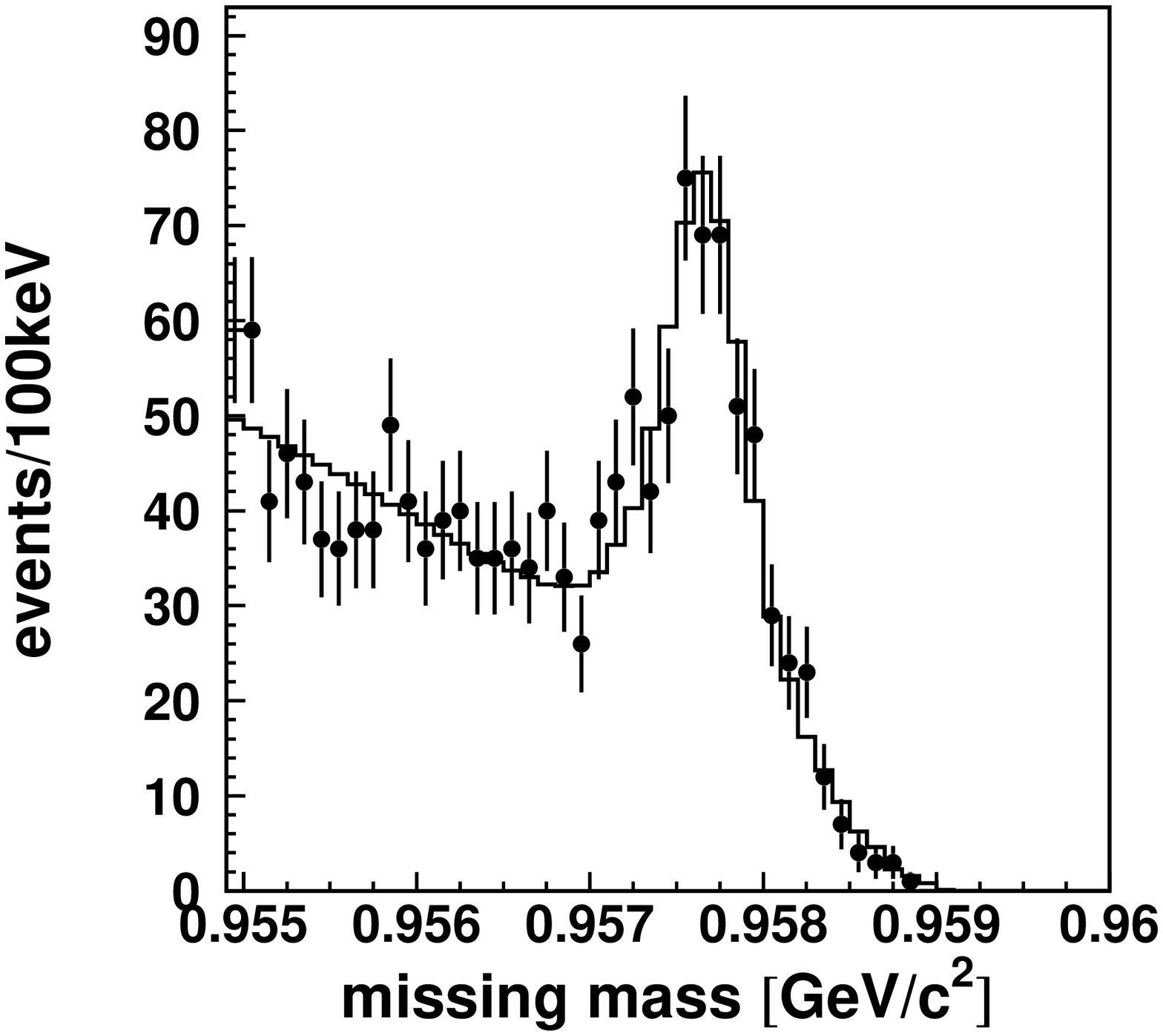}
    \includegraphics[width=0.28\textwidth]{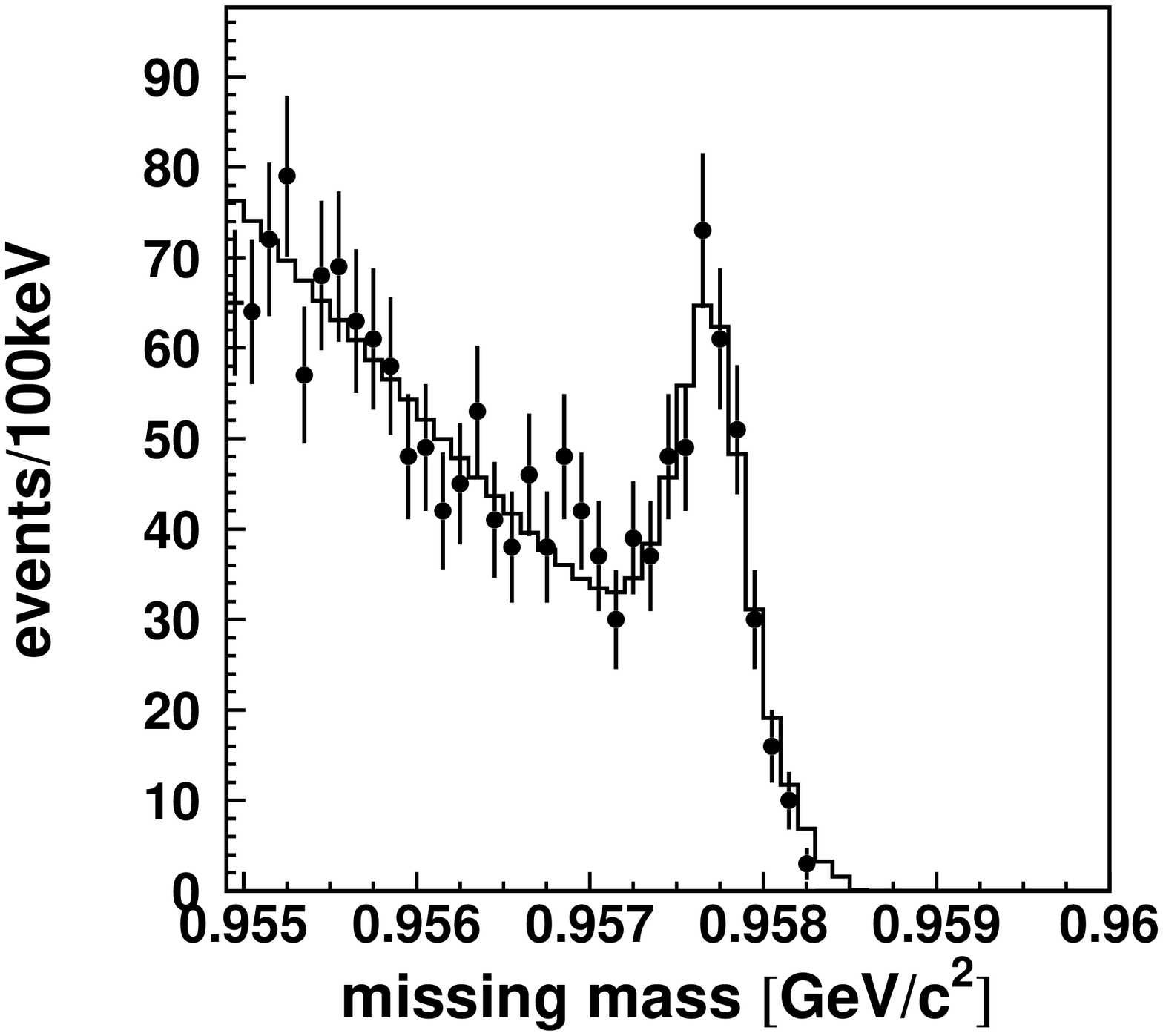}
    \includegraphics[width=0.28\textwidth]{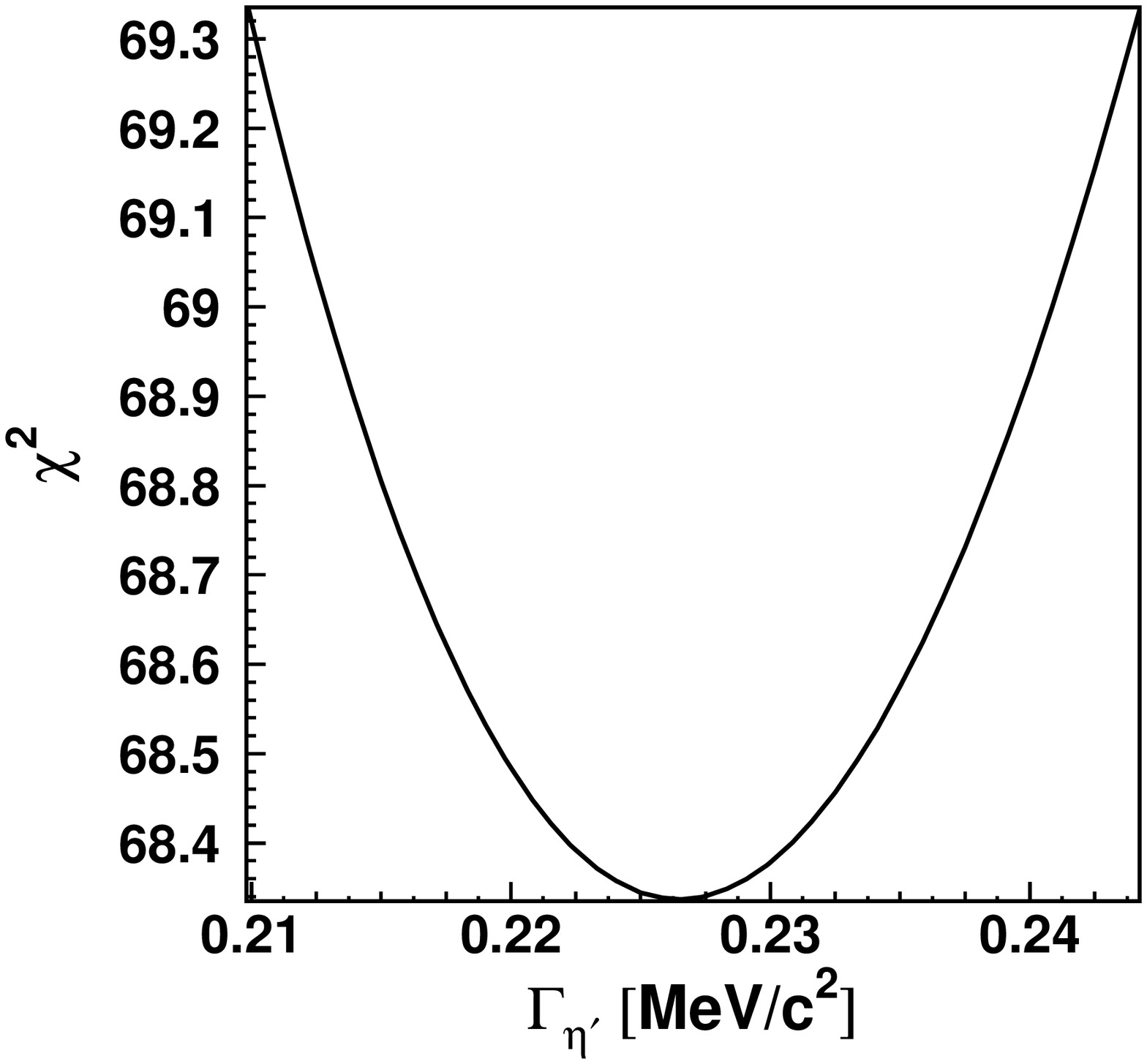}
  \end{center}
  \begin{minipage}[c]{0pt}
    \raisebox{0pt}[0pt][0pt]{
      \mbox{
        \raisebox{5.80cm}{\hspace{1.90cm}\footnotesize{\textsf{Q=4.8~MeV}}}
        \raisebox{5.80cm}{\hspace{2.68cm}\footnotesize{\textsf{Q=2.8~MeV}}}
        \raisebox{5.80cm}{\hspace{2.66cm}\footnotesize{\textsf{Q=1.7~MeV}}}
        \raisebox{1.60cm}{\hspace{-10.30cm}\footnotesize{\textsf{Q=1.4~MeV}}}
        \raisebox{1.60cm}{\hspace{2.65cm}\footnotesize{\textsf{Q=0.8~MeV}}}
      }
    }
  \end{minipage}
 \caption{
         The missing-mass spectra for the \ppx\ reaction.
         The \ep\ meson signal is clearly visible.
         The experimental data are presented
         as points, while in each plot the line corresponds to the sum of the Monte Carlo generated signal
         for the \ppep\ reaction with $\Gamma_{\eta^{\prime}}=0.226$~MeV/c$^2$ and the
         background obtained from another energy.
         The plot at the bottom right of the figure presents  $\chi^2$ as a function of the~\epw.
         }
 \label{mm}
\end{figure}
The total width of the \ep\ meson was extracted from the missing-mass spectra
and amounts to $\Gamma_{\eta'}=0.226\pm0.017(\textrm{stat.})\pm0.014(\textrm{syst.})$~MeV/c$^2$.
The result does not depend on knowing any of the branching ratios or partial decay widths.
The extracted $\Gamma_{\eta'}$ value is in agreement
with both previous direct determinations of this value~\cite{Binnie,Wurzinger},
and the achieved accuracy is similar to that obtained by the PDG~\cite{pdg}.

\coltoctitle{\boldmath Some Open Issues in $\eta$ and $\eta^\prime$ Photoproduction}
\authortochead{K.~Nakayama}
\label{art13}

\maketitle

The current knowledge of most of the nucleon resonances comes primarily from
the study of $\pi N$ scattering and pion photoproduction reactions. In recent
years, the study of other meson production reactions has intensified in search
of so-called missing resonances. These are resonances that are predicted by
quark models but not observed in the existing experimental data. Since it is
conceivable that these resonances escape detection in the usual pion-production
channels simply because they couple only weakly to the pion but might otherwise
couple more strongly to other mesons, it is important to study such competing
meson-production reactions in the hope of revealing at least some of the
``missing'' resonances. In this contribution, some of the open issues in both
the $\eta^\prime$ and $\eta$ photoproduction reactions are discussed. The
theoretical model adopted here for describing these meson photoproduction
reactions is based on an effective Lagrangian approach, whose details are given
in Ref.~\cite{NH12}. It includes the nucleonic, mesonic and nucleon resonance
currents and is fully gauge invariant as demanded by the generalized
Ward-Takahashi identity.

Until recently, not many data existed for $\eta^\prime$
photoproduction, but this situation is now changing rapidly as the new
high-precision data on the free proton as well as on the quasi-free proton and
neutron processes are being measured. The theoretical model of Ref.~\cite{NH12}
reproduces extremely well ($\chi^2/N \cong 0.5$) the
very-high-precision angular-distribution data measured recently
\cite{CLAS09} in the energy range from threshold to $W=2.35$ GeV with a minimum
set of four resonances.  The new high-precision cross-section data certainly
impose a more stringent constraint on the resonance parameters than
compared to earlier cross-section data, however, they are not
sufficient to determine these parameters uniquely. For that one
still requires spin observables. Apart from the insufficient number of
observables (measured experimentally), some of the features that contribute to
our current inability to fix the resonance parameters in $\eta^\prime$
photoproduction are:  (a) the phase of the $\eta^\prime N$ scattering is not
known, (b) no known resonance couples to $\eta^\prime$. Also, as compared to
the more widely studied $\pi$ and/or $\eta$ photoproductions, here, there is
nothing to help fix the phase of the dominant threshold amplitude, such as 
Watson's theorem in $\pi$ photoproduction, or a dominant
resonance like  $S_{11}(1535)$ in  $\eta$ photoproduction.
Furthermore, there is an ambiguity as to how one should treat the
dominant background contribution: in the MAID model \cite{Chiang}, the
reggeized $\rho$ and $\omega$ t-channel contributions have a
destructive interference between them, while in the model of
Ref.~\cite{NH12}, the $\rho$ and $\omega$ meson-exchange contributions are
constructive.  This phase difference in the background contribution --- which
is non-negligible --- certainly leads to a difference in the extracted resonance
parameters from the two models. Data at higher energies beyond the resonance
region may help fix the relative sign of these two contributions when they
become available. Experimentally, on the other hand, there is a noticeable
discrepancy between the recent cross-section data sets from the CLAS
\cite{CLAS09} and CBELSA/TAPS \cite{CBELSA09} Collaborations, especially at high
energies and backward angles.

Unlike $\eta^\prime$ photoproduction, the $\eta$ photoproduction
reaction has been studied extensively.  Apart from the fact that this
reaction serves as an isospin filter since the $\eta$ meson is an isoscalar
meson (as is the $\eta^\prime$ meson), it reveals most clearly the
$S_{11}(1535)$ resonance which couples strongly to the $\eta N$ channel. Our
model reproduces the recent differential cross section and beam asymmetry data
\cite{GRAAL07} with a very reasonable fit quality $(\chi^2/N \cong 1.8)$.
Unlike in $\eta^\prime$ photoproduction, here the resonance parameters are much
better constrained. Neither the present model nor other existing models can
describe the measured target asymmetry \cite{PHOENICS98}, especially at
energies close to threshold. Unlike the beam asymmetry, which involves the real
part of a product of spin matrix elements, the target asymmetry ($T$) involves
the imaginary part of a product of spin matrix elements. Therefore, the latter
observable is expected to be more sensitive to the final-state
interaction. Unfortunately, so far, no calculation exists that
explicitly includes the final-state interaction. The recoil-nucleon
polarization ($P$) involves the same expression in terms of the spin matrix
elements as $T$, except for one of the additive terms which has a
different sign than $T$. In attempting to describe $P$, we, therefore,
expect difficulties similar to those found when describing $T$. In this
regard, the combinations $T\pm P$ may help isolate the source of the problem
with the current models in describing $T$.

In recent years, the $\eta$ photoproduction reaction has received
renewed attention because of an observed bump
structure around $\sqrt{s} \approx 1.67$ GeV in the ratio of the total cross
sections, $\sigma_n/\sigma_p \equiv \sigma( \gamma n \to \eta n) / \sigma(
\gamma p \to \eta p)$ in the quasi-free processes \cite{Kuz07}. This bump
structure was interpreted
as a potential signal of a non-strange member of anti-decuplet
pentaquarks. Here, we address this issue within an SU(3) coupled-channel chiral
unitary approach. Within this approach, the bump structure arises naturally  as
a result of coupled-channel effects due to the strong coupling to the
intermediate $K\Lambda$ and $K\Sigma$ channels \cite{Dor10}. In the quasi-free
proton case, the photon can couple to the charged Kaon in the intermediate
$K^+\Lambda$ state, while in the quasi-free neutron case, this contribution is
absent, for the corresponding intermediate state is $K^0\Lambda$ and the photon
does not couple to a neutral Kaon.  This coupled-channel effect is a
direct consequence of the dynamics driven by the Weinberg-Tomozawa interaction
together with the strong coupling to $K\Lambda$ and $K\Sigma$ channels
resulting from SU(3) symmetry. The present result does not rule out other
explanations for the bump structure, such as narrow resonances. On
the other hand, the prediction of our chiral unitary model for
$\sigma_n/\sigma_p$ in quasi-free $\gamma N \to K \Lambda$ processes on the
deuteron does not exhibit any bump structure, unlike for $\eta$
photoproduction. The measurement of $\sigma_n/\sigma_p$ in
$K\Lambda$ photoproduction, therefore, can shed light on the accuracy
of our chiral unitary model.  Furthermore, it is interesting to note that a
measurement of the longitudinal spin transfer coefficient ($L_z$) in $\gamma N
\to \eta N$, together with the corresponding unpolarized cross sections,
is capable of disentangling the coupled-channel scenario
discussed above from the narrow-resonance scenario in a
model-independent way. Alternatively, measurements of the target and
recoil polarizations with a circularly polarized photon beam can also
distinguish between the two scenarios in a model-independent
way. Of course, such measurements pose enormous experimental
challenges.

\coltoctitle{\boldmath The $pd \rightarrow ^{3}$He$ \eta$ $\pi^0$ Reaction at $T_p$ = 1450 MeV}
\authortochead{K.~Sch\"onning}
\label{art22}

\maketitle

Recently, many important results on photo-production of the $\eta\pi^0$ system have appeared.
Chrystal Ball/TAPS data from the $\gamma p \rightarrow \pi^0 \eta p$ reaction \cite{kashevarov} were
interpreted in terms of a dominant cascade decay of the $D_{33} \Delta(1700)$ isobar through the
$s$-wave $\Delta(1700) \rightarrow \eta\Delta(1232)$ followed by the $p$-wave $\Delta \rightarrow
\pi^0 p$ \cite{doring}. A similar interpretation is possible also for $\gamma d \rightarrow \pi^0
\eta d$ data \cite{krusche}. 

To further investigate the role of resonances like $\Delta(1700)$, $\Delta(1232)$ and $N^*(1535)$
in $\pi^0 \eta$ production and the competition between the $^3$He$\pi^0$ and $^3$He$\eta$
interactions, the $pd \rightarrow ^{3}$He$ \eta \pi^0$ reaction has been studied with the WASA
detector \cite{wasa} at the CELSIUS storage ring in Uppsala, Sweden.

The experiment was carried out using a 1450 MeV proton beam impinging on a deuterium pellet target.
The $^3$He ions were measured in a Forward Detector and the photons from the $\pi^0$ and $\eta$
decays were in a central electromagnetic calorimeter. Only the $\eta \rightarrow \gamma\gamma$ decay
channel was considered (BR=39.3\%). Events with one identified $^3$He, four photons, one $\pi^0$
candidate and one $\eta$ candidate were selected. Furthermore, the missing momentum of the $^3$He
was required to match the total momentum of the four photons. Finally, all events where two good
$\pi^0$ candidates could be formed from the four photons, were rejected. More details about the
selection criteria are given in Ref. \cite{wasaetapi0}. The background from 2$\pi^0$ production is
discussed in some detail in Ref. \cite{karinthesis} and 3$\pi^0$ production was studied in Ref.
\cite{wasa3pi}. The $\eta\pi^0$ events were identified by the peak at the $\eta$ mass in the
$^3$He$\pi^0$ missing mass spectrum, see figure \ref{fig:mmhepi}. The events fulfilling all
selection criteria were background subtracted, using simulated phase space events from 2$\pi^0$ and
3$\pi^0$ production normalised to the cross sections given in Refs. \cite{karinthesis,wasa3pi}. The
data were then corrected for acceptance and normalised, all following the proceedure described in
Ref. \cite{wasaetapi0}. This resulted in a measured cross section of $23.6 \pm 1.6 \pm 2.2$ nb plus
a normalisation uncertainty of 14\%. Details about the normalisation proceedure can be found in
\cite{wasaomega}. 

\begin{figure}[h]
  \begin{minipage}[b]{0.45\textwidth}
    \centerline{\includegraphics[width=0.9\textwidth]{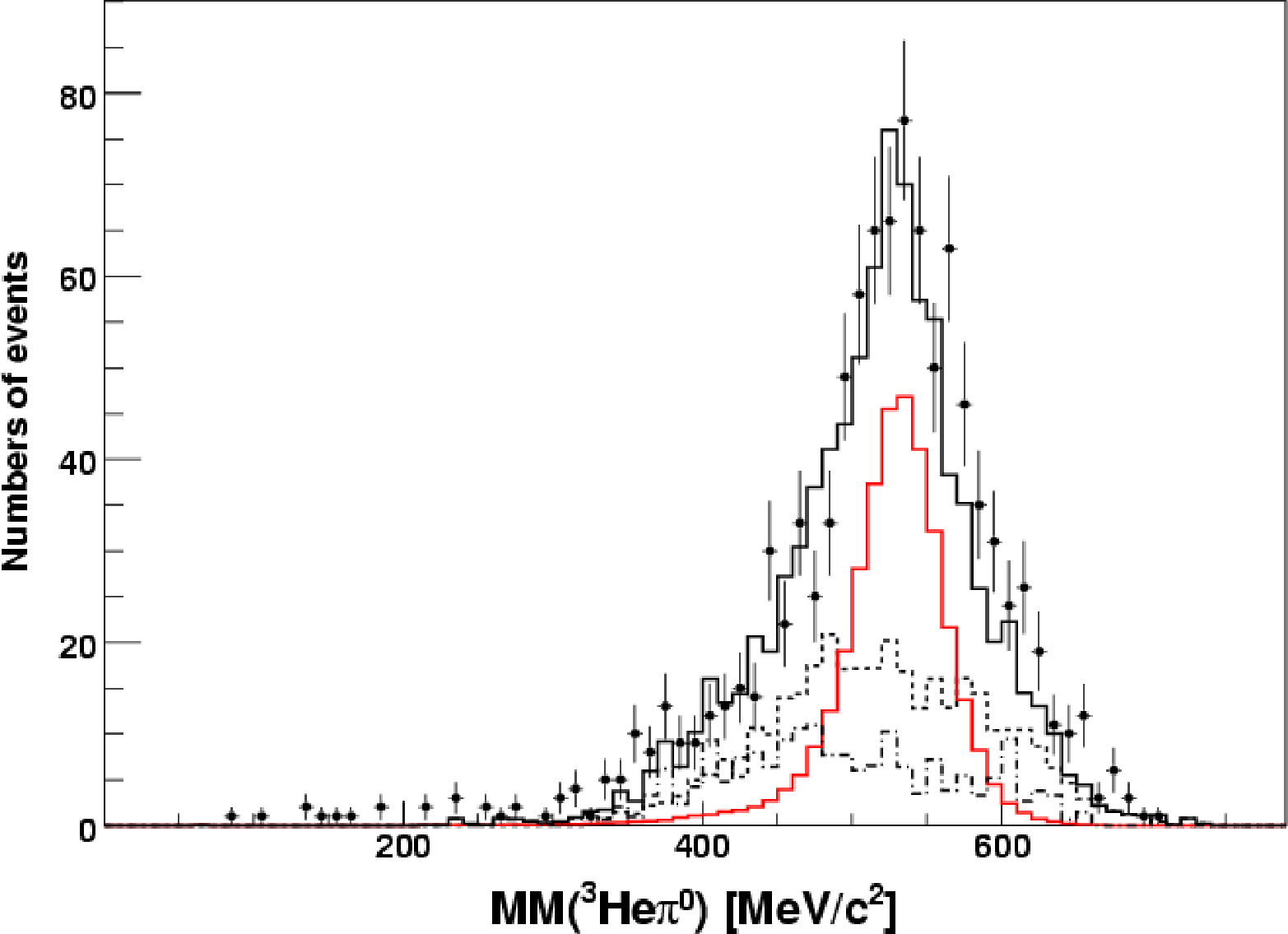}}
    \caption{\label{fig:mmhepi}(colour online)
The missing mass of the $^3$He$\pi^0$-system. The dots are data, the red histogram simulated $pd
\rightarrow ^{3}$He$\eta \pi^0$ events, the dashed histogram phase space $pd \rightarrow
^{3}$He$2\pi^0$ MC data, the dashed-dotted phase space $pd \rightarrow ^{3}$He$3\pi^0$ and the solid
line shows the sum of all simulated events.}
  \end{minipage} \hfill \begin{minipage}[b]{0.45\textwidth}
    \centerline{\includegraphics[width=0.9\textwidth]{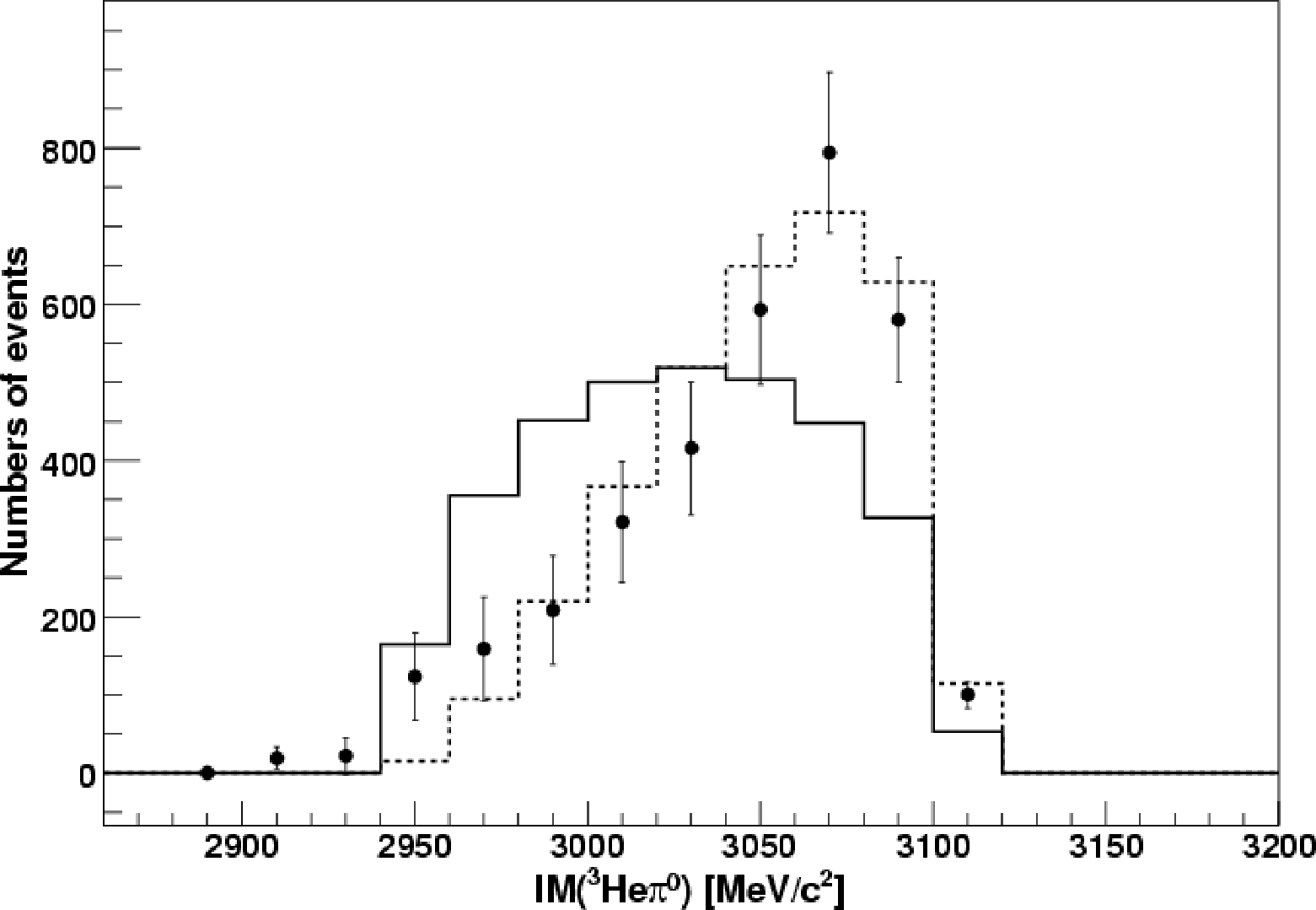}}
    \caption{\label{fig:imhepi} The invariant mass of the $^3$He$\pi^0$-system. The dots are
background subtracted and acceptance corrected data. The solid line histogram phase space simulated
$pd \rightarrow ^{3}$He$\eta \pi^0$ events and the dashed histogram is phase space events weighted
with the momentum squared of the $\pi^0$ in the $^3$He$\pi^0$ rest system.}
  \end{minipage}

  \begin{minipage}[b]{0.45\textwidth}
    \centerline{\includegraphics[width=0.9\textwidth]{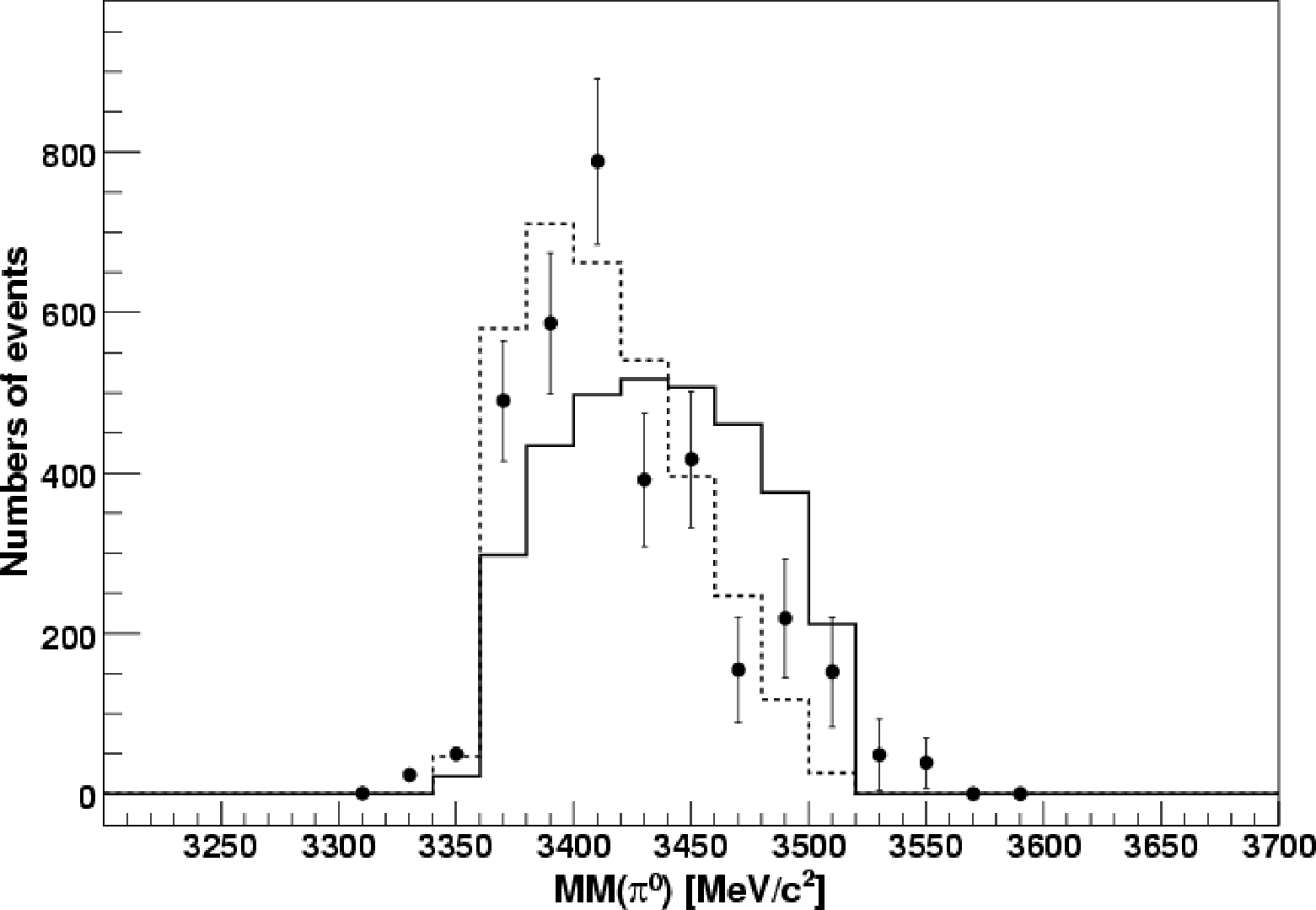}}
    \caption{\label{fig:mmpi} The missing mass of the $\pi^0$, which is equivalent to the invariant
mass of the $^3$He$\eta$-system.}
  \end{minipage} \hfill \begin{minipage}[b]{0.45\textwidth} 
    \centerline{\includegraphics[width=0.9\textwidth]{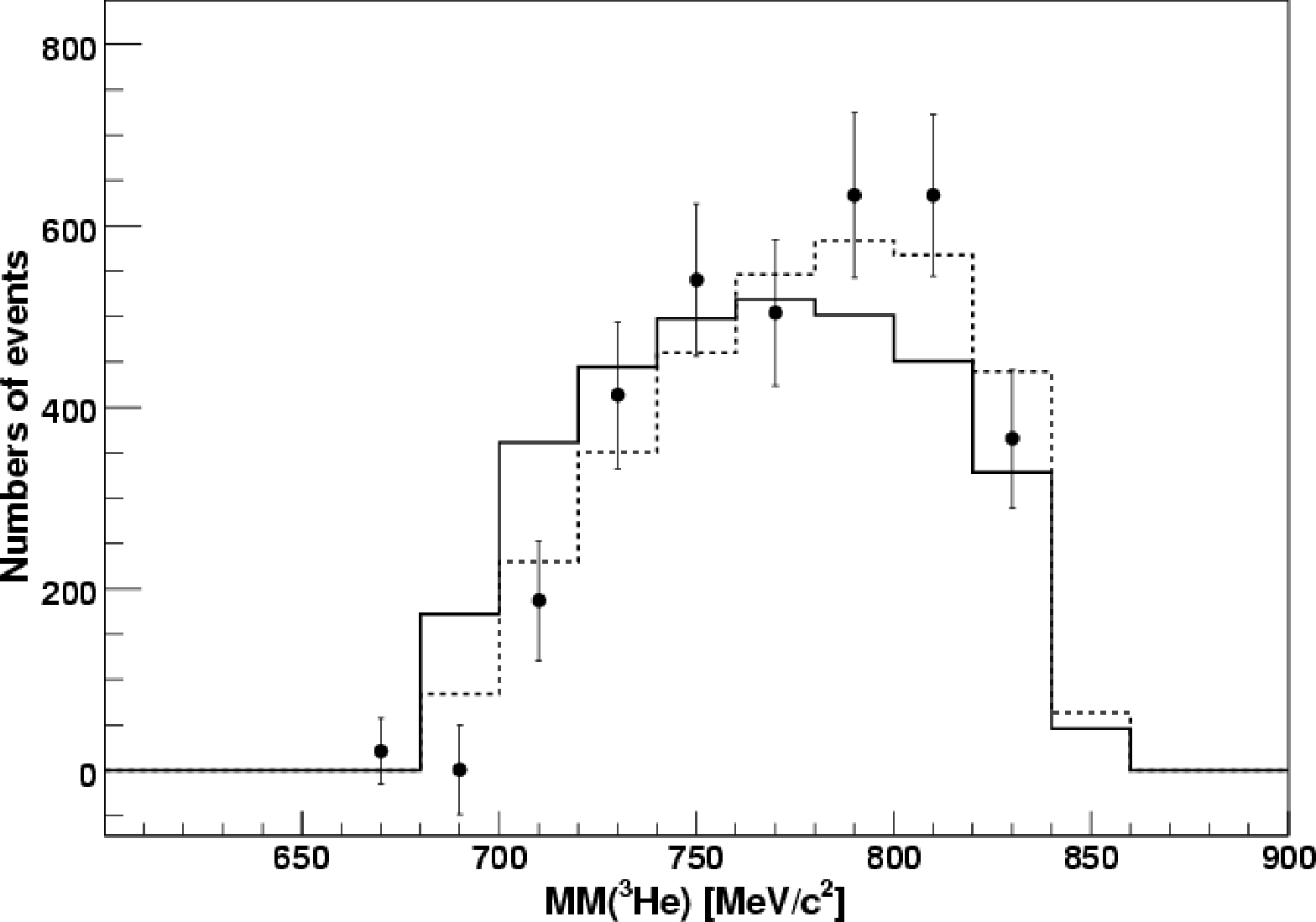}}
    \caption{\label{fig:mmhe}
             The missing mass of the $^3$He, which is equivalent to the invariant mass of the $\eta\pi^0$ system.}
  \end{minipage}

\end{figure}

Invariant mass distributions of the $^3$He$\pi^0$, $^3$He$\eta$ and $\eta\pi^0$ systems were
reconstructed and are shown in figures \ref{fig:imhepi}-\ref{fig:mmhe}. The $^3$He$\pi^0$ invariant
mass is strongly enhanced near $m_p + M_{\Delta(1232)}$ which indicates the importance of the
$\Delta(1232)$ resonance. Since no microscopic model of $\eta\pi^0$ production in $pd$ collisions
exists as yet, the $p$-wave rise due to the $\Delta(1232)$ was simulated by weighting the phase
space Monte Carlo data with the momentum squared $k^2$ of the $\pi^0$ in the $^3$He$\pi^0$ rest
system. These weighted MC data were found to reproduce the data well. The $^3$He$\eta$ invariant
mass is peaked near the $m_p + M_{N(1535)}$ sum, which suggest involvement of the $N^*(1535)$
resonance. However, it turned out that the weighted MC data reproduce also this behaviour. Thus, the
low mass enhancement in the $^3$He$\eta$ system may very well be an effect of the $\Delta(1232)$ in
the $^3$He$\pi^0$ interaction. The $\eta\pi^0$ invariant mass distribution show no sign of any
"ABC-like" effect (see \textit{e.g} Ref. \cite{bashkanov}) nor any significant high mass enhancement
due to the $a_0(980)$ resonance. The latter is not surprising since we are far below the nominal
threshold of $a_0(980)$ production.

For the future it would be interesting with high statistics measurements by \textit{e.g.} WASA at
COSY, at both higher and lower beam energies. The influence from the $\Delta(1232)$ resonance should
have an impact on the energy dependence of the cross section and also on the angular distributions
of the final state particles. Some angular distributions were studied with the CELSIUS/WASA data
presented here and they are consistent with isotropy, though with large error bars. It is also not
clear from this data what the dynamical origin of the $\eta$ meson is. Does it come from the
$N^*(1535)$ in a two-step process or from the subsequent decay of the $\Delta(1700)$? It will be
interesting to see what a new generation of measurements can tell us.

\coltoctitle{\boldmath Investigation of Double Pion Production in pp Collisions at $T_{p}$=1400 MeV}
\authortochead{T.~Tolba}
\label{art23}

\setcounter{chapter}{1}
\setcounter{section}{0}
\maketitle

\Section{Introduction}
The reaction $pp\rightarrow pp\pi^{0}\pi^{0}$ was investigated using the WASA at COSY facility. The
total and differential cross sections of the double pion production at a beam energy of 1400~MeV
have been evaluated. Preliminary results \cite{ex:1} are presented here and compared to theoretical
expectations.\\
Recently, double pion production has been extensively studied in the $pp\rightarrow pp\pi\pi$
reaction from near threshold ($T_{p}$=775~MeV) up to $T_{p}$=1300~MeV~\cite{ex:2,ex:3}. In
contrast, experimental information on the double pion production in $NN$ collisions is scarce at
higher energies.\\
A full reaction model describing the double pion production in $NN$ collisions has been developed
by a theoretical group at the University of Valencia - Spain at the end of the 1990s \cite{ex:4}.
The model involves chiral Lagrangian terms for nucleons and pions (non-resonant terms) plus
amplitudes describing the $\Delta$(1232) and Roper excitations at energy ranges from threshold up to
1400~MeV. This model was slightly modified by a group from T\"ubingen University - Germany
\cite{ex:5} and denoted as Valencia model in the following. \\
Here we report on the total cross section and the reaction mechanism for the
$pp\rightarrow pp\pi^{0}\pi^{0}$ reaction at $T_{p}$ = 1400~MeV where no previous
experimental information is available.

\Section{Experimental Setup and Data Analysis}
The experimental data analyzed in this work were collected using the Wide Angle Shower Apparatus
(WASA) \cite{ex:6} at the COSY accelerator in the Forschungszentrum J\"ulich, Germany. The detector
provides nearly full solid angle coverage for both charged and neutral particles, and allows
multi-body final state hadronic interactions to be studied with high efficiency.\\
The recoil protons are detected in the forward detector and identified using the $\Delta$E-E
method. The two $\pi^{0}$ are reconstructed in the central detector from their two-photon decays.
The criterion used in the selection of the event sample demands 1 or 2 charged tracks in the forward
detector and exactly 4 neutral clusters in the central detector with no restrictions on charged
tracks in the central detector. With these constraints, the geometrical acceptance is found to be
$\sim$ $45\%$.

\Subsection{Results}
The total cross section of the neutral double pion production at $T_{p}$ = 1400~MeV is found to
be \mbox{$\sigma_{tot}$ = (324$\pm$61) $\mu$b}. The data were corrected for the detector acceptance
generated by a Monte Carlo model tuned in order to match the experimental data (explained in detail
in Ref.~\cite{ex:1}). Figure~\ref{fin} shows the new point determined in this work compared with
the previous work at lower beam energies \cite{ex:2,ex:3} and the theoretical expectations
calculated with the Valencia model \cite{ex:4}. Since previous experimental data do not exist for
the $pp\rightarrow pp\pi^{0}\pi^{0}$ reaction at the proton beam energy used here, a direct
comparison between this result and previous data is not possible. However, our result shows the same
rising trend of the previous cross section values starting at $\sim$ 1170~MeV. This new data is
lower by factor of two compared to the total cross section calculated by the theoretical
expectations from the Valencia model (the solid black line in Figure~\ref{fin}), but it matches very
well their expected cross section for the double $\Delta$(1232) process, i.e. in the case that the
Roper excitation process is negligible (the dashed-dotted green line in Figure~\ref{fin}).

\begin{figure}[hbt]
\centering
    \includegraphics[width=.4\textwidth]{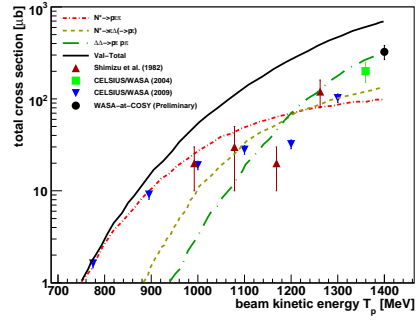}
    \caption{\label{fin} The total cross section for the
$pp\rightarrow pp\pi^{0}\pi^{0}$ reaction as a function of beam kinetic energy. The
result presented here at $T_{p}$ = 1400~MeV denoted by the black dot together with previous data
and theoretical calculations: From near threshold up to 1300~MeV (blue inverted triangles)
\cite{ex:3}, at 1360~MeV (green square) \cite{ex:7}, bubble chamber results (dark red triangles)
\cite{ex:2} and the theoretical expectations \cite{ex:4} (where the long dashed dotted line
represents the production via double $\Delta$(1232), the dashed line represents the production via
Roper decay $N^{*}$(1440)$\rightarrow\Delta\pi^{0}$, the dashed dotted line represents the
production via direct Roper decay $N^{*}$(1440)$\rightarrow$p$\pi^{0}\pi^{0}$ and the solid
black line represents the expected total cross section). }
    
\end{figure}

In order to study the double pion production mechanism, the differential cross section distribution
for different kinematical variables describing the system have been determined. The results are
compared to the expectations based on the uniform population of the available phase space (PS) and
to the theoretical expectations. The theoretical models are normalized to the same total cross
section as the data.

\begin{figure}[hbt]
  \centerline{\includegraphics[width=.38\textwidth]{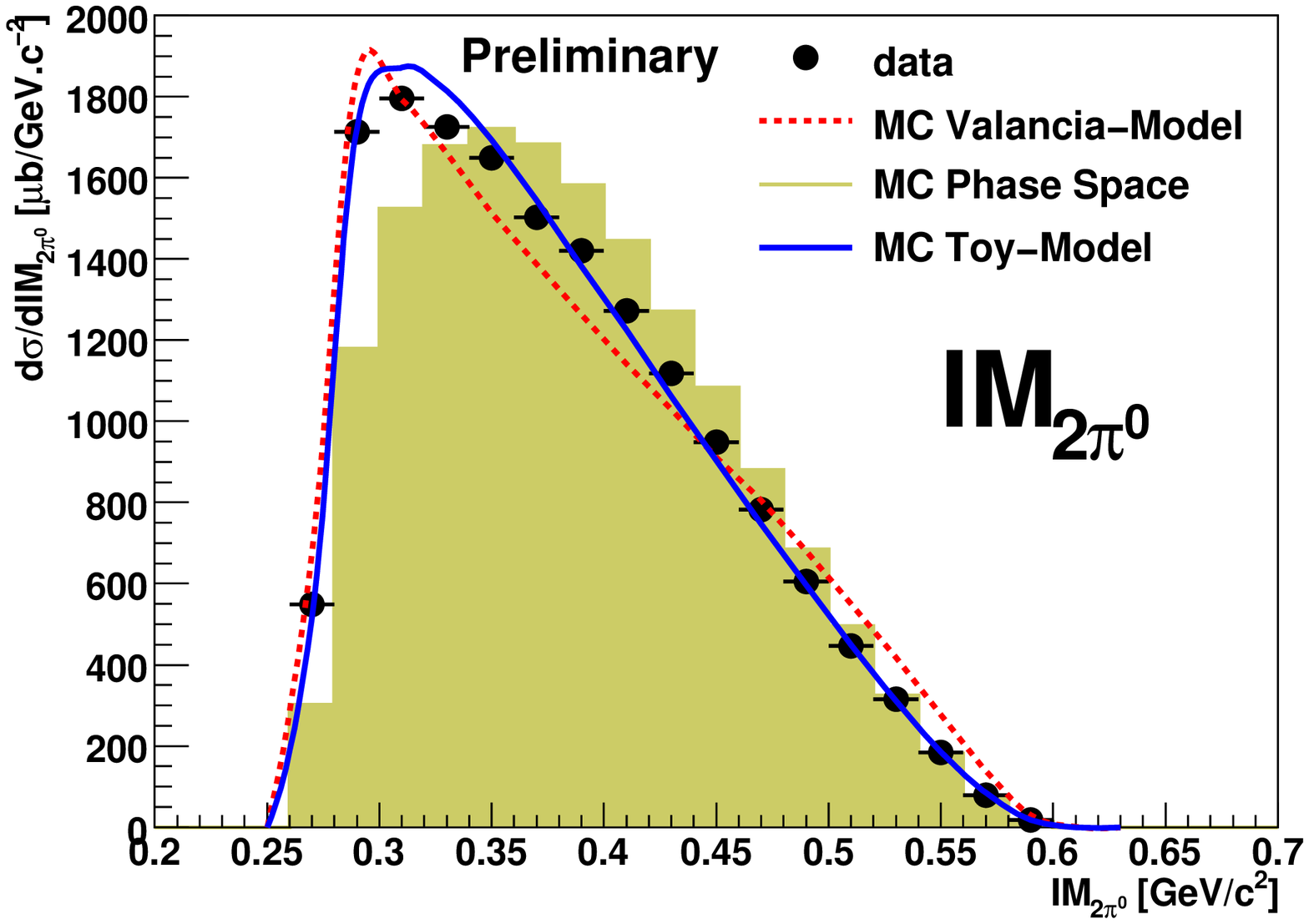}%
              \includegraphics[width=.38\textwidth]{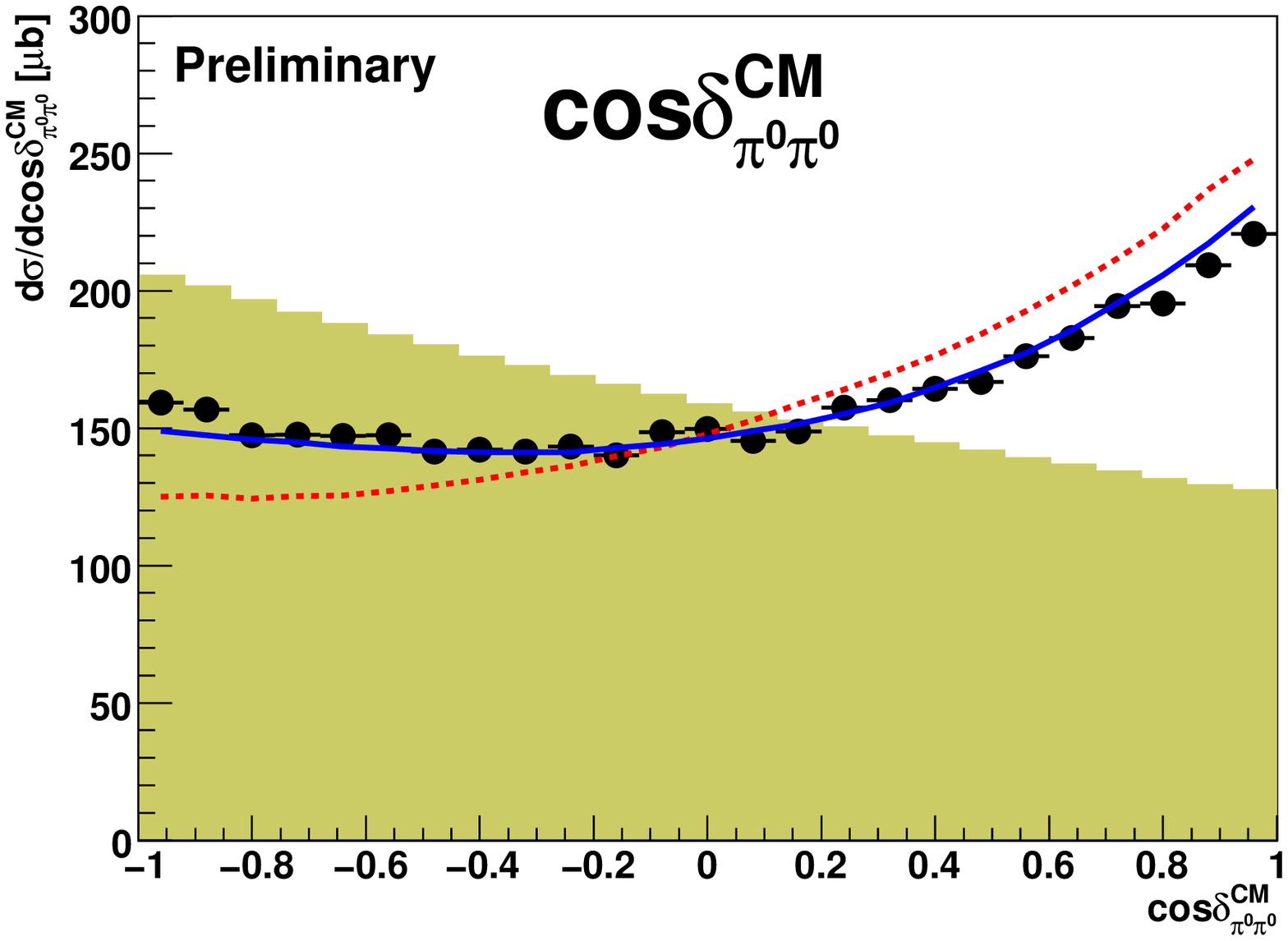}}
  \caption{\label{fina} Comparison of the data (black dots) with theoretical expectations: phase
space (shaded area), the Valencia model (red dashed line) and the tuned Monte Carlo model (blue
solid line), for the differential cross section as functions of the 2$\pi^{0}$ invariant mass
(left plot) and the 2$\pi^{0}$ opening angle (right plot).}
\end{figure}
From Figure~\ref{fina} we see that the 2$\pi^{0}$ invariant mass ($IM_{2\pi^{0}}$) distribution
(left plot) shows a behavior similar to phase space with a systematic enhancement at low
$IM_{2\pi^{0}}$. This indicates the tendency of the two pions to be emitted parallel in the
overall center of mass frame. This can be clearly seen in the two pion opening angle distribution
$\cos\delta^{CM}_{\pi^{0}\pi^{0}}$ (right plot), where the enhancement at
$\cos\delta^{CM}_{\pi^{0}\pi^{0}}$ = 1 corresponds to a small angle between the pions. 

\begin{figure}[hbt]
  \centerline{\includegraphics[width=.38\textwidth]{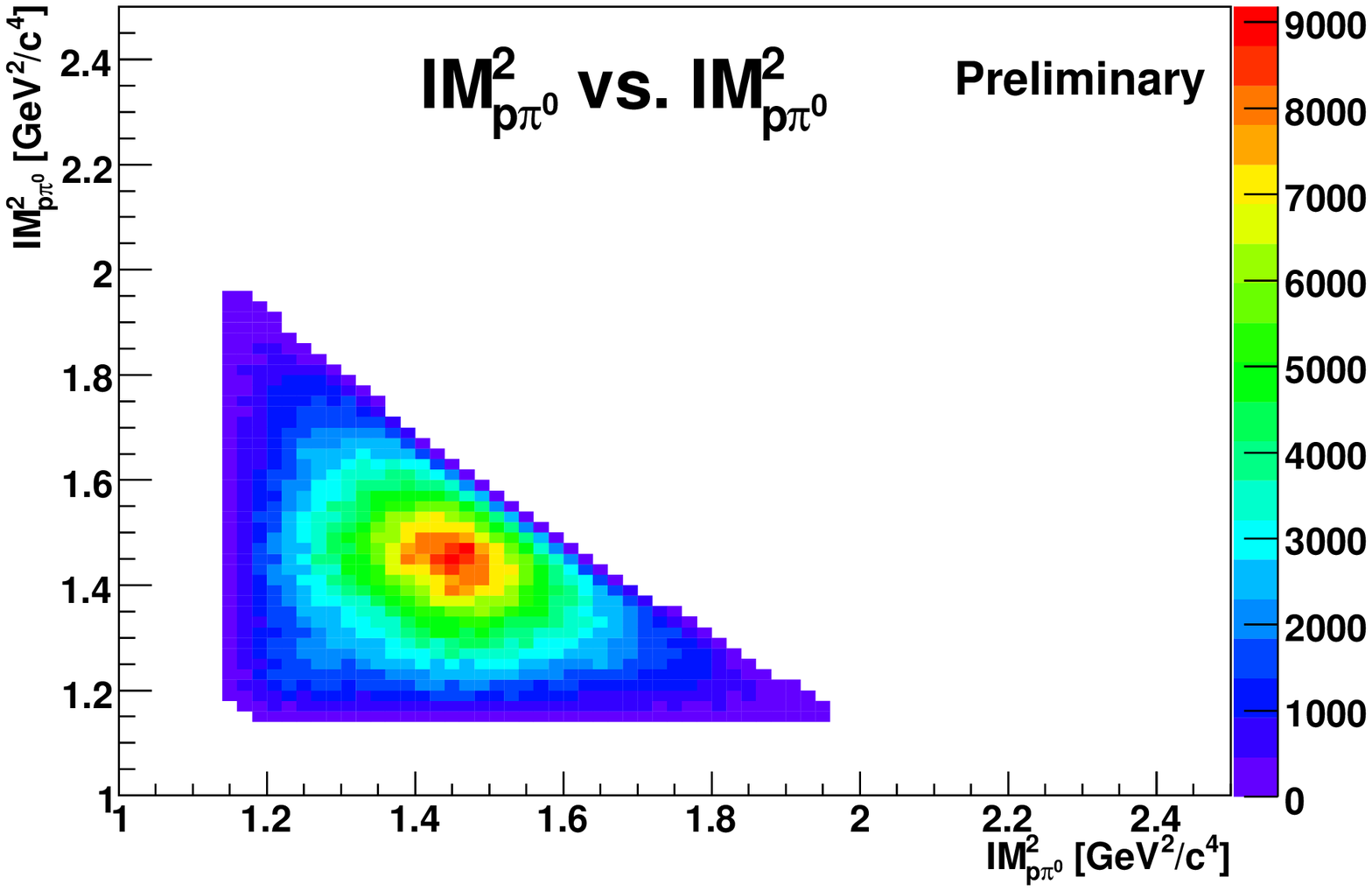}%
              \includegraphics[width=.38\textwidth]{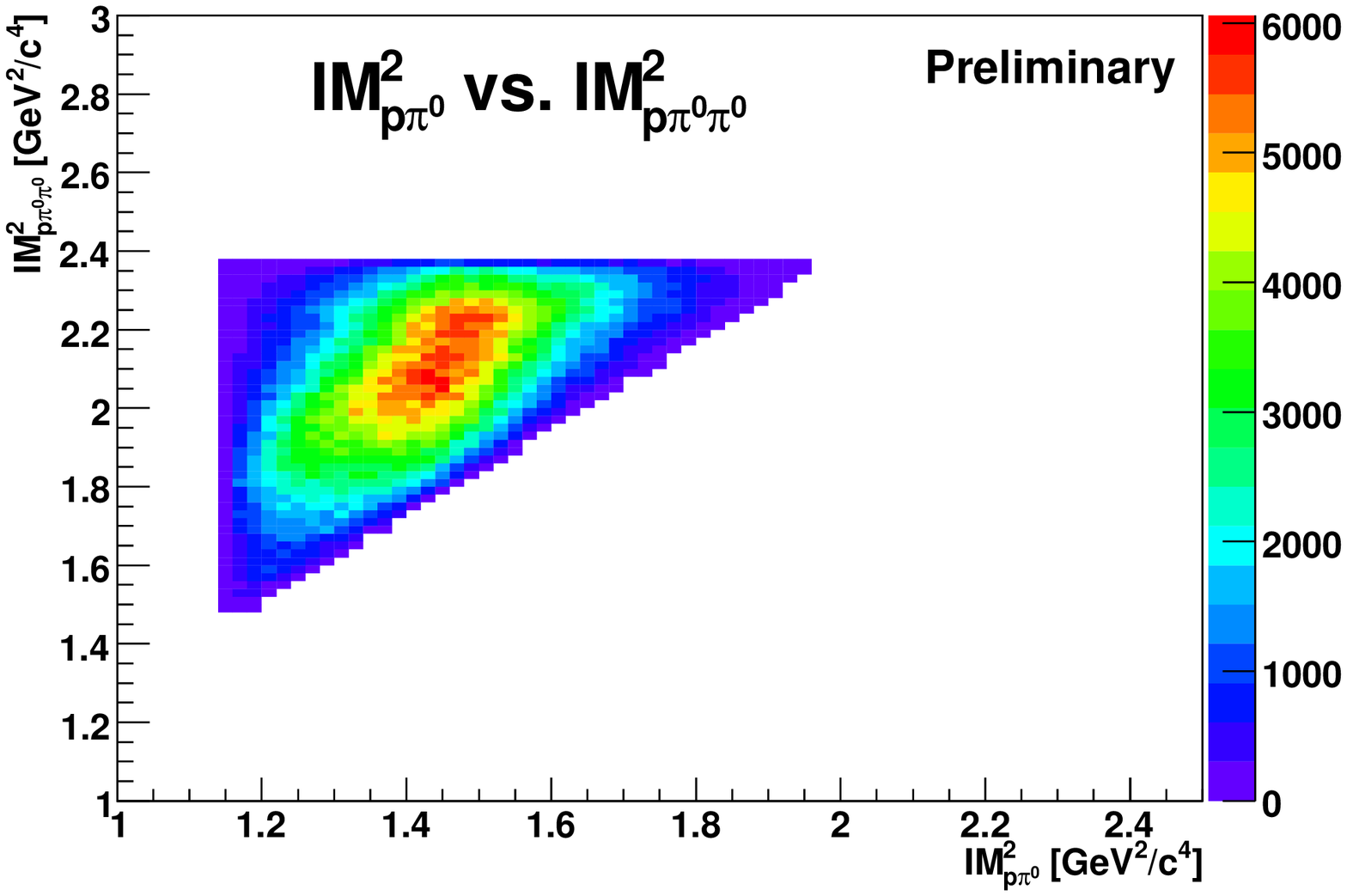}}
  \centerline{\includegraphics[width=.38\textwidth]{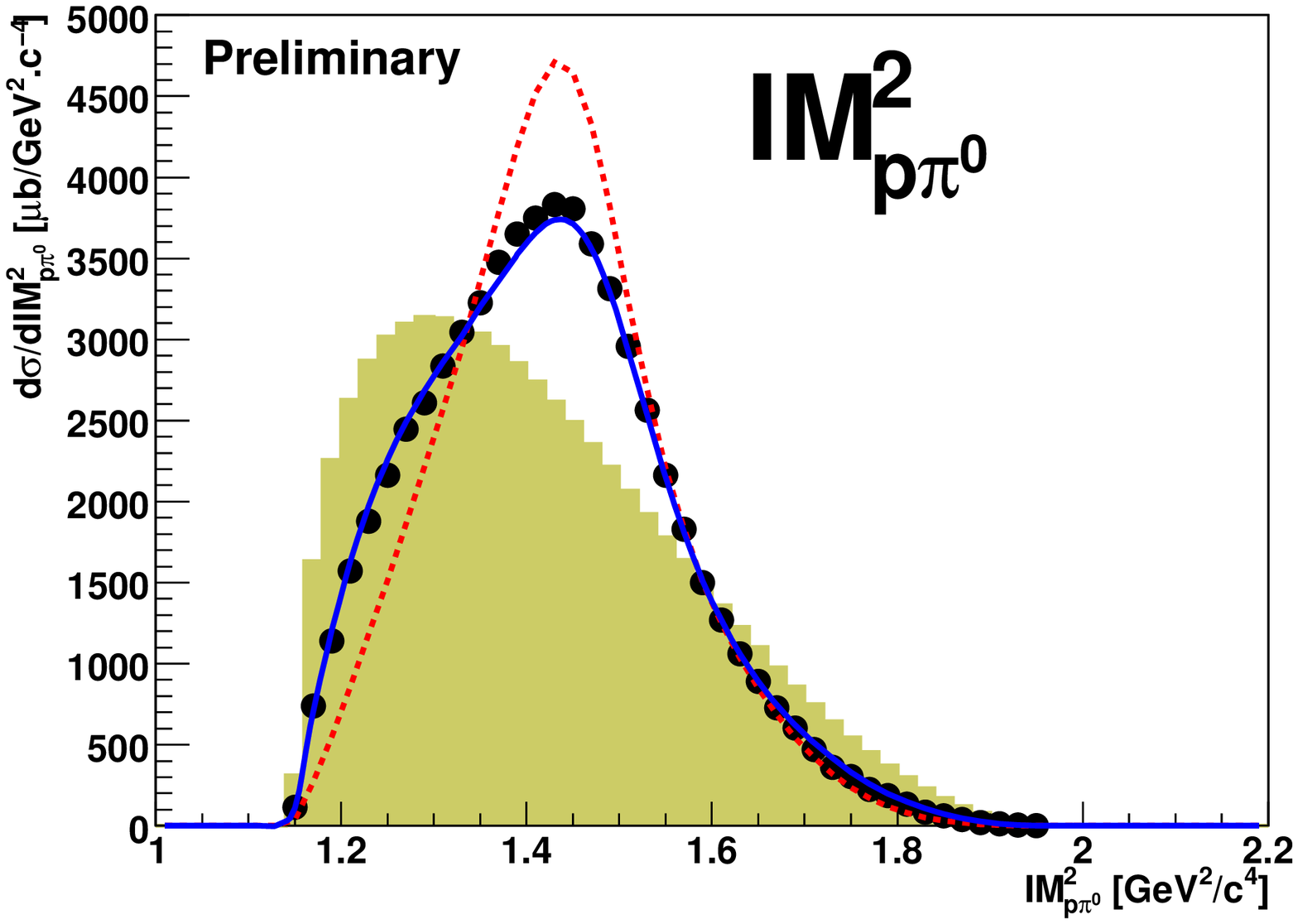}%
              \includegraphics[width=.38\textwidth]{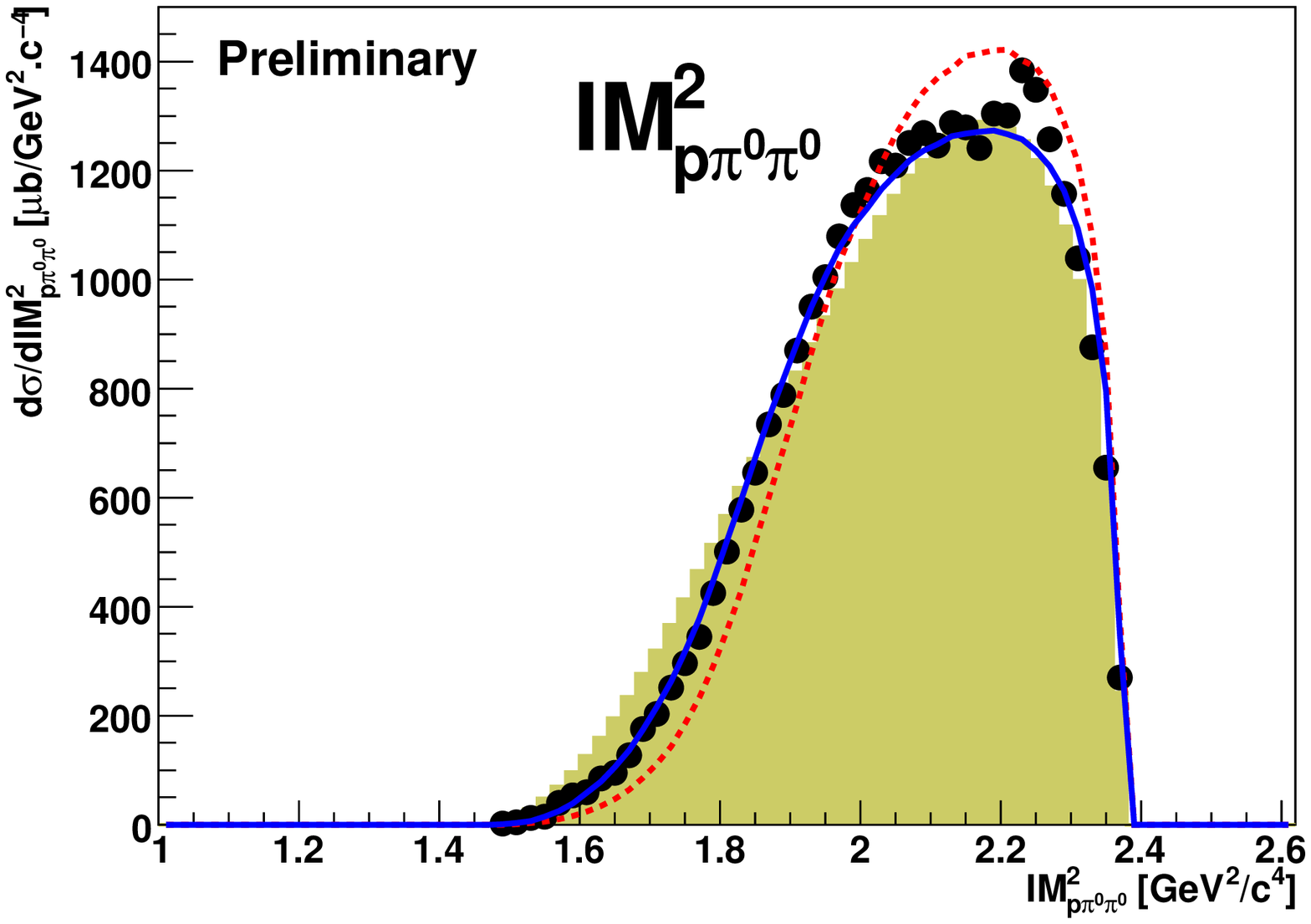}}
 
  \caption{\label{final} The same as Figure~\ref{fina} except for differential cross section
distributions as functions of $IM^{2}_{p\pi^{0}}$ versus $IM^{2}_{p\pi^{0}}$ (upper left, data
only), $IM^{2}_{p\pi^{0}}$ versus $IM^{2}_{p\pi^{0}\pi^{0}}$ (upper right, data only),
$IM^{2}_{p\pi^{0}}$ (lower left), and $IM^{2}_{p\pi^{0}\pi^{0}}$ (lower right).}
\end{figure}

The upper left frame of Figure~\ref{final} shows the correlation of the pairs of the invariant
mass square of the p$\pi^{0}$ system ($IM^{2}_{p\pi^{0}}$). Evidence for the $\Delta$(1232)
resonance can be seen as a strong enhancement at $IM^{2}_{p\pi^{0}}$ = 1.52 GeV$^{2}/c^{4}$, which
is the square of the $\Delta$(1232) mass. This observation is supported by the projection onto the
$IM^{2}_{p\pi^{0}}$ axis (lower left plot) where the enhancement at m$^{2}_{\Delta}$ is
observed.\\
In contrast to the clear evidence of the $\Delta$(1232), there is no significant evidence for the
presence of the Roper resonance $N^{*}$(1440) in the y-axis of the $IM^{2}_{p\pi^{0}}$ versus
$IM^{2}_{p\pi^{0}\pi^{0}}$ distribution, as shown in Figure~\ref{final} (upper right plot). This
observation is supported by the one-dimensional projection onto the $IM^{2}_{p\pi^{0}\pi^{0}}$ axis
(lower right plot). 

\section{Conclusion and Outlook}\label{con}
The total as well as the differential cross sections of the $pp\rightarrow pp\pi^{0}\pi^{0}$
reaction at $T_{p}$ = 1400~MeV have been determined and compared to the previous experimental data
and theoretical expectations. The results show a good agreement with the expectations that the
$\pi^{0}\pi^{0}$ production, at the energy studied here, is dominated by $\Delta\Delta$ excitations
as the intermediate state, with no significant contribution from the Roper resonance.\\
The next step is to investigate the production of charged pions ($\pi^{+}\pi^{-}$) in order to
study the complete double pion production mechanisms, of which the vector mesons (e.g.
$\rho^{0}$(770)) are expected to play an important role.

\coltoctitle{\boldmath $\gamma\gamma$ physics with the KLOE experiment}
\authortochead{C.~Taccini}
\label{art26}

\maketitle

The term ``$\gamma\gamma$ physics'' stands for the study of the reaction $e^{+} e^{-} \to e^{+}
e^{-} \gamma^{*} \gamma^{*} \to e^{+} e^{-} X$, where $X$ is some arbitrary final state allowed by
conservation laws.
$\gamma\gamma$ scattering at $e^{+} e^{-}$ colliders can give access to states with
$J^{PC} = 0^{\pm+};2^{\pm+}$, not directly coupled to one photon ($J^{PC} = 1^{--}$).
These processes, of order $\alpha^4$, depend on the logarithm of
the center of mass energy $E$, so that for $E$ greater than a few GeV they dominate hadronic
production at $e^+e^-$ colliders.
For quasi-real photons the number of produced events can be estimated from the expression
\begin{equation}
N = L_{ee} \int dW_{\gamma\gamma}\frac{dL}{dW_{\gamma\gamma}} \sigma(\gamma \gamma \to X)
\end{equation}
where $L_{ee}$ is the $e^+e^-$ luminosity, $W_{\gamma\gamma}$ the photon-photon center of mass
energy ($W_{\gamma\gamma}=M_X$), $dL/dW_{\gamma\gamma}$ the photon-photon flux and $\sigma$ the
cross section into a given final state. By
neglecting single powers of $\mbox{ln}(E/m_e)$, when the scattered leptons are undetected in the
final state one has
\begin{equation}
\frac{dL}{dW_{\gamma\gamma}}=
\frac{1}{W_{\gamma\gamma}}\left(\frac{\alpha}{2\pi}\mbox{ln}\frac{E}{m_e}\right)^2 f(z) \qquad z=\frac{W_{\gamma\gamma}}{2E}
\end{equation}
with $f(z)$ (Low function)~\cite{BKT,Low} given by
\begin{equation}
f(z)= (2+z^2)^2\mbox{ ln}\frac{1}{z}-(1-z^2)(3+z^2) 
\end{equation}
The two $\gamma\gamma$ processes studied with KLOE are $e^{+} e^{-} \to e^{+} e^{-} \eta$ and
$e^{+} e^{-} \to e^{+} e^{-} \pi^0\pi^0$.
DA$\Phi$NE is an $e^+e^-$ collider operating at $\sqrt{s}\simeq M_\phi=$ 1.02 GeV. The KLOE
detector consists of a large volume drift chamber surrounded by a lead and scintillating fibers
calorimeter. Charged particle momenta are reconstructed with resolution $\sigma_p/p\simeq0.4\%$ for
large angle tracks. Calorimeter clusters are reconstructed with energy and time resolution of
$\sigma_E/E=5.7\%/\sqrt{E({\rm GeV})}$ and $\sigma_t=57\mbox{ ps}/\sqrt{E({\rm GeV})}\oplus 100$ ps.
The sample used for the analyses consists of data taken at $\sqrt{s}$ = 1 GeV, which allows
reduction of the background from $\phi$ decays, with an integrated luminosity of 240 pb$^{-1}$.
Data are processed with a dedicated $\gamma\gamma$ filter asking for
at least two photons (prompt clusters not associated to tracks) with energy $E>15$ MeV and polar
angle $20^\circ<\theta<160^\circ$, the most energetic with $E>50$ MeV, the fraction of energy
carried by photons $R=(\sum_\gamma E_\gamma)/E_{cal}>0.3$ and the total energy in the calorimeter
$100<E_{cal}<900$ MeV. The latter requirement rejects low energy background and the high rate
processes $e^{+} e^{-} \to e^{+} e^{-}$, $e^{+} e^{-} \to \gamma \gamma$. \\
A search for the $e^+e^-\to e^+e^-\eta$ process is performed, with $\eta\to\pi^+\pi^-\pi^0$. 
The selection of these events asks for two photons constrained to originate from a $\pi^0$ decay
and two tracks with opposite curvature coming from the collision point.
The charged pion mass is assigned to the two tracks and a least squares function based on Lagrange
multipliers imposes that $\pi^+\pi^-\pi^0$ come from an $\eta$ decay. Therefore most background
events are suppressed, except for the irreducible process
$e^+e^-\to\eta\gamma\to\pi^+\pi^-\pi^0\gamma$, with the monochromatic photon, $E_\gamma$ = 350 MeV,
lost in the beam pipe. For these background events, however, the correlation between the squared
missing mass, $M_{miss}^2$, and the longitudinal momentum, $p_L$, of the $\pi^+\pi^-\pi^0$ system is
rather different than for the signal. Further criteria are applied to suppress processes with
photons and $e^+e^-$ in the final state. 
The MC generator used for the signal~\cite{Nguyen:2006sr} allows the full phase space generation of
$e^+e^-\to e^+e^-\gamma^*\gamma^*\to e^+e^-\eta$ events.
Fig.~\ref{f:MPIPIGGpi+pi-pi0} (left) shows the distribution of $M_{miss}^2$ for data fitted with
the superposition of MC shapes for signal and background. An independent fit is performed with the
distribution of $p_L$. Both fits show the same yields for the background processes and more than 600
signal events. 
The cross section of the irreducible process $e^+e^-\to\eta\gamma\to\pi^+\pi^-\pi^0\gamma$ is
evaluated using the same sample of data and asking for three photons in the final state. A kinematic
fit is performed, requiring energy and momentum conservation, and the improved variables are used to
fit the distribution of the energy of the recoil photon (center plot of
Fig.~\ref{f:MPIPIGGpi+pi-pi0}), which yields $\sigma(e^+
e^-\to\eta\gamma\to\pi^+\pi^-\pi^0\gamma,1\mbox{ GeV}) = (198 \pm 2_{\mbox{stat}})$ pb. Systematics
are under evaluation.
\\
Finally, a search for $e^+e^-\to e^+e^-\pi^0\pi^0$ events is performed, motivated by the interest
in the $\gamma\gamma\to\sigma$ dynamics~\cite{PDG:sigma}. The main requirements of the data analysis
are: four photons originated from 2$\pi^0$ decays, no tracks in the drift chamber, photon energy
fraction $R > 0.75$, transverse momentum ${p_T}_{4\gamma} < 80$ MeV. Other cuts on the photon
energies are applied to improve the signal to background ratio.
The MC used allows for full phase space generation and treats the process $e^+e^-\to
e^+e^-\pi^0\pi^0$ including a $\sigma$ particle as a Breit-Wigner resonance~\cite{Nguyen:2006sr}.
The spectrum in the $4\gamma$ invariant mass compared with the expected backgrounds is shown in
Fig.~\ref{f:MPIPIGGpi+pi-pi0} (right). From the plot, an excess is evident at low $M_{4\gamma}$
values, consistent in shape with expectations from $e^+e^-\to e^+e^-\pi^0\pi^0$ events.
This is the only new measurement of $\gamma\gamma\to\pi^0\pi^0$ reaction in the region of the
$\sigma$ meson after~\cite{CrystalBall}. 
Systematic study of the cross section and understanding of the $\sigma\to\pi^0\pi^0$ contribution
are in progress.
\newline
\begin{figure}[htbp]
\psfrag{Miss}{\hspace*{-2.5cm}\tiny M$_{miss}^2$ (GeV$^2$)}
\psfrag{e3afp}{\tiny $E_\gamma$ (MeV)}
\centering
\includegraphics[width = 4.3cm,height=4.5cm]{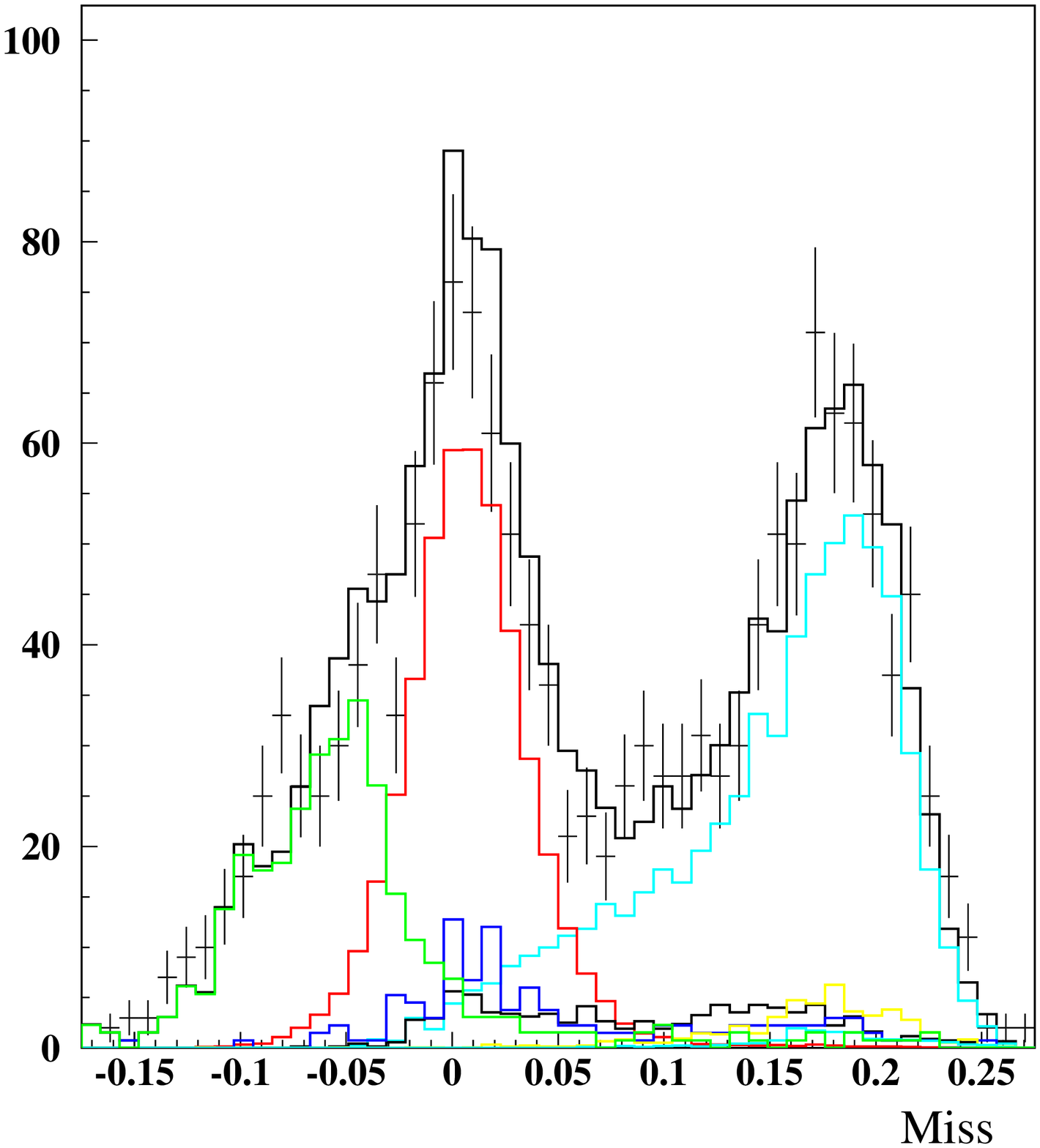}%
\hfill
\includegraphics[width = 4.3cm,height=4.5cm]{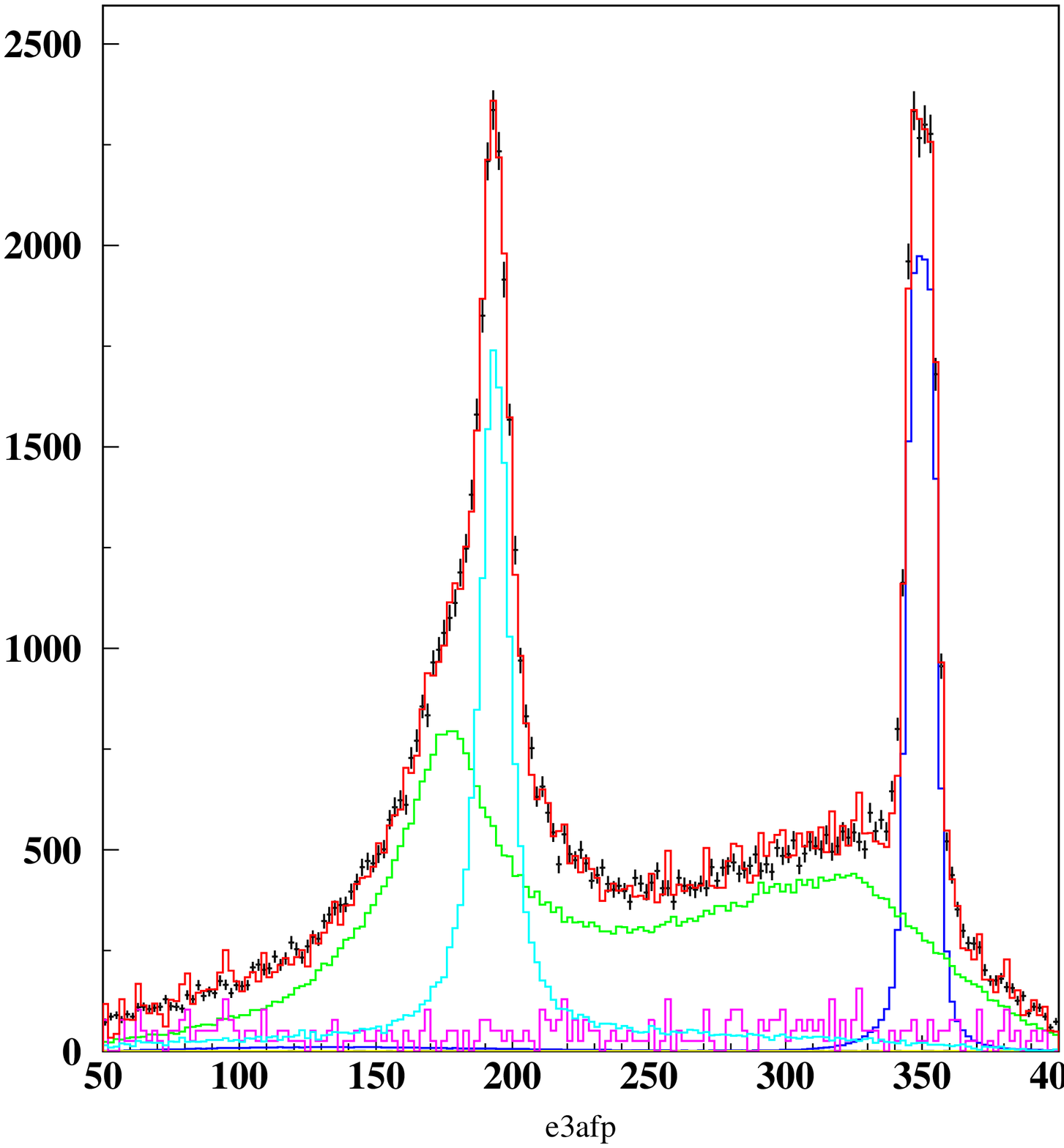}%
\hfill
\includegraphics[width = 4.3cm,height=4.5cm]{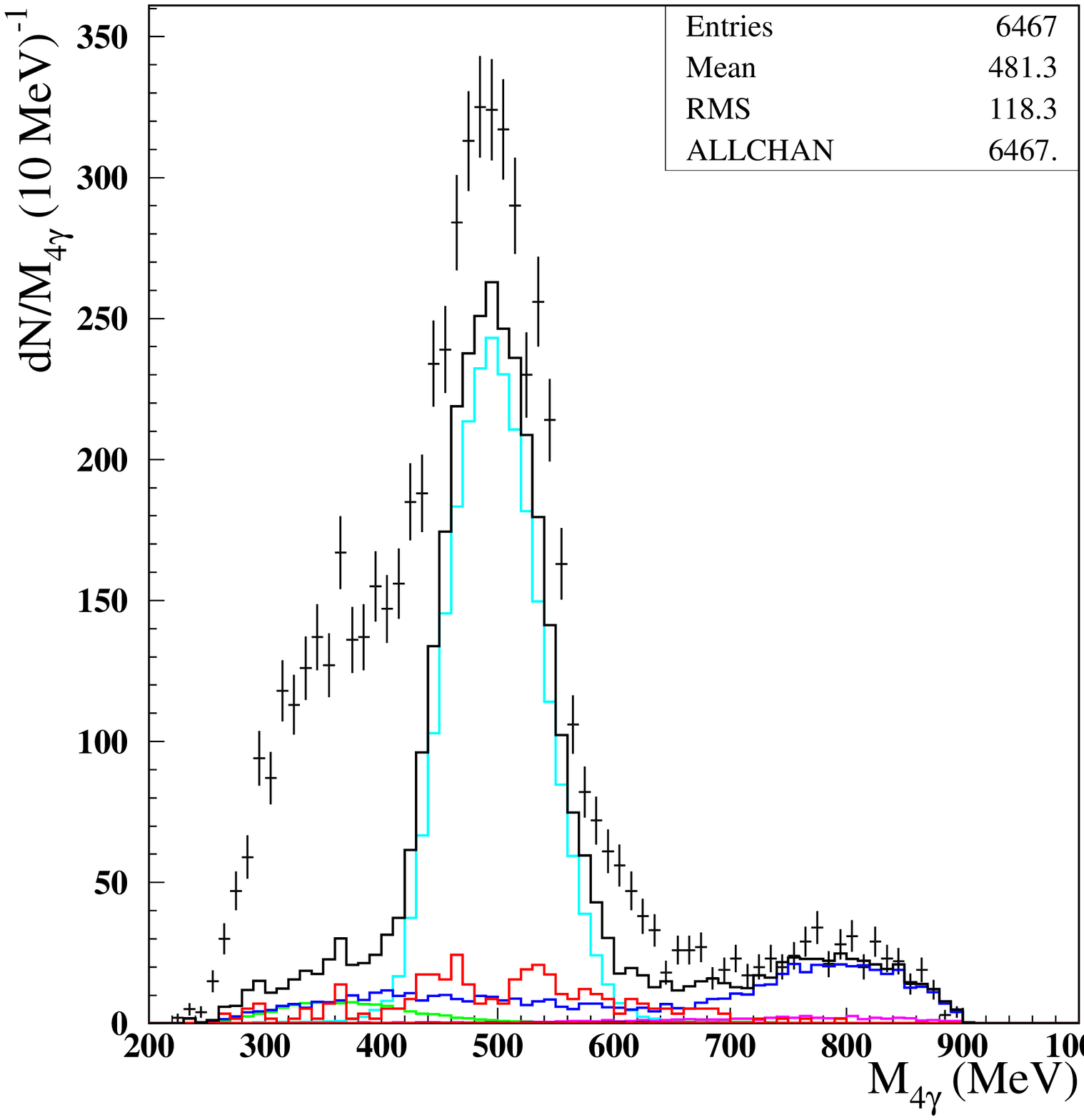}
\caption{Left: fit of the $M_{miss}^2$ distribution for the $e^+e^-\to e^+e^-\eta$ analysis. Main
contributions are: $e^+e^-\gamma$ (green) at negative $M_{miss}^2$ values due to the pion mass
assigned to $e^+e^-$ tracks, $\eta\gamma$ (red), signal events (ligt blue). Middle: fit of the
recoil photon energy spectrum for the $e^+e^-\to\eta\gamma$ analysis. The $\eta\gamma$ peak (blue)
and the $\omega\gamma$ peak (light blue) are visible; the broad distribution (green) is due to
$\omega\pi^0$ events. Right: $M_{4\gamma}$ spectrum for events selected in the $e^+e^-\to
e^+e^-\pi^0\pi^0$ data analysis, compared with the sum of the expected backgrounds from MC. Main
background contributions are: $K_S\to\pi^0\pi^0$ (light blue), $\omega(\to\pi^0\gamma)\pi^0$ (blue)
and $\gamma\gamma$ (red).}
 \label{f:MPIPIGGpi+pi-pi0}
\end{figure}

These results are encouraging in view of the forthcoming data taking campaign of the KLOE-2
project~\cite{KLOE2}, when analyses of the on-peak sample will be possible with the information
coming from two different tagging detectors: the low energy taggers (LET), located in the region
between the two quadrupoles inside KLOE (about 1 m from the IP), and the high energy taggers (HET),
located at the exit of the first bending magnet (about 11 m from the IP). The energy range covered
will be (160-230) MeV for LET and (425-490) MeV for HET.

\coltoctitle{\boldmath A covariant formalism for the $\gamma^\ast N \to N^\ast$ transitions}
\authortochead{G.~Ramalho}
\label{art27}

\maketitle

In the covariant spectator quark model
the $\gamma^\ast N \to B$ transition current 
is determined by the (constituent) quark current and by the
covariant wave functions $\Psi_N$ for the nucleon and $\Psi_B$ 
for the baryon $B$ \cite{ExclusiveR,Nucleon}.
The quark current is parameterized using 
vector meson dominance and 
the wave functions  $\Psi_B$ represented 
in terms of the quark and diquark (quark pair) 
states of flavor, spin, orbital angular momentum 
and radial excitations, corresponding to the quantum numbers 
defining each baryon \cite{ExclusiveR}.

Our goal was to study the role of the 
valence quark degrees of freedom and also 
the effect of the meson cloud effects, 
in particular for the inelastic reactions.
In general, one can decompose the form factor $G_X$
as $G_X= G_X^B + G_X^{mc}$, where 
$G_X^B$ is the bare contribution (from valence quarks)
and $G_X^{mc}$ the meson cloud contribution \cite{ExclusiveR}.

The first application of the model was for the case $B=N$, 
the nucleon elastic form factors,  
where the nucleon was taken as a quark-diquark S-wave system \cite{Nucleon}. 
In that study the quark current and nucleon scalar wave function 
were calibrated by the experimental data,
defining the nucleon wave function and the quark current 
for future applications.
In the nucleon case no pion cloud structure was considered.
The parametrization proved to be consistent 
also with lattice QCD simulations, once 
the model was extended for the lattice QCD 
regime where meson cloud mechanisms are negligible 
\cite{Lattice}.

The next application was for the case $B=\Delta$, 
where the wave function is represented by an admixture 
of the dominant S-wave with two distinct smaller D-wave components 
\cite{NDeltaD,LatticeD}.
The calibration of the $\Delta$ wave function 
was done using lattice QCD data,  and 
an excellent fit was achieved \cite{LatticeD}.
To describe the physical data, 
a phenomenological pion cloud contribution for the 
magnetic dipole transition form factor 
and a pion cloud contribution  for the quadrupole 
form factors 
based on large $N_c$ limit \cite{LatticeD}, 
were added to the extrapolation of lattice to the physical region.
The final result is in agreement with the experimental data  \cite{LatticeD}.

The covariant spectator model can also be applied for the radial excitations 
of the nucleon and $\Delta(1232)$ systems \cite{Roper,Delta1600}.
Because in those cases the wave function 
of the radial excited state is 
defined in terms of the lower order states, nucleon or $\Delta(1232)$, 
there are no additional parameters involved.
For the nucleon radial excitation $N(1440)$, 
also known as Roper resonance, no explicit meson cloud was 
added. The results are consistent with the 
experimental data for large $Q^2$ ($Q^2 >$ 1.5 GeV$^2$) \cite{Roper}.
[See the top of Fig.~\ref{fig1}].
As for the $\Delta(1600)$, 
by considering  
valence quark degrees of freedom only overshoots the data for low $Q^2$.
However, a simple estimation of the pion cloud 
effects based on the analogy between the reaction 
with $\Delta(1232)$ and $\Delta(1600)$, 
makes the results closer to  the data \cite{Delta1600}
[See the bottom of Fig.~\ref{fig1}].

\newpage

Other possible applications of the covariant spectator quark model 
are the determination of the octet and decuplet 
electromagnetic form factors \cite{Octet,Omega,DeltaFF2}.
A reaction of particular interest for the Workshop 
corresponds to the $B=N(1535)$, the  S11 resonance. 
The calculation of the $\gamma^\ast N \to N(1535)$, 
transition form factors is in progress  \cite{S11}.

\begin{figure}
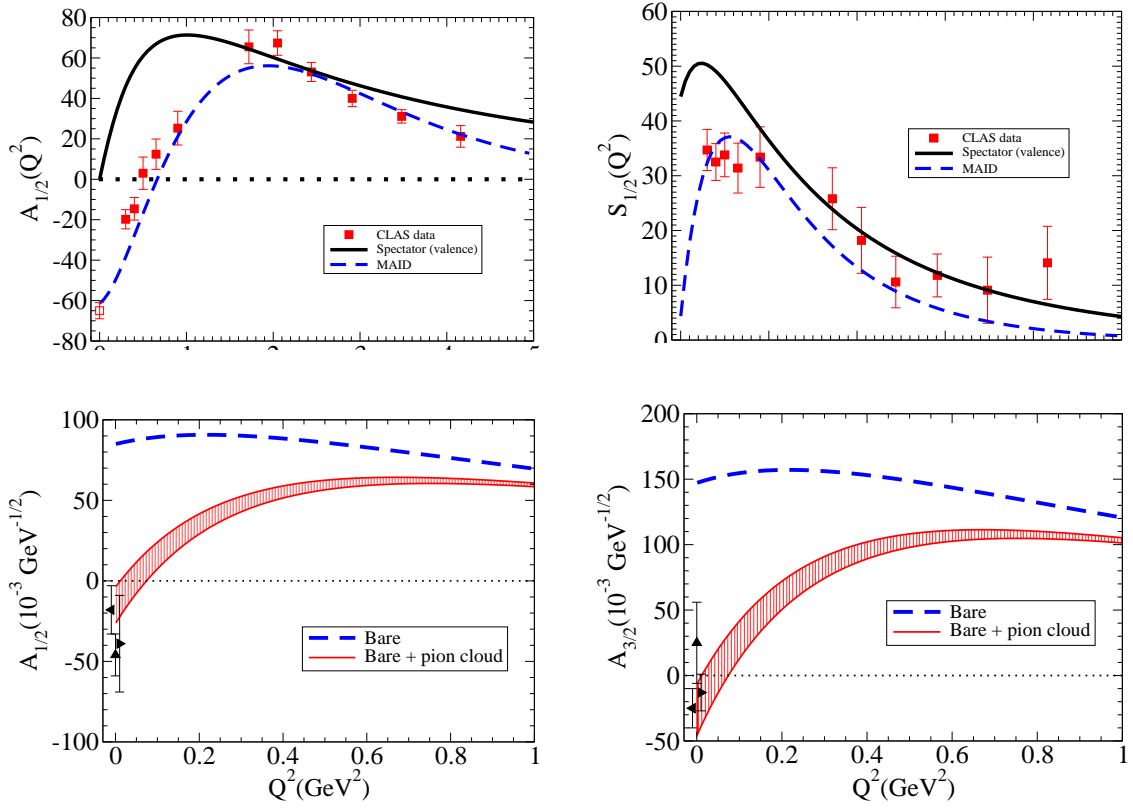

\mbox{
\includegraphics[width=7.0cm]{./RamalhoGilberto/Ramalho1.eps}  \hspace{.5cm}
\includegraphics[width=7.0cm]{./RamalhoGilberto/Ramalho2.eps}
\vspace{.3cm}
}
\mbox{
\vspace{.3cm}
\includegraphics[width=7.0cm]{./RamalhoGilberto/Ramalho3.eps} \hspace{.5cm}
\includegraphics[width=7.0cm]{./RamalhoGilberto/Ramalho4.eps} }
\caption{
Helicity amplitudes for the 
$\gamma^\ast N \to N(1440)$ \cite{Roper} (top);
and $\gamma^\ast N \to \Delta(1600)$ \cite{Delta1600} 
(bottom) reactions, determined with the spectator quark model.
For the $\Delta(1600)$ the pion cloud contribution is also represented. 
}
\label{fig1}
\end{figure}

\coltoctitle{What do we learn from the ABC?}
\authortochead{H.~Clement}
\label{art31}

\setcounter{chapter}{1}
\setcounter{section}{0}
\maketitle

\Section{Introduction}

The ABC effect, an intriguing low-mass enhancement in the $\pi\pi$
invariant mass spectrum, is known from inclusive measurements \cite{abc}
of the production of an isoscalar pion pair in fusion reactions to
light nuclei. Its nature has been a puzzle now for 50 years. 

In an effort to solve this long-standing problem by exclusive and
kinematically complete high-statistics experiments, we have measured the
fusion reactions to {\it d} and $^4$He with WASA at COSY \cite{wasa}. These
measurements cover the full energy region, where the ABC effect has been
observed previously in inclusive reactions. They also complement the systematic
measurements of nucleon-nucleon induced two-pion production
\cite{deldel,iso,Roper,WB,JP,jan,MB,FK,mb,SK}
carried out at CELSIUS-WASA.

\Section{Experimental Results}
From the present data base covering all ${\bf pp}$ induced two-pion production
channels -- including the fusion processes $pp \to d\pi^+\pi^0$ to the deuteron
and to quasi-bound $^2$He -- we find that the $t$-channel Roper, $\Delta\Delta$
and $\Delta(1600)$ excitations are the dominant processes and sufficient for
explaining all data for the two-pion production in {\bf isovector} $NN$
collisions.   

The situation changes dramatically for the {\bf isoscalar} $NN$ channel. The
most 
basic fusion reaction in this channel $pn \to d\pi^0\pi^0$  -- measured as
quasifree process in $pd$ collisions -- exhibits not only a
pronounced ABC effect, {\it i.e.} a low-mass enhancement in the $\pi\pi$
invariant mass spectrum, but in correlation with it also a sharp
Lorentzian-shaped structure in the total cross section (Fig. 1). Its peak
energy is about 80 MeV below 
the nominal $\Delta\Delta$ threshold of 2 $m_{\Delta}$ and its width of only
70 MeV is much less than the width of $2\Gamma_{\Delta} \approx$ 230 MeV
expected from the 
conventional $t$-channel $\Delta\Delta$ process. At the same time the peak
cross section is about five times larger than that of the $t$-channel
$\Delta\Delta$ process.

\begin{figure}
\begin{center}
\includegraphics[width=0.8\textwidth]{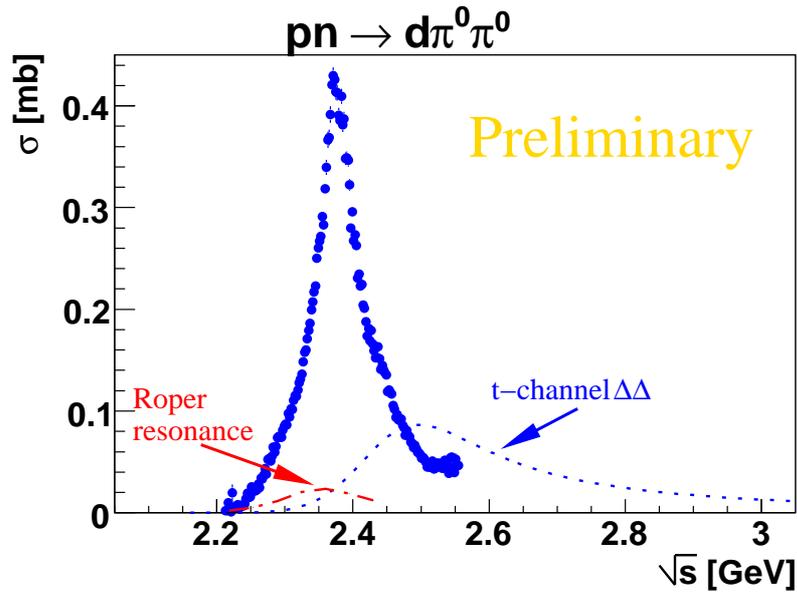}
\caption{Preliminary results \cite{panic,MBmeson10} from WASA-at-COSY
  for the total cross section of  
  the $pn \to d\pi^0\pi^0$ reaction in comparison to the {\bf conventional}
  contributions originating from the excitation of the Roper resonance and of
  the $\Delta\Delta$ system ($t$-channel process).
}
\label{fig1}
\end{center}
\end{figure}

The Dalitz plots of the data exhibit at all energies within this
Lorentzian-shaped structure the ABC effect, {\it i.e.} the low-mass
enhancement in $M_{\pi\pi}$, but simultaneously also the excitation of the
$\Delta\Delta$  system, though this excitation is below the nominal
$\Delta\Delta$ threshold of 2 $m_{\Delta}$. At energies above the
Lorentzian-shaped structure the Dalitz plot 
changes to what is expected from the
conventional $t$-channel $\Delta\Delta$ process.

From the angular distributions we tentatively assign the quantum numbers
$I(J^P) = 0(3^+)$ to this structure in the total cross section. At present no
conventional process is known, which could at least qualitatively explain this
phenomenon. We note, however, that quark-model calculations, notably those of 
Ref. \cite{ping} predict a state with exactly these quantum numbers at about
this mass. However, the estimated width is too large by far. 

In a second experiment at COSY we have measured the double pionic fusion to
$^4$He over the full energy region of the ABC effect by use of the reaction
$dd \to ^4$He$\pi^0\pi^0$ in the range $T_d$ = 0.8 - 1.4 GeV \cite{APmeson10}.

Our findings for the double-pionic fusion to $^4$He are in accordance
with the observations for the basic $pn \to d \pi^0\pi^0$ reaction pointing to
the same fundamental mechanism. In both  
reactions we observe the ABC effect to be correlated with a peak in the 
energy dependence of the total cross section exhibiting a peak cross section
about 90 MeV below the mass of $2m_{\Delta}$ and a width, which is several times
smaller than that of the conventional $\Delta\Delta$ process. If attributed to
a resonance in 
$pn$ and $\Delta\Delta$ systems, then it apparently is robust
enough to survive even in nuclei. In the $^4$He case the peak structure
in the total cross section is considerably broader than in the
deuteron case. This is consistent with both Fermi motion of the basic $pn$
system and collision broadening - as is well-known, {\it e.g.} from
excitations of the $\Delta$ resonance in nuclei. 

\Section {Acknowledgments}
We acknowledge valuable discussions with E. Oset, A. Sibirtsev and C. Wilkin
on this issue. This work has
been supported by BMBF 
(06TU9193), COSY-FFE (Forschungszentrum J\"ulich) and 
DFG (Europ. Graduiertenkolleg 683).

\end{papers}
\cleardoublepage

\pagestyle{plain}
\begin{center}
 \vspace*{0.3\textheight} {\Huge\textbf{\boldmath Interaction of $\eta$ and $\eta^\prime$}} \\
 \vspace{0.5cm} {\Huge\textbf{with Nucleons and Nuclei}} \\
 \vspace{0.7cm} {\Huge\textbf{\boldmath including $\eta$ bound states}}
 \addcontentsline{toc}{part}{\boldmath Interaction of $\eta$ and $\eta^\prime$ with Nucleons and
Nuclei}
\end{center}
\cleardoublepage
\pagestyle{fancy}

\begin{papers}
\coltoctitle{\boldmath Use of $\pi^+ d \to\eta pp$ to Study the $\eta N$ Amplitude Near Threshold}
\authortochead{H.~Garcilazo and M.~T.~Pe\~na}
\label{art1}

\setlength{\parindent}{0pt}
\setlength{\parskip}{10pt}
\setcounter{chapter}{1}
\setcounter{section}{0}
\maketitle

\Section{Introduction}
The $\eta N$ scattering length $a_{\eta N}$ is not well known since it must be
obtained indirectly from the combined analysis of $\pi N - \pi N$ and $\pi N
-\gamma N$ together with the available differential and total  cross sections
data of $\pi N\to\eta N$. Although for the various analyses 
\cite{BATI,GRE1,GRE2,SIBI,GASP} the imaginary part of the $\eta N$ scattering
length comes out quite stable at $\sim$ 0.26 fm, basically as a consequence of
the optical theorem, the values for Re $a_{\eta N}$ run from 0.4 to 1.07 fm. A
recent approach \cite{GATE} using the process $np\to\eta d$ with a three-body 
treatment of the final state concluded that $0.4 \le {\rm Re\,\, } a_{\eta N}
\le 0.6$ fm.

In this study we follow a suggestion by Fujioka and Itahashi \cite{FUJI} who
pointed out that the process $\pi d\to\eta pp$ at incident pion momentum
$p_{lab}=903$ MeV/c ($T_{lab}=776$ MeV) allows for the $\eta$ and one of the 
protons to be at rest in the laboratoy frame and therefore the rescattering
process $\eta p\to\eta p$ near threshold could give an important contribution to
the cross section. The model of $\pi d\to\eta pp$ that we will use includes single- and
double-scattering terms and it has been described in Refs. 
\cite{GATE2,GATE3}. The dominant diagram is the single-scattering term where the
neutron in the deuteron undergoes the elementary process $\pi^+ n\to\eta p$
while the proton remains as spectator. The double-scattering term where a
nucleon is exchanged which leads to the $NN$ final-state-interaction (FSI) will
not be considered here. This term is important only when the relative momentum
of the two final nucleons is close to zero, i.e., when the $\eta$ gets the
maximum energy and the two nucleons recoil with equal momenta so that the $NN$
relative momentum is zero \cite{GATE2}. In the kinematical region of the $\eta
N$ FSI the $\eta$ and one nucleon are left at rest while the other nucleon moves
with maximal energy so that the $NN$ relative momentum is very large.

\Section{Results}
We calculated the diferential cross section $d\sigma/d\Omega$ in the backward
direction ($\theta =180^\circ$) without the $\eta N$ FSI as well as with three
models of the $\eta N$ amplitude that have been proposed in the literature,
namely, Re $a_{\eta N}$ = 0.407 fm \cite{GASP}, Re $a_{\eta N}$ = 0.717 fm
\cite{BATI}, and Re $a_{\eta N}$ = 1.07 fm \cite{GRE2}. The results for the
diffecrential cross section using these four models are, respectively 0.67
$\mu$b, 1.43 $\mu$b, 2.65 $\mu$b, and 4.33 $\mu$b. Thus, in the backward
direction the $\eta N$ FSI can increase the cross section by a factor of 6.5 for
the model with Re $a_{\eta N}$ = 1.07 fm and by factors of 4 and 2 for the
other two models.

As we have shown the $\eta N$ FSI can produce large effects in the $\pi d\to\eta
pp$ cross section in the backward direction at $p_{lab}$ = 903 MeV/c. Therefore,
a measurement of it will allow us to determine the $\eta N$ scattering length.

\coltoctitle{\boldmath Search for $\eta$-mesic nuclei at J-PARC}
\authortochead{H.~Fujioka}
\label{art28}

\setcounter{chapter}{1}
\setcounter{section}{0}
\maketitle

\Section{Introduction}
The J-PARC facility, whose construction was completed recently, supplies secondary particles by
bombarding $30\,\mathrm{GeV}$\footnote{It will be upgraded to $50\,\mathrm{GeV}$ in near future.}
protons accelerated by the main ring onto a production target. Whereas strangeness nuclear physics
with intense kaon beam is one of the main issues, experiments with pion beam are also possible.

Looking back at past experimental search for $\eta$-mesic nuclei, the ($\pi^+$, $p$) reaction was
already investigated by Chrien \textit{et al.}~\cite{Chrien88}, who reported non-observation of any
narrow peak from $\eta$-mesic nuclei. However, we consider the experimental conditions have room for
improvement. The momentum transfer in their measurement was as large as $\sim 200\,\mathrm{MeV}$
because of finite scattering angle they adopted, which can be almost zero by choosing the proper
beam momentum around the magic momentum and detecting scattered particles in very forward direction.
Furthermore, the detection of the decay particles of $\eta$-mesic nuclei will improve the
signal-to-noise ratio drastically, as demonstrated by recent observations by
TAPS/MAMI~\cite{Pfeiffer04} and COSY-GEM~\cite{COSYGEM09}.

\Section{\boldmath Search for $\eta$-mesic nuclei}
\subsection{\boldmath($\pi$, $N$) reaction on nuclei}
Among already approved experiments at J-PARC, the E15 experiment aims to search for
antikaon-nuclear bound states, $K^-pp$, by use of the ($K^-$, $n$) reaction on $^3\mathrm{He}$
target~\cite{E15Prop}. Since its experimental principle can be applied to $\eta$-mesic nuclei
search, the case when we will use its setup with a small modification after the experiment is
discussed here.  For details, please refer to Refs.~\cite{Itahashi_LoI, Fujioka10}.

The E15 experiment will use the K1.8BR beamline, which can provide secondary particles with their
momenta up to $1.1\,\mathrm{GeV}/c$. It should be noted that the magic momentum for the $\eta$-mesic
nuclei with the ($\pi$, $N$) reaction on a heavy nucleus is around $0.7$ -- $1.0\,\mathrm{GeV}/c$,
which depends on the binding energy of an $\eta$ meson. In addition, the scattered neutron and
proton\footnote{As well as the ($K^-$, $n$) reaction, the ($K^-$, $p$) reaction will be measured
simultaneously in the E15 experiment~\cite{P28Prop}.} in forward direction (up to $6^\circ$ degrees)
can be detected by TOF counters. Therefore, the missing-mass spectroscopy for the ($\pi^-$, $n$) or
($\pi^+$, $p$) reaction can be performed there.

As for the coincidence measurements, the E15 experiment is preparing a Cylindrical Detector System
to detect charged particles from a $K^-pp$ system ($K^-pp\to p+\Lambda\to p+p+\pi^-$). An
$\eta$-meson and an nucleon in an $\eta$-mesic nucleus couple strongly with the $N^*(1535)$
resonance, and the $\eta$-mesic nucleus can emit a back-to-back pair of proton and $\pi^-$ in its
decay. An exclusive measurement will be feasible by detecting these charged particles by the
Cylindrical Detector System.

By the way, a theoretical study related with this experimental plan has been developed by Nagahiro
\textit{et al.}~\cite{Nagahiro09} They argue that a possible change of the property of in-medium
$N^*(1535)$ resonances would affect $\eta$-nucleus interaction. In the chiral doublet model, which
assumes the resonance as the chiral partner of the nucleon, their mass gap is expected to reduce as
a function of nuclear density, while the mass of $N^*(1535)$ will not change largely according to
the chiral unitary model.

\Subsection{Pilot experiment with deuteron target}
Even after the coincidence of the decay particles, some of events which do not associate with an
$\eta$ meson will survive as a background, and it might be too huge to observe a peak structure from
$\eta$-mesic nuclei production.

In order to estimate the background level, a pilot experiment using deuteron target is under
consideration. The signal corresponds to the following reaction:
\begin{equation}
\pi^+ + d\to p+p^*(1535) \to p+p+\eta.\label{reaction1}
\end{equation}
It should be emphasized that the final state of this reaction is the same as that of a quasi-free
reaction ($\pi^++\mbox{``}n\mbox{''}\to p+\eta$) with a proton spectator, but the former can be
distinguished from the latter by detecting two protons fast enough ($\gtrsim 200\,\mathrm{MeV}/c$).
The reaction (\ref{reaction1}) is a two-step process: $\pi^+ +\mbox{``}n\mbox{''}\to p+\eta$,
followed by $\eta+\mbox{``}p\mbox{''}\to p^*(1535) \to p+\eta$. The rescattering in the second step
can be seen as low-energy $\eta N$ scattering, which is difficult to investigate experimentally in a
direct way.

It is interesting to compare the cross section of this reaction ($p+p+\eta$ final state) with that
of a different reaction with $p+p+\pi^0$ final state. Since not only $N^*(1535)$ but also other
resonances ($N^*$ and $\Delta^*$) can decay into $p+\pi^0$, this channel corresponds to
``signal+background''. If its cross section is much larger, it may be difficult to observe signals
from $\eta$-mesic nuclei in a coincidence measurement. of the ($\pi$, $N$) reaction on finite
nuclei.

\subsection{\boldmath ($p$, $^3\mathrm{He}$) reaction}
Inspired by the result by the COSY-GEM collaboration~\cite{COSYGEM09}, an experimental plan to
utilize the SKS spectrometer, constructed for hypernuclear spectroscopy, at the K1.8 beamline, and
the ENSTAR detector for detection of decay particles. One of the candidate of the target could be
lithium-6, which can be regarded as an $\alpha$-$d$ cluster, to produce the $^4\mathrm{He}-\eta$
system.

\coltoctitle{\boldmath Search for $\eta$ Bound States in Nuclei}
\authortochead{H.~Machner}
\label{art29}

\setcounter{chapter}{1}
\setcounter{section}{0}
\maketitle

\Section{Introduction}
The existence of bound states between $\eta$ mesons and nuclei via the strong interaction was
predicted by Liu and coworkers \cite{Haider_Liu86}. Their finding that nuclei with mass number
$A>12$ can bind was later extended to even lighter nuclei \cite{Rakityansky96}. While there a a lot
of theoretical predictions concerning the binding energy and the width of such a state, experimental
evidence is scarce. Here we will concentrate on searches performed at COSY in J\"{u}lich, mainly by
the GEM collaboration.

\Section{Recoil Free Production}
It was shown in the study of hypernuclei and pionic atoms that recoil free meson production is
essential to see a signal of the bound state. This means that in a transfer reaction a particle
carries away all the beam momentum and the meson (or hyperon) is produced at rest. Then the meson
wave function has maximal overlap with the nuclear wave function. This condition was not obeyed in
previous searches. The $(d,^3$He) reaction transfers only one nucleon and has large cross section.
On the contrary the cross section for deuteron break up is large and the break-up protons habe the
same magnetic rigidity as the $^3$He ions, thus making the identification of a state almost
impossible. We, therefore, choose the $(p,^3$He) reaction. Although the cross section for the two
nucleon transfer reaction is small, the background is much, much smaller than in the previous case.
In summary we searched for the bound state in a two step process:
\begin{gather}
p+^{27}\text{Al}\to ^3\text{He}+X\label{gat:step1}\\
X\to \pi^-+p+Y.\label{gat:step2}
\end{gather}
Suppose $X=\eta\bigotimes^{25}$Mg. Then the chain $\eta+N\rightleftarrows N^*(1535)\to \pi+N$ will
occur. This is the elementary process of step (\ref{gat:step2}). The $(d,^3$He) ions were detected
at forward angles by the magnetic spectrograph Big Karl \cite{Drochner98} while the proton and pion
were detected by the detector ENSTAR \cite{Betigeri07}. The details of the measurements were given
in \cite{Budzanowski09a}. The data are shown in Fig. \ref{Fig:Binding-spectra} as cross sections in
the laboratory system. A peak was found with
\begin{equation}
BE+i\sigma =(-13.13 \pm 1.64)+i( 4.35\pm 1.27)\text{ MeV}.
\end{equation}
Its significance is 5$\sigma$. The rather small width was attributed by Haider and Liu
\cite{Haider_Liu10} due to an interference between the resonant and the non-resonant process leading
to the same final state.
\begin{figure}[!h]
\begin{center}
\parbox[c]{0.46\textwidth}{
\includegraphics[width=0.35\textwidth]{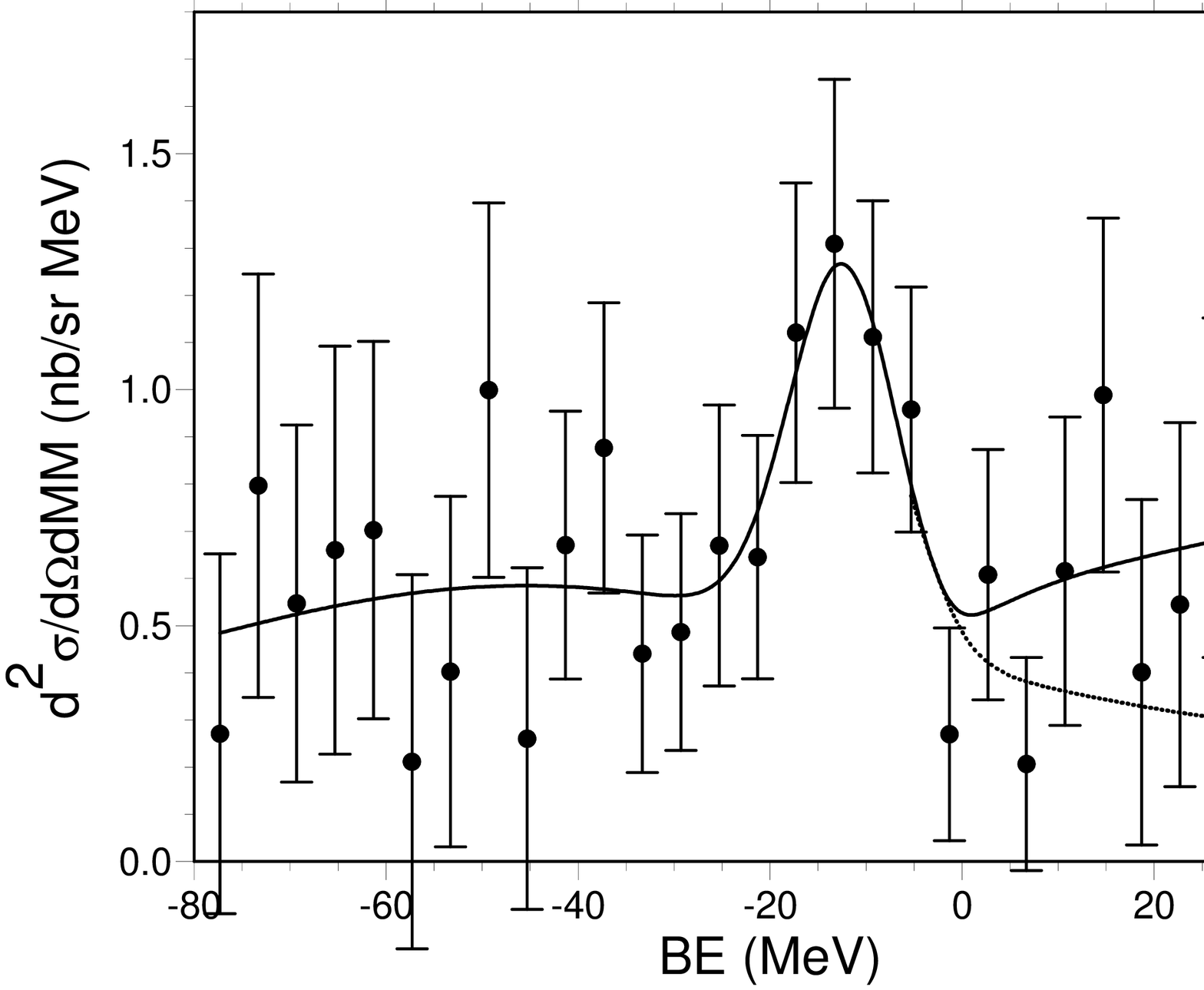}
\caption{Missing mass spectrum converted to binding energy $BE$ of a bound system
$^{25}Mg\otimes\eta$ as measured with the magnetic spectrograph. All cuts and background subtraction
have been applied. The solid curve is a fit with a constant background, two Gaussians and a phase
space behavior for the unbound system. Dotted curve: same as solid curve but without phase space
contribution.\label{Fig:Binding-spectra}}}
\parbox[c]{0.46\textwidth}{
\includegraphics[width=0.45\textwidth]{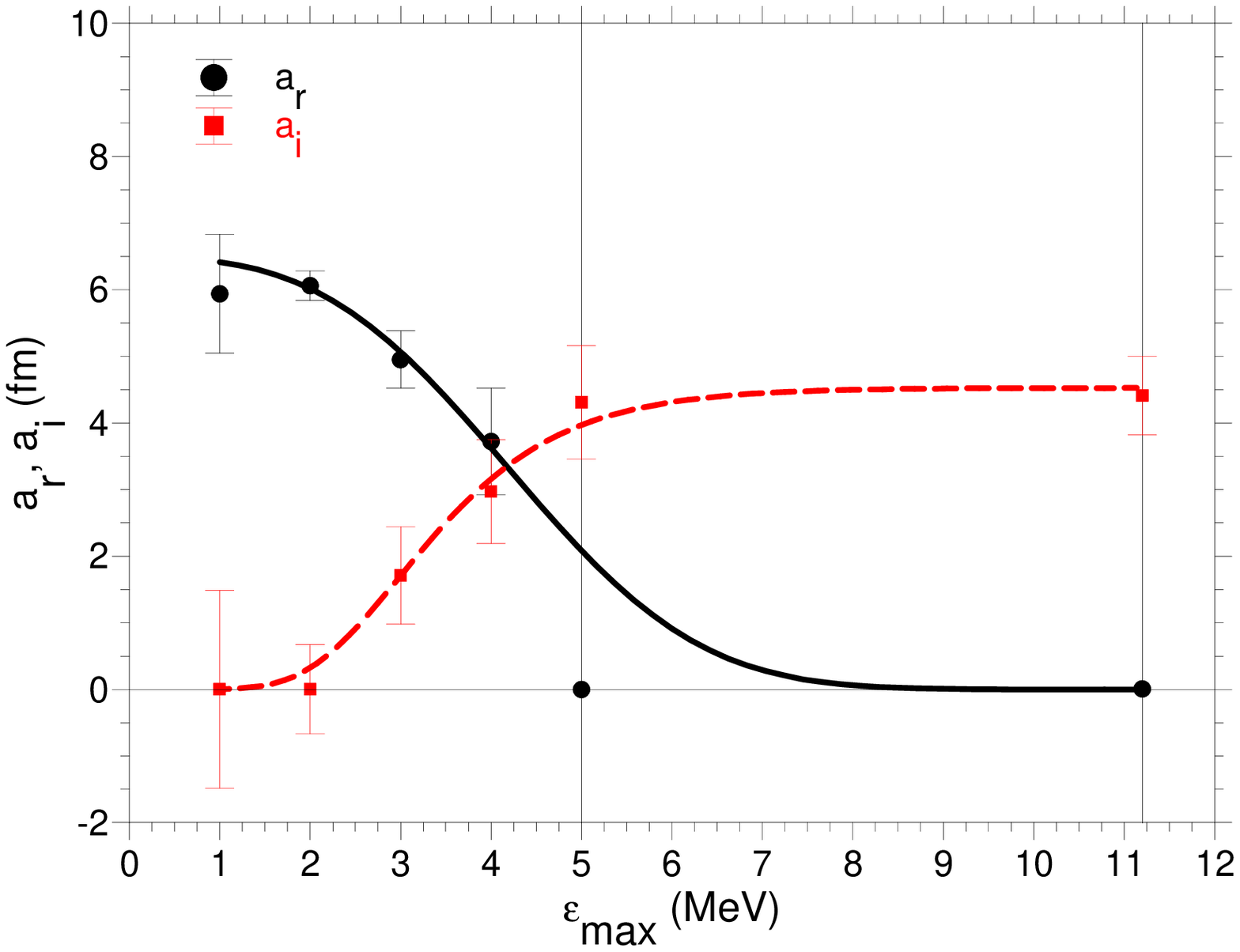}
\caption{Real and imaginary part of the scattering length for the $\eta^3$He interaction as deduced
from the data of ref. \cite{Mersmann07} as function of the energy bin width.}}
\label{fig:scattering_length}
\end{center}
\end{figure}
Another possibility might be that the peak is not the $1s$ state but the $1p$ state. Then the $1s$
state is deeper bound. A fit with two Gaussians and phase space behavior for the unbound part is
also shown in Fig. \ref{Fig:Binding-spectra}. Of course only the narrow peak is statistical
significant.

\Subsection{Search in Two Body Final States}
The usual approach is to factorize the $s$-wave reaction amplitude, $f_s$, near threshold in the
form
\begin{equation}\label{equ:FSI}
f_s=\frac{f_B}{\frac{1}{a}+\frac{r_0}{2}p^2-ip}
\end{equation}
where $p$ is the $\eta$ c.m. momentum, $a$ the complex scattering length and $r_0$ the effective
range. The unperturbed production amplitude $f_B$ is assumed to be slowly varying and is
often taken to be constant in the near-threshold region.

Unitarity demands that the imaginary part of the scattering length be
positive, \emph{i.e.}, $ a_i>0$. In addition, to have binding, there
must be a pole in the negative energy half-plane, which requires
that\cite{Haider_Liu02}
\begin{equation}\label{equ:Cond2}
\left.{|a_i|}\right/{|a_r|}<1\,.
\end{equation}
Finally, in order that the pole lie on the bound- rather than the virtual-state plane, one needs
also $a_r<0$. In a recent experiment \cite{Budzanowski09b} the $dd\to\eta\alpha$ reaction was
studied employing a tensor polarized beam. Therefore, the s-wave amplitude could be extracted. From
these results, assuming $r_0=0$ in Eq. \ref{equ:FSI},  a scattering length could be deduced
\begin{equation}
a_{{}^4He\eta }  = \left[ { \pm \left( {3.1 \pm 0.5} \right) + i \cdot \left( {0.0 \pm 0.5}
\right)} \right]{\rm{fm}}.
\end{equation}
yielding a pole position
\begin{equation}
|Q_{^4He\eta }|\approx 4 \text{MeV}.
\end{equation}
We now want to compare this result with the scattering length for the $pd\to\eta ^3$He reaction.
Although there are high precise data close to threshold \cite{Smyrski07}, \cite{Mersmann07} their
results differ to to different analysis methods. In addition both experiments found a $s-p$
interference close to threshold. Therefore, we have reanalyzed the data from \cite{Mersmann07}
applying the same formalism as in the $dd\to\eta\alpha$ case. In order to study where the onset of
the interference starts the analysis is performed by truncating the data above $\epsilon_{max}$. The
results are shown in Fig. \ref{fig:scattering_length}. For $\epsilon_{max}> 2$ MeV the real part
starts to decrease while the imaginary part stars to increase. Therefore, we stick to the range
below and have
\begin{equation}
a_{{}^3He\eta }  = \left[ { \pm \left( {6.06 \pm 0.22} \right) + i \cdot \left( {0.0 \pm 0.7}
\right)} \right]{\rm{fm}}
\end{equation}
quite similar to the $dd\to\eta\alpha$ case. This gives
\begin{equation}
|Q_{^3He\eta }|\approx 1 \text{MeV}.
\end{equation}
These above-threshold data are, of course,
insensitive to the sign of $a_r$ so that they could never tell whether the system is quasi-bound or
virtual. Since $|Q_{^3He\eta }|<|Q_{^4He\eta }|$ indicates that $^3$He$\eta $ is indeed quasi-bound.

GEM has the extended the study to heavier systems. In the $p^6$Li$\to\eta^7$Be reaction the
recoiling Be was measured at 11.3 MeV above threshold \cite{Budzanowski10a}. A differential cross
section of $\frac{d\sigma}{d\Omega}=(0.69\pm{0.20}\text{( stat.)}\pm 0.20\text{( syst.)})\text{
nb/sr}$ for the ground state plus the 1/2$^-$ was measured.
\begin{SCfigure}[1][h]
\label{exfu}
\includegraphics[width=7cm]{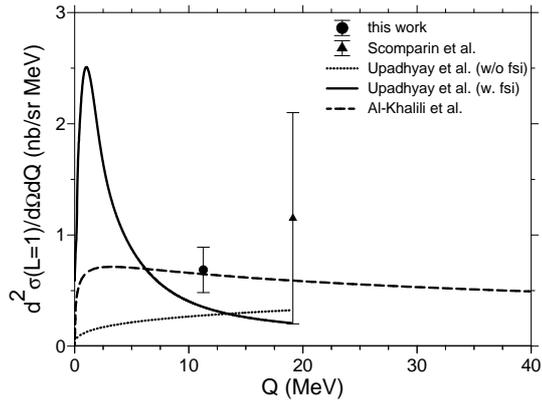}
\caption{Excitation function of the reaction $p+^6\text{Li}\to\eta+^7\text{Be}$ with Be in its
ground state and first excited state. The data are the GEM measurement \cite{Budzanowski10a} (full
dot) and the for these states corrected result from \cite{Scomparin93} (triangle). The calculations
based on the model of \cite{Khalili93} are shown as dashed curve. Those performed by
\cite{Upadhyay09} for the total cross section were divided by 4$\pi$. The calculation with a strong
fsi is shown as solid curve while the one without fsi is shown as dotted curve.}
\end{SCfigure}
Assuming isotropic emission yield a total cross section shown in Fig. \ref{exfu} and compared with
a previous measurement and two model calculations. The quality of the data is not sufficient to
extract a scattering length. More measurements closer to threshold are necessary.

\coltoctitle{\boldmath Search for $\eta$-mesic ${^4\mbox{He}}$ with WASA-at-COSY}
\authortochead{W.~Krzemie\'n}
\label{art30}

\setcounter{chapter}{1}
\setcounter{section}{0}
\maketitle

\Section{Introduction}
Recently, the progress in the spectroscopy of pionic and kaonic atoms, as well as pionic and kaonic
nuclei has permitted to obtain deeper insights on the meson-nucleus interaction and the in-medium
behaviour of spontaneous chiral symmetry breaking~\cite{hiren}. 

Analogically to the other exotic nuclear systems, the investigation of the $\eta$-mesic nuclei
would provide many interesting informations about the $\eta$-N interaction, N* in-medium
properties~\cite{jido}  and would deepen our knowledge of the fundamental structure of $\eta$
meson~\cite{basssymposium}. The $\eta$ meson is electrically neutral, therefore such a system can be
formed only via the strong interaction which distinguishes it qualitatively from pionic atoms where
the binding is the effect of the sum of the attractive electromagnetic force and the repulsive
strong interaction. 

The search of the $\eta$~-~mesic nucleus was performed in many experiments in the
past~\cite{lampf,lpi,jinr,gsi,gem,mami} and is being continued at COSY~\cite{moskalsymposium},
JINR~\cite{jinr}, J-PARC~\cite{fujiokasymposium} and MAMI~\cite{mami}. Many promising indications
where reported, however, so far there is no direct experimental confirmation of the existence of
mesic nucleus. In the region of the light nuclei systems such as $\eta$-$\mbox{He}$ or
$\eta$-$\mbox{T}$, the observation of the strong enhancement in the total cross-section and the
phase variation in the close-to-threshold region provided strong evidence to the hypothesis of the
existence of a pole in the scattering matrix which can correspond to the bound state~\cite{wilkin}. 
However, as it was stated by Liu~\cite{liu3}, the theoretical predictions of width and binding
energy of the $\eta$-mesic nuclei is strongly dependent on the not well known  subtreshold
$\eta$-nucleon interaction. Therefore, the direct measurements which could confirm the existence of
the bound state, are mandatory.

\Section{Experiment}
In June 2008 we performed a search for the $^4\mbox{He}-\eta$ bound state by measuring the
excitation fun\-ction of the $dd \rightarrow ^3\mbox{He} p\pi^-$  reaction near the eta production
threshold using the WASA-at-COSY detector. During the experimental run the momentum of the deuteron
beam was varied continuously within each acceleration cycle from  2.185~GeV/c to 2.400~GeV/c,
crossing the kinematic threshold for the $\eta$ production in the $dd \rightarrow
{^4\mbox{He}}\,\eta$ reaction at 2.336~GeV/c. This range of beam momenta corresponds to the
variation of $^4\mbox{He}-\eta$  excess energy  from -51.4~MeV to 22~MeV. The experimental method is
based on measuring the excitation function for chosen decay channels of the ${^4\mbox{He}}-\eta$
system and a search for a resonance-like structure below the ${^4\mbox{He}}-\eta$ threshold.  The
relative angle between the outgoing $p - \pi^-$ pair which originates from the decay of the N*(1535)
resonance created via  absorption of the $\eta$ meson on a nucleon in the $^4\mbox{He}$ nucleus, is
180$^{\circ}$ in the N* reference frame and is smeared by about 30$^{\circ}$ in the center-of-mass
frame due to the Fermi motion of the nucleons inside the $^4\mbox{He}$ nucleus. The center-of-mass
kinetic energies of the $p$ and $\pi^- $ originate from the mass difference $m_{\eta}-m_{\pi}$ and
are around 50 MeV and 350 MeV, respectively.

The Figure \ref{openangle_cut_div_p} presents the preliminary excitation function in 20 degrees
intervals in $\Theta_{cm}(p-\pi)$ angle. 
\begin{figure}[h]
    \begin{center}
        {\includegraphics[scale=0.290]{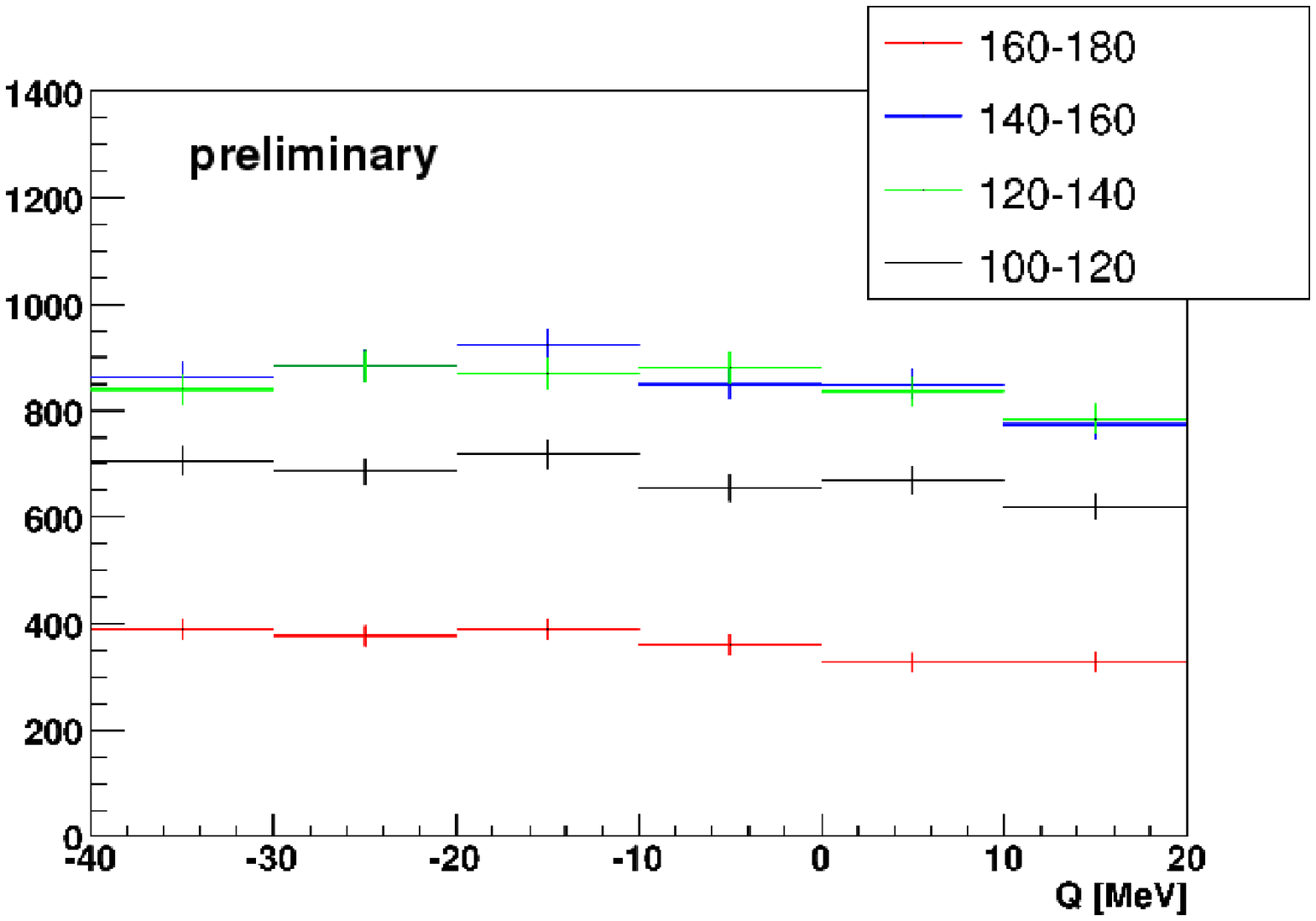}} \hfill
        {\includegraphics[scale=0.300]{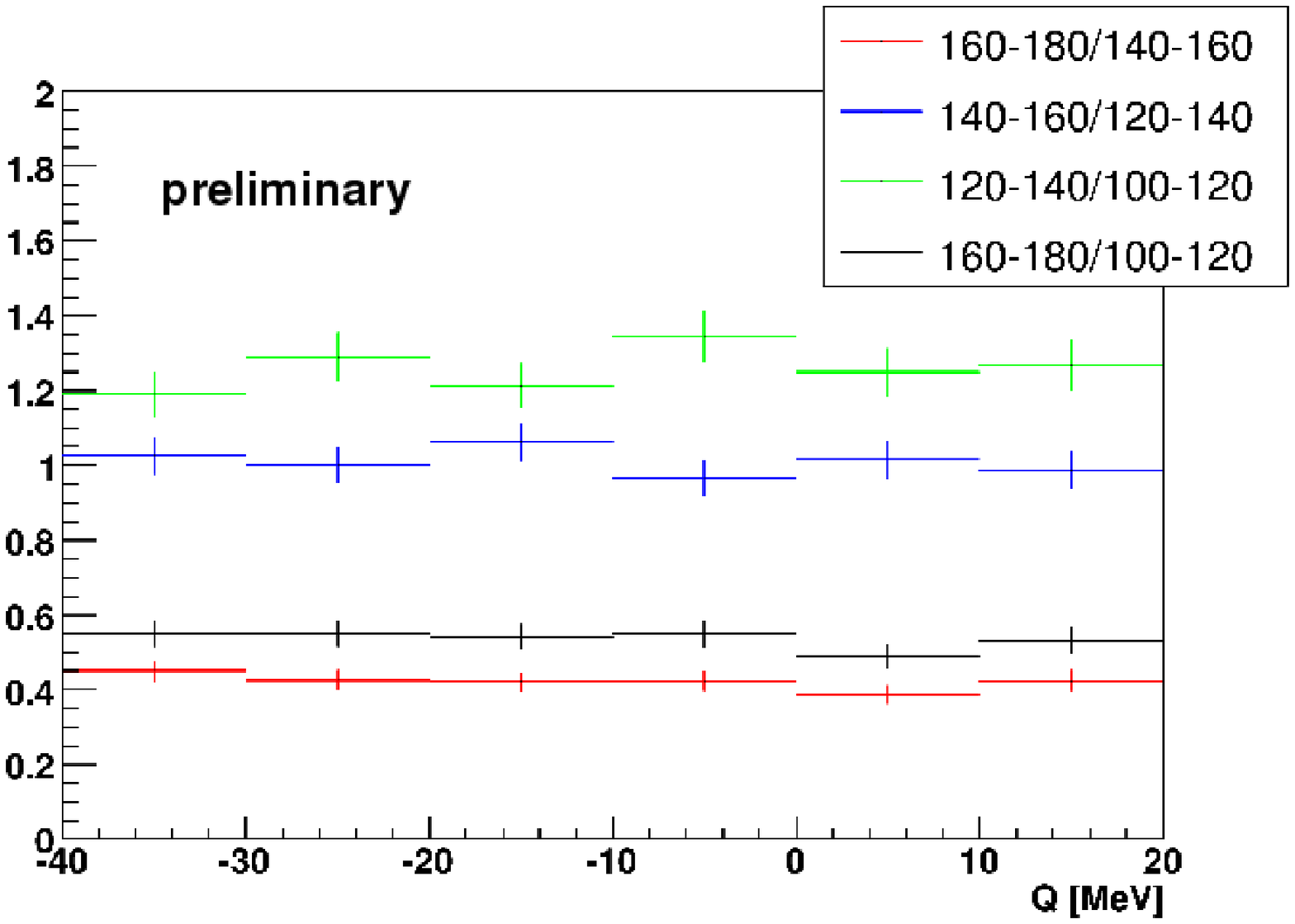}} 
        \caption{(left) Excitation function for the  $dd\to{^3\mbox{He}}p\pi$ reaction measured in
                 the 20 degrees intervals of the  $\Theta_{cm}(p-\pi)$ angle. (Right) Ratio of
                 excitation functions angular ranges as indicated in the figure. 
        \label{openangle_cut_div_p}} 
    \end{center} 
\end{figure} 

The figure indicates no structure in the angular range close to the 180 degree where the signal is
expected. The ratio of excitation functions from various angular ranges is also constant as
indicated in the right panel of Fig.~\ref{openangle_cut_div_p}. Therefore, taking into account the
luminosity and the detector acceptance we preliminary estimated that an upper limit for the
$\eta$-mesic  production via the $dd \to (\eta\,{^4\mbox{He}})_{b.s.} \to {^3\mbox{He}}\, p\, \pi^-$
reaction is equal to  about 20~nb on a one $\sigma$ level.

\Section{Outlook}
The research program of $\eta$-mesic $\mbox{He}$ with WASA-at-COSY will be continued. A two-week
measurement with WASA-at-COSY for the $dd \rightarrow {^3\mbox{He}} p \pi^-$ channel is scheduled
for November 2010. After two weeks of measurement with a luminosity of $4 \cdot 10^{30}$ cm$^{-2}$
s$^{-1}$, we expect a statistical sensitivity of a few nb ($\sigma$). A non-observation of this
signal would significantly lower the upper limit for the existence of the bound state.

\end{papers}
\cleardoublepage
\pagestyle{plain}

\section*{List of Participants}\addcontentsline{toc}{part}{List of Participants}
 \begin{center}
   \begin{tabular}{l|l|c}
     Name                   &      Affiliation                                & Contribution \\
    \hline
   Adlarson, Patrick      &      Uppsala University                         & Page~\pageref{art24}\\
   Bashkanov, Mikhail     &      University Tuebingen                       &                     \\
   Berlowski, Marcin      &      Soltan Institute Nuclear Studies           &                     \\
   Bijnens, Johan         &      Lund University                            &                     \\
   Chatterjee, Banhi      &      Forschungszentrum Juelich                  &                     \\
   Clement, Heinz         &      Physikalisches Institut, Univ. Tuebingen   & Page~\pageref{art31}\\
   Coderre, Daniel        &      Forschungszentrum Juelich                  &                     \\
   Colangelo, Gilberto    &      University of Bern                         & Page~\pageref{art16}\\
   Czerwinski, Eryk       &      Jagiellonian University Cracow             & Page~\pageref{art12}\\
   Denig, Achim           &      University Mainz                           & Page~\pageref{art10}\\
   Ditsche, Christoph     &      HISKP, Bonn University                     &                     \\
   Escribano, Rafel       &      Universitat Aut\`{o}noma Barcelona         &                     \\
   Fransson, Kjell        &      Uppsala University                         &                     \\
   Froehlich, Ingo        &      Goethe University Frankfurt                &                     \\
   Fujioka, Hiroyuki      &      Kyoto University                           & Page~\pageref{art28}\\
   Garcilazo, Humberto    &      Instituto Politecnico Nacional             & Page~\pageref{art1} \\
   Gauzzi, Paolo          &      Sapienza Universit\'{a} di Roma / INFN     &                     \\
   Giovannella, Simona    &      INFN Frascati                              & Page~\pageref{art5} \\
   Goldenbaum, Frank      &      Forschungszentrum Juelich                  &                     \\
   Hodana, Malgorzata     &      Jagiellonian University Cracow             &                     \\
   H\"{o}istad, Bo        &      Uppsala University                         &                     \\
   Jany, Anna             &      Jagiellonian University Cracow             &                     \\
   Jany, Benedykt         &      Jagiellonian University Cracow             &                     \\
   Kampf, Karol           &      Lund University                            & Page~\pageref{art14}\\
   Klaja, Joanna          &      Forschungszentrum Juelich                  & Page~\pageref{art21}\\
   Klaja, Pawel           &      Forschungszentrum Juelich                  & Page~\pageref{art11}\\
   Krusche, Bernd         &      University of Basel                        & Page~\pageref{art3} \\
   Krzemien, Wojciech     &      Jagiellonian University Cracow             & Page~\pageref{art30}\\
   Kubis, Bastian         &      HISKP, Bonn University                     & Page~\pageref{art18}\\
   Lanz, Stefan           &      Bern University                            & Page~\pageref{art17}\\
   Leupold, Stefan        &      Uppsala University                         & Page~\pageref{art19}\\
   Machner, Hartmut       &      Forschungszentrum Juelich                  & Page~\pageref{art29}\\
   Masjuan, Pere          &      University Vienna                          & Page~\pageref{art9} \\
   Nakayama, Kanzo        &      University of Georgia                      & Page~\pageref{art13}\\
   Ostrick, Michael       &      University Mainz                           &                     \\
   Passeri, Antonio       &      INFN Roma                                  & presentation only   \\
   Pe\~{n}a, Teresa       &      Instituto Superior T\'{e}cnico             &                     \\
   Petri, Thimo           &      Forschungszentrum Juelich                  & Page~\pageref{art20}\\
   Pricking, Annette      &      Physikalisches Institut, Univ. Tuebingen   &                     \\
   Ramalho, Gilberto      &      Instituto Superior T\'{e}cnico             & Page~\pageref{art27}\\
   Ramstein, Beatrice     &      Institut de Physique Nucleaire d Orsay     & Page~\pageref{art7} \\
   Redmer, Christoph      &      Uppsala University                         & Page~\pageref{art25}\\
   Rio, Elena Perez del   &      Physikalisches Institut, Univ. Tuebingen   &                     \\
   Roy, Ankhi             &      Indian Institute of Technology Indore      &                     \\
   Schadmand, Susan       &      Forschungszentrum Juelich                  &                     \\
   Schneider, Sebastian   &      HISKP, Bonn University                     & Page~\pageref{art15}\\
   Sch\"{o}nning, Karin   &      Uppsala University                         & Page~\pageref{art22}\\
   Starostin, Aleksandr   &      University of California                   & Page~\pageref{art8} \\
   Stepaniak, Joanna      &      Soltan Institute Nuclear Studies           &                     \\
   Taccini, Cecilia       &      INFN Roma                                  & Page~\pageref{art26}\\
   Tengblad, Ulla         &      Uppsala University                         &                     \\
   Tolba, Tamer           &      Forschungszentrum Juelich                  & Page~\pageref{art23}\\
   Unverzagt, Marc        &      University of Mainz                        & Page~\pageref{art4} \\
   Wilkin, Colin          &      University College London                  & Page~\pageref{art2} \\
   Wirzba, Andreas        &      Forschungszentrum Juelich                  &                     \\
   Wurm, Patrick          &      Forschungszentrum Juelich                  &                     \\
   Zdr\'{a}hal, Martin    &      Charles University Prague                  &                     \\
   \hline
  \end{tabular}
\end{center}
\cleardoublepage
\end{document}